\documentclass[11pt,oneside,english]{amsart}
\usepackage[LGR,T1]{fontenc}
\usepackage[latin9]{inputenc}
\usepackage{geometry}
\usepackage{amstext}
\usepackage{amsthm}
\usepackage{amssymb}

\makeatletter

\DeclareRobustCommand{\greektext}{%
  \fontencoding{LGR}\selectfont\def\encodingdefault{LGR}}
\DeclareRobustCommand{\textgreek}[1]{\leavevmode{\greektext #1}}
\ProvideTextCommand{\~}{LGR}[1]{\char126#1}

\numberwithin{equation}{section}
\numberwithin{figure}{section}
\theoremstyle{plain}
\newtheorem{thm}{\protect\theoremname}[section]
\theoremstyle{plain}
\newtheorem{cor}[thm]{\protect\corollaryname}
\theoremstyle{remark}
\newtheorem{rem}[thm]{\protect\remarkname}
\theoremstyle{plain}
\newtheorem{prop}[thm]{\protect\propositionname}
\theoremstyle{plain}
\newtheorem{lem}[thm]{\protect\lemmaname}
\theoremstyle{definition}
\newtheorem{example}[thm]{\protect\examplename}
\theoremstyle{definition}
\newtheorem{defn}[thm]{\protect\definitionname}

\def\makebbb#1{
    \expandafter\gdef\csname#1\endcsname{
        \ensuremath{\Bbb{#1}}}
}\makebbb{R}\makebbb{N}\makebbb{Z}\makebbb{C}\makebbb{H}\makebbb{E}\makebbb{H}\makebbb{P}\makebbb{B}\makebbb{Q}\makebbb{E}

\usepackage{babel}

\usepackage{babel}

\makeatother

\usepackage{babel}

\makeatother

\usepackage{babel}
\providecommand{\corollaryname}{Corollary}
\providecommand{\definitionname}{Definition}
\providecommand{\examplename}{Example}
\providecommand{\lemmaname}{Lemma}
\providecommand{\propositionname}{Proposition}
\providecommand{\remarkname}{Remark}
\providecommand{\theoremname}{Theorem}

\begin{document}
\title{Priors leading to well-behaved Coulomb and Riesz gases versus zeroth-order
phase transitions - a potential-theoretic characterization }
\author{Robert J. Berman}
\begin{abstract}
We give a potential-theoretic characterization of measures $\mu_{0}$
which have the property that the Coulomb gas, defined with respect
to the prior $\mu_{0},$ is ``well-behaved'' and similarly for more
general Riesz gases. This means that the laws of the empirical measures
of the corresponding random point process satisfy a Large Deviation
Principle with a rate functional which depends continuously on the
temperature, in the sense of Gamma-convergence. Equivalently, there
is no zeroth-order phase transition at zero temperature, in the mean
field regime. This is shown to be the case for the Hausdorff measure
on a compact Lipschitz hypersurface, as well as Lesbesgue measure
on a bounded Lipschitz domain. We also provide constructions of priors
$\mu_{0},$ absolutely continuous with respect to Lebesgue measure
on a smoothly bounded domain, such that the corresponding 2d Coulomb
exhibits a zeroth-order phase transition. This is based on relations
to Ullman's criterion in the theory of orthogonal polynomials and
Bernstein-Markov inequalities. 
\end{abstract}

\maketitle

\section{Introduction}

In broad terms, the main aim of the present work is to study the interplay
between fine potential-theoretic properties of a measure $\mu_{0}$
in $\R^{d}$ (the ``prior'') and properties of the corresponding
Coulomb gas in $\R^{d},$ in the mean-field regime. The pair-interaction
potential of this gas (also known as the one-component plasma in the
physics literature) is the fundemental solution of the Laplacian $\Delta$
in $\R^{d}$ and yields a family of random point processes on the
support of $\mu_{0}$ in $\R^{d},$ parametrized by temperatures $T_{N}\in]0,\infty].$
We will, in particular, give a potential-theoretic characterization
of measures $\mu_{0}$ for which the corresponding Coulomb gas is
``well-behaved at zero-temperature'', which is equivalent to the
absence of a zeroth-order phase transition. In fact, the main results
will be shown to hold in a more general setting involving a Riesz
gas in $\R^{d},$ where the role of the Laplacian is played by the
fractional Laplacian $-(-\Delta)^{\alpha/2},$ assuming that $\alpha\in]0,2].$
Complex-geometric analogs of the results, where the role of the Laplacian
is played by the complex Monge-Ampère operator, are described in \cite{berm12}.
Our results yield new probabilistic constructions of capacities, equilibrium
measures etc, using random point processes, in contrast to the usual
probabilistic approach based on Markov processes (and their hitting
probabilities \cite{ge,ch}). More precisely, the setting of a Riesz
gas with $\alpha\in]0,2]$ corresponds precisely to the class of symmetric
stable Levy processes in $\R^{d}$ \cite{bo} (i.e. Brownian motion
in the ``Coulomb case'' $\alpha=2).$ 

For concreteness we will introduce the main results in the Coulomb
case ($\alpha=2),$ where the \emph{energy} $E(\mu)$ of a compactly
supported measure $\mu$ in $\R^{d},$ with $d\geq2,$ is defined
by 
\[
E(\mu)=\frac{1}{2}\int_{\R^{d}}W\mu\otimes\mu,
\]
 where $W(x,y)$ denotes the standard Green function of the Laplacian
$\Delta,$ i.e. $W(x,y)$ is proportional to $|x-y|^{2-d}$ when $d\geq3$
and to $-\log|x-y|$ when $d=2.$ The\emph{ potential} $\psi_{\mu}$
of $\mu$ is the subharmonic function on $\R^{d}$ defined by 
\[
\psi_{\mu}(x):=-\int_{\R^{d}}W(x,y)\mu(y)
\]
 (using the opposite sign convention compared to the standard convention
in physics). A bounded subset $S$ of is said to be \emph{polar} if
there exists a potential $\psi_{\mu}$ such that $S\Subset\{\psi_{\mu}=-\infty\}.$
We will be particularly interested in measures $\mu_{0}$ not charging
polar subsets (for example, this is the case if $\mu_{0}$ has finite
energy or if $\mu_{0}$ is absolutely continuous wrt Lebesgue measure).
We will denote by $\mathcal{P}(S)$ the space of all probability measures
on a closed subset $S\subset\R^{d},$ endowed with the weak topology. 

\subsection{\label{subsec:Energy-approximation-and}Energy approximation and
determining measures}

The main analytical result may be formulated in terms of potential
theory and approximation theory as follows. Assume given a measure
$\mu_{0}$ on $\R^{d}$ and denote by $S_{0}$ its support. We will
say that $\mu_{0}$ has the \emph{Energy Approximation  Property}
if for any measure $\mu$ supported on $S_{0}$ there exists a sequence
$\mu_{j}$ converging weakly towards $\mu$ such that 
\begin{itemize}
\item $\mu_{j}$ is absolutely continuous with respect to $\mu_{0}$
\item $\lim_{j\rightarrow\infty}E(\mu_{j})=E(\mu)$ 
\end{itemize}
Note that, by the lower semi-continuity of $W,$ the second point
is equivalent to 
\begin{equation}
\limsup_{j\rightarrow\infty}E(\mu_{j})\leq E(\mu)\label{eq:bound on limsup E}
\end{equation}
Theorem \ref{thm:EA vs det intr} below  relates the Energy Approximation
Property to the potential-theoretic notion of determining measures\emph{.
}Given a weighted set $(S,\phi)$ consisting of a subset $S$ of $\R^{d}$
and a continuous function $\phi$ on $S,$ a measure $\nu$ on $\R^{d}$
is said to be \emph{determining for $(S,\phi)$ }if for all potentials
$\psi$ on $\R^{d}$ 

\begin{equation}
\psi\leq\phi\,\,\text{almost everywhere wrt }\nu\implies\psi\leq\phi\,\,\text{on\,\ensuremath{S}}\label{eq:det intro}
\end{equation}
We will say that $\nu$ is\emph{ determining for $S$} if $\nu$ is
determining for $(S,0)$ and\emph{ strongly determining for $S$}
if $\nu$ is determining for $(S,\phi)$ for all $\phi\in C(S).$
Similarly we will say that $\nu$ is (strongly) determining if it
is (strongly) determining for its support. For example, Lebesgue measure
$1_{\Omega}dx$ on a bounded domain $\Omega$ in $\R^{d}$ is strongly
determining if $\Omega$ is\emph{ non-thin} at all boundary points,
in the classical sense (Prop \ref{prop:top domain}). In general,
if $\nu$ is strongly determining and does not charge polar subsets
then the support of $\nu$ is automatically\emph{ locally regular
}(see Section \ref{subsec:Regularity} for the potential-theoretic
notions of regularity). 
\begin{thm}
\label{thm:EA vs det intr}Let $\mu_{0}$ be a measure on $\R^{d}$
which does not charge polar subsets and assume that the support $S_{0}$
of $\mu_{0}$ is compact and locally regular. Then $\mu_{0}$ has
the Energy Approximation Property iff $\mu_{0}$ is strongly determining. 
\end{thm}

The main virtue of the property of beeing strongly determining is
that it can often be verified using maximum (/domination) principle
type arguments. For example, using \cite{da}, we will show that the
$(d-1)-$dimensional Hausdorff measure on a Lipschitz hypersurface
is strongly determining (Theorem \ref{thm:Lip}). 

Theorem \ref{thm:EA vs det intr} appears to be new even in the simplest
case when $S_{0}$ is an interval in $\R\subset\R^{2},$ which is
the classical setting where the notion of determining measures was
first introduced by Ullman (as discussed in Section \ref{subsec:Relations-to-Bernstein-Markov}
below). In this special case a measure $\mu_{0}$ is determining iff
it is strongly determining. 

\subsection{The zero-temperature limit of the Coulomb gas in the mean-field regime}

The main motivation for Theorem \ref{thm:EA vs det intr} above comes
from the study of the large deviations of the Coulomb gas on a measure
$\mu_{0}$ in $\R^{d},$ where the Energy Approximation Property has
previously appeared as a technical hypothesis \cite{c-g-z,d-l-r,berm10,gz}.
To give some background, assume given a continuous function $\phi$
on $\R^{d}$ and consider the corresponding mean field $N-$particle
Hamiltonian
\[
H_{\phi}^{(N)}(x_{1},...x_{N}):=\frac{1}{(N-1)}\frac{1}{2}\sum_{i\neq j}W(x_{i},x_{j})+\sum_{i=1}^{N}\phi(x_{i})
\]
describing the Coulomb energy of $N-$particles in the exterior potential
$\phi$ (with the divergent self-energies removed) with a mean field
scaling. In physical terms this means that each particle is subject
to the average of the Coulomb potentials created by the other particles,
plus the exterior potential $\phi.$ We recall that the mean-field
scaling is often used for systems with long-range interactions, i.e.
such that $W(x,y)\leq C1/|x-y|^{d}$ (which includes general Riesz
gases) \cite{b-b-d-r,clmp,k2}. Given a measure $\mu_{0}$ on $\R^{d}$
(the ``prior'') with compact support $S_{0}$ and a sequence of
numbers $T_{N}\in]0,\infty[$ the corresponding\emph{ (mean field)
Coulomb gas at temperature $T_{N}$ on $\mu_{0}$ }is defined as the
probability space (canonical ensemble) $\left((\R^{d})^{N},\mu_{T_{N}}^{(N)}\right),$
where 
\begin{equation}
\mu_{\phi,T_{N}}^{(N)}:=\frac{1}{Z_{N,\phi,T_{N}}}e^{-T_{N}^{-1}H_{\phi}^{(N)}}\mu_{0}^{\otimes N},\,\,\,\,Z_{N,\phi,T_{N}}:=\int_{(\R^{d})^{N}}e^{-T_{N}^{-1}H_{\phi}^{(N)}}\mu_{0}^{\otimes N}\label{eq:def of Gibbs measure intro}
\end{equation}
The normalizing constant $Z_{N,\phi,T_{N}}$ is called the\emph{ partition
function} and
\[
F_{N,\phi,T_{N}}:=-\frac{T_{N}}{N}\log Z_{N,\phi,T_{N}}
\]
 is called th\emph{e $N-$particle free energy at temperature $T_{N}.$
}It extends continuously to $T_{N}=0$ by setting 
\[
F_{N,\phi,0}:=\frac{1}{N}\inf_{S_{0}^{N}}H_{\phi}^{(N)},
\]
 where $S_{0}$ is the support of $\mu_{0}$ (using that $e^{-H_{\phi}^{(N)}}$
is continuous; see the beginning of the proof of Lemma \ref{lem:commute}).
In the case when $\phi=0$ we will simply drop the subscripts $\phi.$
The mean field scaling $1/(N-1)$ appearing in the definition of $H_{\phi}^{(N)}(x_{1},...x_{N})$
could alternatively have been absorbed by the temperature $T_{N}$.
Anyhow, throughout the paper we will employ the mean field scaling
above and assume that the corresponding temperatures $T_{N}$ have
a limit $T$ as $N\rightarrow\infty:$
\[
T:=\lim_{N\rightarrow\infty}T_{N}\in[0,\infty[.
\]
 For example, if, with the present mean field scaling, $T_{N}$ is
taken to be proportional to $1/(N-1),$ so that $\mu_{T_{N}}^{(N)}$
is the Gibbs measure for a Coulomb gas without a mean field scaling
at a fixed positive temperature \cite{se}, then the corresponding
limiting temperature $T$ above vanishes (compare Section \ref{subsec:Relations-to-Bernstein-Markov}).
But here it will be important to allow non-vanishing limiting temperatures
$T,$ where entropy enters the picture.

As first shown in \cite{clmp,ki-s}, in the case when $T>0$ and $\mu_{0}$
is equal to Lebesgue measure on a compact domain $S_{0},$ the empirical
measure 
\begin{equation}
\delta_{N}:=\frac{1}{N}\sum_{i=1}^{N}\delta_{x_{i}},\label{eq:def of empir measur}
\end{equation}
 viewed as a random measure on $\left((\R^{d})^{N},\mu_{T_{N}}^{(N)}\right),$
converges in probability, as $N\rightarrow\infty,$ towards a deterministic
measure $\mu_{\phi,\beta}$ 
\[
\lim_{N\rightarrow\infty}\delta_{N}=\mu_{\phi,T},
\]
 where $\mu_{\phi,T}$ is the unique minimizer of the following \emph{free
energy functional} $F_{\phi,T}$ on the space $\mathcal{P}(S_{0}):$
\[
F_{\phi,T}(\mu)=E_{\phi}(\mu)+TD_{\mu_{0}}(\mu),\,\,\,E_{\phi}(\mu):=E(\mu)+\int\phi\mu
\]
 where $D_{\mu_{0}}$ denotes the entropy of $\mu$ relative to $\mu_{0},$
using the sign convention making $D_{\mu}$ non-negative (see formula
\ref{eq:def of rel entropy}). In particular,
\[
F_{\phi,0}:=E_{\phi}
\]
 and we denote by $\mu_{(S,\phi)}$ the \emph{equilibrium measure
}of a non-polar compact weighted set $(S,\phi),$ i.e. the unique
minimizer of $E_{\phi}$ on $\mathcal{P}(S).$ It should be stressed
that the presence of the entropy term in the free energy $F_{\phi,T}(\mu)$
entails that the minimizer $\mu_{T}$ is absolutely continuous wrt
$\mu_{0}$ for any $T>0,$ while this is often not the case for the
equilibrium measure $\mu_{(S_{0},\phi)}.$ For example, if $\mu_{0}$
is Lesbesgue measure on a compact domain $S_{0}$ with smooth boundary
and $\phi=0,$ then $\mu_{(S_{0},\phi)}$ is supported in the boundary
of $S_{0}$ (in the Coulomb case).

Under appropriate regularity assumptions on $\mu_{0}$ it is shown
in \cite{c-g-z,d-l-r,berm10,gz} that the convergence of $\delta_{N}$
towards $\mu_{\phi,T}$ is, in fact, exponential in the sense of large
deviation theory. More precisely, the laws of the empirical measures
$\delta_{N}$ satisfy a\emph{ Large Deviation Principle (LDP) }at
speed $T_{N}^{-1}N,$ whose rate functional $I_{\phi,T}$ coincides
with $F_{\phi,T},$ up to an additive constant. In symbolic notation
this may be expressed as 
\[
(\delta_{N})_{*}\left(e^{-T_{N}^{-1}H_{\phi}^{(N)}}\mu_{0}^{\otimes N}\right)\sim e^{-T_{N}^{-1}NF_{\phi,T}},\,\,N\rightarrow\infty
\]
as measures on $\mathcal{P}(S_{0}).$We recall that, in general, the
rate functional $I$ for an LDP is a proper lower-semicontinuous function
(the precise meaning of the LDP is recalled in Section \ref{sec:Large-deviations}).

\subsubsection{The LDP in the zero-temperature limit}

When $T>0$ the LDP for the Coulomb gas holds for\emph{ }any measure
$\mu_{0}$ not charging polar subsets (Theorem \ref{thm:LDP general}).
A natural question is thus what further conditions on $\mu_{0}$ need
to be imposed in order to ensure that the LDP also holds for $T=0?$
As shown in \cite{c-g-z,d-l-r,berm10,gz}, the Energy Approximation
Property is a sufficient condition. However, as will be shown below,
this condition is not necessary, but rather equivalent to a ``well-behaved''
LDP. The starting point is the basic observation that the Energy Approximation
Property is equivalent to a certain continuity property of the free
energy functional $F_{T},$ namely that $F_{T}$ be continuous with
respect to Gamma-convergence of functionals, as $T\rightarrow0.$
Recall that the notion of Gamma-convergence plays a prominent role
in variational calculus (see definition \ref{def:gamma}) and corresponds
to the Fell topology on the space of lower-semicontinuous functions
\cite{dal}. It implies, in particular, that the minimizer $\mu_{T}$
of $F_{T}$ converges towards the minimizer of $F_{0}.$ In the present
setting the Gamma-convergence of the functional $F_{T}$ is, by a
duality argument, equivalent to the continuity of 
\[
f_{\phi}(T):=\inf_{\mu\in\mathcal{M}(S_{0})}F_{\phi,T}(\mu)
\]
 as $T\rightarrow0$ for all exterior potentials $\phi.$ The number
$f_{\phi}(T)$ is usually called the \emph{free energy at temperature
$T$ }(wrt the exterior potential $\phi).$ The main result may now
be formulated as the following
\begin{thm}
\label{thm:det vs gamma intro}Let $\mu_{0}$ be a measure on $\R^{d}$
which does not charge polar subsets and assume that the support $S_{0}$
of $\mu_{0}$ is compact and locally regular. Then the following is
equivalent:
\begin{itemize}
\item $\mu_{0}$ is strongly determining
\item For any given continuous function $\phi,$ the corresponding free
energy $f_{\phi}(T)$ at temperature $T,$ 
\begin{equation}
f_{\phi}(T):=\inf_{\mu\in\mathcal{M}(S_{0})}F_{\phi,T}(\mu)\label{eq:f as inf F in Thm intr}
\end{equation}
is continuous wrt $T\in[0,\infty]$ 
\item For any given potential $\phi$ the minimizer $\mu_{\phi,T}$ of the
functional $F_{T,\phi}$ on $\mathcal{P}(S_{0})$ converges, as $T\rightarrow0,$
towards the equilibrium measure of $(S_{0},\phi),$ i.e. towards the
minimizer $\mu_{(S_{0},\phi)}$ of $F_{\phi,0}:$ 
\begin{equation}
\lim_{T\rightarrow0}\mu_{\phi,T}=\mu_{(S_{0},\phi)}\label{eq:converg towards equi in thm intro}
\end{equation}
 in the weak topology on $\mathcal{P}(S_{0}).$
\item The LDP for the Coulomb gas on $\mu_{0}$ holds for all exterior potentials
$\phi$ and all $T\in[0,\infty[$ with a rate functional which is
continuous wrt $T\in[0,\infty[$ in the sense of Gamma-convergence
\end{itemize}
\end{thm}

In fact, in the present setting the LDP in the previous theorem is
equivalent to the free energy asymptotics
\begin{equation}
\lim_{N\rightarrow\infty}F_{N,\phi,T_{N}}=f_{\phi}(T),\,\,\,T_{N}\rightarrow T,\label{eq:free energy as in intro}
\end{equation}
 for any potential $\phi$ and sequence $T_{N}\in[0,\infty[$ such
that $T_{N}\rightarrow T\in[0,\infty[$ for a function $f_{\phi}(T)$
which is continuous on $[0,\infty[$ (and a posteriori of the form
\ref{eq:f as inf F in Thm intr}). The continuity of $f_{\phi}(T)$
when $T>0$ is automatic and, as discussed in Section \ref{sec:Relations-to-phase},
a discontinuity at $T=0$ can be interpreted as a \emph{zeroth-order
phase transition. }Another equivalent formulation is obtained by taking
$T_{N}=T\in]0,\infty[$ and demanding that the limits $T\rightarrow0$
and $N\rightarrow\infty$ of $F_{N,\phi,T}$ commute (see Lemma \ref{lem:commute}).
The existence of discontinuities at $T=0$ should be contrasted with
the fact that 
\[
\lim_{T\rightarrow\infty}\mu_{\phi,T}=\mu_{0}
\]
 always holds (see Section \ref{subsec:The-limit beta zero}). This
means that if $\mu_{0}$ is strongly determining, then the curve $\mu_{T,\phi}\in\mathcal{P}(S_{0})$
interpolates between $\mu_{0}$ at $T=\infty$ and the weighted equilibrium
measure $\mu_{(S_{0},\phi_{0})}$ at $T=0.$

The previous theorem will be deduced from the following result concerning
the case when the potential $\phi$ is fixed:
\begin{thm}
\label{thm:fixed phi intr}Let $\mu_{0}$ be a measure on $\R^{d}$
which does not charge polar subsets and assume that the support $S_{0}$
of $\mu_{0}$ is compact and locally regular. For a given continuous
function $\phi$ on $S_{0}$ the following is equivalent
\begin{itemize}
\item $\mu_{0}$ is determining for $(S_{0},\phi)$ 
\item The following convergence of free energies holds:
\begin{equation}
\lim_{T\rightarrow0}\lim_{N\rightarrow\infty}F_{N,\phi,T_{N}}=\inf_{\mathcal{P}(S_{0})}E_{\phi}\label{eq:lim inf F =0000E4r inf E in theorem text-1-1}
\end{equation}
\item The following weak convergence of the expectations $\E_{T,\phi}(\delta_{N})$
of the empirical measure $\delta_{N}$ holds:
\begin{equation}
\lim_{T\rightarrow0}\lim_{N\rightarrow\infty}\E_{T,\phi}(\delta_{N})=\mu_{(S_{0},\phi)}\label{eq:lim inf F =0000E4r inf E in theorem text-1}
\end{equation}
\end{itemize}
\end{thm}

We recall that the inverse of the infimum of the functional $E$ on
$\mathcal{P}(K),$ for a given compact set $K$ in $\R^{d},$ is usually
called the \emph{(Wiener) capacity of $K$ }(see Section \ref{subsec:Capacities-and-determining}
for the general weighted setting). 

Coming back to the energy approximation property in Section \ref{subsec:Energy-approximation-and}
we will also show, in Section \ref{subsec:Constructive-approximations-usin},
that the approximating sequence in question can be constructed quasi-explicitly.

\subsection{\label{subsec:Relations-to-Bernstein-Markov}Relations to Bernstein-Markov
measures, orthogonal polynomials and Ullman's criterion}

Now specialize to the two-dimensional case and identify $\R^{2}$
with $\C.$ Then the Gibbs measure \ref{eq:def of Gibbs measure intro}
may, for $\phi=0,$ be expressed as
\begin{equation}
\mu_{\beta_{N}}^{(N)}:=\frac{1}{Z_{N,\beta_{N}}}\left|D^{(N)}\right|^{p_{N}}\mu_{0}^{\otimes N},\,\,\,p_{N}:=2\frac{1}{T_{N}(N-1)}\label{eq:Gibbs as Vanderm intro}
\end{equation}
 where $D^{(N)}(z_{1},...,z_{N})$ denotes the\emph{ Vandermonde determinant,
}i.e. the polynomial on $\C^{N}$ defined by\emph{ }
\[
D^{(N)}(z_{1},...,z_{N_{k}})=\prod_{1\leq i<j\leq N}(z_{i}-z_{j})=\det_{1\leq i,j\leq N}(z_{i}^{j-1})
\]
Accordingly, the corresponding partition function $Z_{N,T_{N}}$ is
equal to the $L^{p_{N}}-$norm of $D^{(N)}.$ More generally, introducing
an exterior potential $\phi$ corresponds to replacing the $L^{p_{N}}-$
norms with weighted norms, i.e replacing $\mu_{0}$ with $e^{-T_{N}^{-1}\phi}\mu_{0}.$
Note that for a fixed $p,$ the corresponding $T_{N},$ tend to zero
as $N\rightarrow\infty,$ i.e. a fixed $p$ induces a vanishing limiting
temperature $T$ (in terms of the mean field setup in the previous
section). A notion of a phase transition for the Coulomb gas with
respect to the parameter $p$ - from a liquid to a crystalline phase,
as $p$ is increased towards a critical value - has been discussed
extensively in the physics literature, supported by numerical studies
\cite{st} (and related to microscopic large deviation principles
in \cite[Section 4.6]{se}). But this notion is different than the
zeroth-order phase transitions discussed here, where the parameter
$T$ is decreased towards zero.

The Coulomb gas has been studied extensively in connection to Random
Matrix Theory, in particular in the case when $\mu_{0}$ is Lesbegue
measure on $\R\subset\R^{2}$ with $\phi$ of sufficent growth at
infinity and $p=2$. The corresponding Gibbs measure $\mu_{\beta_{N}}^{(N)}$
then arises as the eigenvalue distribution of a Hermitian matrix \cite{me,ki-s,fo}
(see also \cite{d-e,a-b-g} for the case of general $p$). The asymptotics
of the corresponding free energies, as $T_{N}\rightarrow0$ was established
in \cite[Thm 2.1]{jo} and the LDP in \cite{ben-g}. 

We recall that a measure $\mu_{0}$ in $\C$ is said to satisfy a
\emph{Bernstein-Markov inequality with weight $\phi$} if, for any
given $\epsilon>0$ there exists a constant $C$ such that 
\begin{equation}
\sup_{S_{0}}|p_{k}|^{2}e^{-k\phi}\leq Ce^{\epsilon k}\int_{\C}|p_{k}|^{2}e^{-k\phi}\mu_{0}\label{eq:BM prop intro}
\end{equation}
 for all polynomials $p_{k}$ on $\C,$ where $k$ denotes the degree
of $p_{k}$ (see the survey \cite{b-l-p-w}). For such a measure the
existence of the limit $f_{\phi}(0)$ of the corresponding $N-$particle
free energies at $T=0$  was established in \cite{b-l0}. In fact,
if the Bernstein-Markov inequality holds for all weights then the
LDP holds at zero temperature, by the results in \cite{berm 1 komma 5}
(see also \cite{b-l} for a different approach and \cite{z-z} for
relations to random polynomials). In view of Theorem \ref{thm:fixed phi intr}
this means that the Coulomb gas on a measure $\mu_{0}$ with compact
support $S_{0},$ which satisfies the weighted Bernstein-Markov inequality
for $(S_{0},\phi)$ - but which is not determining for $(S_{0},\phi)$
- exhibits a zeroth-order phase transition at $T=0.$ A general procedure
for constructing such measures is explained in \cite{b-l-p-w}, where
a concrete example of Totik on the interval is reported (see the appendix).
We thus arrive at the following corollary, exhibiting a zeroth-order
phase transition:
\begin{cor}
\label{cor:phase intro}Let $K$ be a compact domain in $\C$ with
smooth boundary or equal to a disjoint finite union of intervals in
$\R.$ For any given continuous function $\phi$ on $K$ there exists
a measure $\mu_{0}$ with support $K$ such that $\mu_{0}$ is absolutely
continuous wrt $dx$ and such that the corresponding 2d Coulomb gas
satisfies a LDP for any $T\in[0,\infty[$ with a rate functional which
is discontinuous at $T=0$ in the sense of Gamma-convergence. More
precisely, in the case $\phi=0,$ the function
\[
f(T):=-\lim_{N\rightarrow\infty}\frac{T_{N}}{N}\log\int_{\R^{N}}\left|D^{(N)}\right|^{\frac{2}{T_{N}N}}\mu_{0}{}^{\otimes N}\,\,\,\,T=\lim_{N\rightarrow\infty}T_{N}
\]
is well-defined on $[0,\infty[,$ continuous on $]0,\infty[,$ but
discontinuous at $T=0$ (and similarly for a general $\phi$). 
\end{cor}

The property of being determining can, for a non-polar measure $\mu_{0},$
be viewed as a potential-theoretic refinement of the Bernstein-Markov
inequality, where a polynomial $p_{k}$ of degree $k$ is replaced
by $e^{T^{-1}\psi_{\mu}}$ for a general measure $\mu$ and positive
number $T^{-1},$ playing the role of $k.$ This was first shown in
\cite{b-b-w} in a general complex geometric setting. In Section \ref{sec:Large-deviations}
we will extend these notions to the case of general pair interaction
potentials (the notion of Bernstein-Markov measures for Riesz interaction
was introduced in \cite{b-l-w}). 

\subsubsection{Orthogonal polynomials and Ullman's criterion}

We recall that the Bernstein-Markov-inequality has it roots in the
theory of orthogonal polynomials on $\R.$ In fact, for a measure
$\mu_{0}$ on $\R$ with compact and regular support the Bernstein-Markov
inequality is equivalent to the notion of \emph{regular measures }on
$\R$ introduced in  \cite{st-t}, whose definition involves the asymptotics
of the degree $N$ orthogonal polynomials $p_{N}$ associated to $\mu_{0}$
(see the proof of Prop \ref{prop:totik}). In the case when the support
of $\mu_{0}$ is $[-1,1]$ this notion goes back to Ullman. He also
introduced the notion of determining measures on $[-1,1]$ to get
a sufficient condition for regularity, known as \emph{Ullman's criterion}
in the general setting of measures on $\R$ \cite{st-t}. In view
of Theorem \ref{thm:det vs gamma intro} Ullman's criterion naturally
fits into the probabilistic setting of the Coulomb gas, since it is
equivalent to the continuity properties discussed above. It should
be pointed out that Ullman originally used a different, but equivalent,
capacity formulation of determining measures on $[-1,1]$ (see Prop
\ref{prop:capac crit for det} for the relation to the present setting).
The definition of determining measures in the present general potential-theoretic
setting on $\R^{d}$ mimics the definition used in the complex-geometric
setting of \cite{b-b-w}, which goes back to \cite{Le0}.

\subsection{\label{subsec:General-pair-interactions}Towards the case of general
pair interactions}

Finally, let us make some remarks about the case when the Gibbs measure
\ref{eq:def of Gibbs measure intro} is defined by  a general proper
lsc function $W.$ Then it essentially follows from \cite{d-l-r,berm10,gz}
that the corresponding LDP holds for $T>0$ if and only if the corresponding
free energy functional $F_{T}$ is a proper lsc functional (see Theorem
\ref{thm:LDP general}). However, it seems challenging to find a general
potential-theoretic characterization of measures $\mu_{0}$ such that
the corresponding LDP also holds at $T=0.$ In view of Theorem \ref{thm:det vs gamma intro}
(and its generalization to Riesz gases \ref{thm:Riesz gas}) the problem
of characterizing measures $\mu_{0}$ such that the LDP is ``well-behaved
at $T=0"$ should be more accessible also in the general case. It
seems likely that the answer should be given by determining measures
for rather general interactions $W,$ but we shall not pursue this
here and only point out that in the case when $W(x,y)=W_{\alpha}(x,y)+K(x,y),$
where $W_{\alpha}$ denote the Riesz kernel and the perturbation $K(x,y)$
is continuous on $S_{0}$ the present results (concerning the case
$K=0)$ directly generalize. Indeed, since $K(x,y)$ is continuous
it influences neither the energy approximation property, nor the determining
property appearing in Theorem \ref{thm:EA vs det intr}. For example,
this situation naturally appears in electrostatics in $\R^{3}$ with
the screened Coulomb interaction (Yukawa potential) $W(x,y)=e^{-m|x-y|}/|x-y|$
for a given positive number $m.$ 

\subsection{Further relations to previous results}

The idea of studying the Gibbs measures corresponding to a general
lsc pair interaction potential $W,$ in the case $T=0$, by letting
$T\rightarrow0$ goes back to \cite{ki-s}, which builds on the variational
approach introduced in \cite{m-s} (where the case of a continuous
$W$ was considered). In the main result of \cite{ki-s} it is claimed
that, in general, any limit point $\mu$ in $\mathcal{P}(\R^{d})$
of the empirical measure $\delta_{N}$ minimizes the corresponding
energy functional $E.$ However, in the case of the Coulomb gas this
is contradicted by the example in Theorem \ref{thm:BM iff LDP on R},
where $\mu_{0}$ has support $[-1,1]$ and is absolutely continuous
wrt $dx$ (see also Example \ref{exa:counter 2D coul with weight}
for a simple counter-example in the weighted setting). The mistake
in \cite{ki-s} appears to be the claimed inequality \cite[3.12]{ki-s},
which, in general, requires assumptions on $\mu_{0}.$ 

The ``only if'' direction in Theorem \ref{thm:fixed phi intr} was
first shown in the complex-geometric setting of compact Kähler manifolds
$X$ in \cite{berm12} and generalized to the non-compact setting
pluripotential setting of $\C^{n}$ in \cite{berm12}, using a compactification
argument. See also \cite{glz} for a far-reaching generalization of
\cite[Theorem 2.1]{berm11} to general measures not charging pluripolar
subsets and bounded weights.

It should also be pointed out that there are particular situations
where the\emph{ rate} of the convergence of $\mu_{T,\phi}$ towards
the equilibrium measure $\mu_{(S_{0},\phi)}$ can be quantified. In
the setting of a compact Kähler manifold $X,$ studied in \cite{berm11},
the rate of convergence of the $L^{\infty}-$norms of the corresponding
potentials $\psi_{\mu_{T,\phi}}$ was shown to be of the order $O(T)\log T^{-1}.$
In particular, the case when $X$ is the Riemann sphere implies (by
a compactification argument, as in \cite{berm12}) that such a rate
holds for the Coulomb gas in $\R^{2}$ with $\mu_{0}=dx$ if $\phi$
has strictly super logarithmic growth. More general and precise quantitative
convergence results for Coulomb gases in any dimension have very recently
been established in \cite{a-s} (see also \cite[formula 18]{a-b-g}
for an explicit integral formula for $\mu_{T,\phi}$ in the case of
the Coulomb gas on $\R$ with $\phi(x)=x^{2}$ and $\mu_{0}$ the
Gaussian measure, so that $\mu_{T,\phi}$ converges, as $T\rightarrow0,$
to Wigner's semi-circle law). However, it should be stressed that
in the general setting of determining measures $\mu_{0},$ studied
in the present paper, one can not hope for quantitative rates, unless
regularity assumptions on $\mu_{0}$ are imposed.

\subsection{On the proofs}

The core analytic result is Theorem \ref{thm:fixed phi intr} and
its general form \ref{thm:determ for phi iff inf converges} (which
applies to any Riesz interaction with $\alpha\leq2),$ saying, in
particular, that the measure\emph{ $\mu_{0}$ is determining iff the
corresponding free energy $f(T)$ is continuous as $T\rightarrow0.$}
The proof of the ``if'' direction'' mimics the variational proof
of a similar result in the complex geometric setting on a compact
Kähler manifold $X$ \cite[Theorem 2.1]{berm11} (which applies, in
particular, to the case $d=\alpha=2$ by taking $X$ to be the Riemann
sphere). An important ingredient is the potential-theoretic analog
of the Bernstein-Markov property for determining measures in Prop
\ref{prop:det as BM} (proved in \cite{b-b-w} in the complex geometric
setting). In the present setting we also have to deal with the non-compactness
of $\R^{d}$ and the non-local properties of the fractional Laplacian
$-(-\Delta)^{\alpha/2}$ (for $\alpha<2)$. In the case when $d=\alpha=2$
the ``only if'' direction could alternatively be deduced from the
generalization in \cite{glz} of \cite[Theorem 2.1]{berm11} to arbitrary
measures $\mu_{0},$ not charging pluripolar subsets. However, the
proof in \cite{glz} (which is not variational) appears to exploit
some special local features of the complex geometric setting, which
do not seem to apply when $\alpha\neq2.$ Here we instead use a variational
approach, which has the virtue of only demanding some rather general
axioms of potential theory (compare Remark \ref{rem:max}). 

\subsection{Acknowledgments}

It is a pleasure to thank Sébastien Boucksom, David Witt-Nyström,
Vincent Guedj and Ahmed Zeriahi for the stimulating collaborations
\cite{b-b,b-b-w,bbgz}, which provided important motivation for the
present paper and Norm Levenberg for helpful comments. Thanks also
to the referees for very helpful comments that improved the exposition.
This work was supported by grants from the ERC and the KAW foundation.

\subsection{Organization}

In Section \ref{sec:Weighted-potential-theory} we introduce the weighted
potential theory needed for the proofs of the main analytic results.
In particular, a dual representation of the energy $E(\mu)$ as a
Legendre transform is given. In Section \ref{subsec:Energy-approximation-and}
we reformulate the Energy-Approximation Property in terms of Gamma-convergence
of the free energy functional $F_{T},$ which in turn is a given a
dual formulation using Legendre transforms. Then in Section \ref{subsec:Determining-measures-vs}
the proofs of the main analytic results for Riesz interactions are
given, by relating Gamma-convergence of $F_{T}$ to determining measures.
The connections to large deviation principles is studied in Section
\ref{sec:Large-deviations} and connections to Bernstein-Markov inequalities
are explored. The results are then reformulated in terms of phase
transitions in Section \ref{sec:Relations-to-phase}. In the appendix
some construction of measures $\mu_{0}$ are provided, which illustrate
the sharpness of the main results. 

\subsection{General notation}

We will denote by $\mathcal{P}_{c}(\R^{d})$ the space of all compactly
supported probability measures on $\R^{d}$ and by $\mathcal{P}(K)$
the subset consisting of measures supported on a compact subset $K$
of $\R^{d}.$ We endow the space $\mathcal{P}(K)$ with the weak topology.
Throughout the paper we fix a probability measure $\mu_{0}$ with
compact support, denoted by $S_{0}.$ Given a compact subset $K$
of $\R^{d}$ we will denote by $\mathcal{C}(K)$ the space of all
continuous functions on $K.$

We recall that a function $f$ on a topological space $X,$ taking
values in $]\infty,\infty]$ is\emph{ }lower semi-continuous (lsc)
if $\{f\leq\alpha\}$ is closed for any $\alpha\in]\infty,\infty].$
We will say that $f$ is\emph{ proper lower-semicontinuous }under
the further assumption that $f$ is not identically equal to $\infty$
(following standard terminology in convex analysis). If $X$ is compact
and $f$ is lsc the latter condition equivalently means that $\inf_{X}f$
is finite. Hence, under the map $f\mapsto f-\inf_{X}f$ the space
of proper lsc functions on $X$ corresponds to the space of\emph{
rate functionals,} in the sense of large deviation theory. Finally,
it will be convenient to work with inverse temperatures $\beta_{N}:=T_{N}^{-1}$
and $\beta:=T^{-1}$ rather than temperatures.

\section{\label{sec:Weighted-potential-theory}Weighted potential theory and
Legendre transforms }

In this section we develop the weighted potential theory needed for
the proofs of the main results. The key result is the Legendre transform
representation of the energy in Theorem \ref{thm:Legendre}. The presentation
is inspired by the complex-geometric framework in \cite{b-b,b-b-w,bbgz},
which covers in particular the Coulomb case in $\R^{2}$ (see also
\cite{s-t,c-g-z} for different points of view).

\subsection{\label{subsec:Potential-theoretic-preliminarie}Potential-theoretic
preliminaries}

We start by recalling some basic potential-theoretic results. We follow
the classical reference \cite{la}, but with a different sign convention
for the kernels and the potentials (ensuring that the potentials are
subharmonic in the Coulomb case).

We will denote by $W_{\alpha}(x,y)$ the Riesz kernel with parameter
$\alpha\in]0,d[,$ i.e. the lsc function on $\R^{d}\times\R^{d}$
defined by
\[
W_{\alpha}(x,y):=\frac{1}{|x-y|^{d-\alpha}}.
\]
 When $d=2$ we will allow the case $\alpha=d=2,$ by setting 
\[
W_{2}(x,y)=-2\log|x-y|,
\]
 The definition ensures that when $d\geq2$ the function $W_{2}(x,y)$
is a (up to multiplication by a negative constant) a Green's kernel
for the Laplacian $\Delta$ on $\R^{d}.$ Accordingly, we will refer
to the case $\alpha=2$ as the ``Coulomb case'' and the special
case $\alpha=2=d$ the ``logarithmic case''.

The \emph{energy} $E(\mu)$ of a measure $\mu\in\mathcal{P}_{c}(\R^{d})$
is defined by 
\begin{equation}
E(\mu)=\frac{1}{2}\int_{\R^{d}}W_{\alpha}\mu\otimes\mu\in]-\infty,\infty]\label{eq:def of E in prel}
\end{equation}

Given a measure $\mu$ on $\R^{d}$ we will denote by $\psi_{\mu}$
its\emph{ potential:}

\[
\psi_{\mu}(x):=-\int_{\R^{d}}W_{\alpha}(x,y)\mu(y),
\]

Since $W_{\alpha}$ is symmetric the following \emph{symmetry property}
holds:
\begin{equation}
\int\psi_{\nu}\mu=\int\nu\psi_{\mu}\label{eq:recipr}
\end{equation}
if $\mu$ and $\nu$ are in $\mathcal{P}_{c}(\R^{d})$ and of finite
energy. Moreover, $W_{\alpha}$ defines a\emph{ strictly positive
definite bilinear form} in the following sense:
\begin{equation}
-\int(\psi_{\mu}-\psi_{\nu})(\mu-\nu)\geq0\label{eq:pos def}
\end{equation}
 with equality iff $\nu=\mu.$ This implies that the map $\mu\mapsto\psi_{\mu}$
is injective and we will denote the inverse operator by $\Delta_{\alpha},$
which coincides with the ordinary Laplacian when $\alpha=2.$ \footnote{In general, $\Delta_{\alpha}$ is a fractional Laplacian $\Delta_{\alpha}:=-(-\Delta)^{\frac{\alpha}{2}}$
in the sense of functional calculus, which in the case $\alpha\in]0,2]$
corresponds precisely to the generator of a symmetric stable Levy
process \cite{b-g-r,bo}. } A bounded set $S$ is said to be \emph{polar} if $S\subset\{\psi_{\mu}=-\infty\}$
for some measure $\mu$ (equivalently, $S$ has vanishing outer capacity)
\footnote{This terminology is standard, but different from the one in \cite{la},
where the terminology polar is used for the sets which are precisely
equal to some $\{\psi_{\mu}=-\infty\}$ (such sets are called completely
polar in modern terminology). } . If $S$ is compact then $S$ is polar iff $E(\mu)=\infty$ for
any measure $\mu\in\mathcal{P}(S).$ A property is said to hold \emph{quasi-everywhere
(q.e) }if it holds on the complement of a polar set. 

If $\mu_{j}\in\mathcal{P}(K)$ for a compact subset $K$ and $\mu_{j}\rightarrow\mu$
weakly, then, for any given $x\in\R^{d}$
\[
\limsup_{j\rightarrow\infty}\psi_{\mu_{j}}(x)\leq\psi_{\mu}(x),
\]
 as follows directly from the lower semi-continuity of $W.$ Moreover,
for q.e. $x$ in $\R^{d},$ 
\begin{equation}
\limsup_{j\rightarrow\infty}\psi_{\mu_{j}}(x)=\psi_{\mu}(x)\label{eq:limsup is psi qe}
\end{equation}
(see \cite[Therem 3.8, page 190]{la}). As a consequence \cite[page 191, Remark 2]{la},
\begin{equation}
\left(\limsup_{j\rightarrow\infty}\psi_{\mu_{j}}\right)^{*}=\psi_{\mu}\label{eq:limsup reg is psi}
\end{equation}
 on all of $\R^{d},$ where the limsup is defined point-wise and $f^{*}$
denotes the upper semi-continuous regularization of a function $f$
on $\R^{d}:$ 
\[
f^{*}(x)=\sup\left\{ f(x_{j}):\,\,\,x_{j}\rightarrow x\right\} ,
\]
 where the sup runs over all sequences $x_{j}$ converging to $x.$
We will be mainly interested in the case when $\alpha\leq2,$ since
the following \emph{domination principle} then applies \cite[Thm 1.29]{la}:
for a given constant $C$

\begin{equation}
\psi_{\nu}\leq\psi_{\mu}+C\,\,\,\mu-\text{a.e}\,\,\,\implies\psi_{\nu}\leq\psi_{\mu}+C,\label{eq:dom princ}
\end{equation}
 assuming that $\mu$ has finite energy. 
\begin{rem}
\label{rem:max}When $\alpha\leq2$ it is also known that the space
of potentials is preserved under the max operation \cite[Thm 1.31]{la}
(the case $\alpha=2$ follows directly from subharmonicity). But for
our purposes it will be enough to use the domination principle. This
should be useful in order to extend Theorem \ref{thm:det vs gamma vs ea}
to more general kernels $W(x,y)$ appearing in axiomatic potential
theory, where the domination principle (aka the second maximum principle)
is often is taken as an axiom \cite[Page 364]{la}. For example, the
domination principle holds when $W(x,y)$ is the potential kernel
of a Markov process satisfying Hunt's hypothesis (H) \cite{f-g} (then
$-\psi_{\mu}$ is called the excessive function associated to $\mu).$
But we shall not go further into this here.
\end{rem}

In order to simplify the notation we we will omit the dependence on
$\alpha$ of the potential-theoretic objects associated to the Riesz
kernel $W_{\alpha}(x,y),$ such as the energy $E(\mu),$ potentials
$\psi_{\mu}$ and the corresponding inverse operator $\Delta$ (which
coincides with the Laplacian when $\alpha=2).$ As $\alpha$ will
be fixed this should not cause any confusion. 

\subsection{Function spaces }

We will use the notation 
\[
\mathcal{L}_{c}(\R^{d}):=\left\{ \psi:\,\,\psi=\psi_{\mu}+C,\,\,\,\mu\in\mathcal{P}_{c}(\R^{d}),\,\,C\in\R\right\} .
\]
We endow the space $\mathcal{L}_{c}(\R^{d})$ with the $L_{loc}^{1}-$topology
induced from the inclusion $\mathcal{L}_{c}(\R^{d})\Subset L_{loc}^{1}(\R^{d}).$
Denote by $\mathcal{E}_{c}(\R^{d})$ the subspace of $\mathcal{L}_{c}(\R^{d})$
satisfying $E(\Delta\psi)<\infty.$ Given a compact subset $K$ of
$\R^{d}$ we will write $\mathcal{L}_{K}(\R^{d})$ and $\mathcal{E}_{K}(\R^{d})$
for the subspaces of $\mathcal{L}_{c}(\R^{d})$ and $\mathcal{E}_{c}(\R^{d}),$
respectively, obtained by demanding that $\Delta\psi$ be a probability
measure supported in $K.$ The definitions are made so that, for any
compact $S,$ we have a bijection (whose inverse is $\Delta):$
\begin{equation}
\mu\mapsto\psi_{\mu}\,\,\,\,\mathcal{P}(S)\longleftrightarrow\mathcal{L}_{S}(\R^{d})/\R\label{eq:bijection}
\end{equation}

\begin{prop}
\label{prop:compact}(Compactness) Let $S$ be a compact subset of
$\R^{d}$ and fix a closed ball $B$ containing $S.$ Then the subspace
of $\mathcal{L}_{S}(\R^{d})$ consisting of all $\psi$ which are
``mean normalized'', i.e. 
\[
\int_{B}\psi dx=0,
\]
is compact wrt the $L_{loc}^{1}-$topology. Moreover, given a compact
subset $K$ there exists a constant $C$ such that 

\begin{equation}
\sup_{K}\psi\leq\frac{\int_{B}\psi dx}{\int_{B}dx}+C\label{eq:bound on sup psi in Prop comp}
\end{equation}
on $\mathcal{L}_{S}(\R^{d}).$
\end{prop}

\begin{proof}
\emph{Step 1: Compactness for ``mean-normalized functions''}

The compactness is a consequence of the following general properties
of the kernel $-W_{\alpha}:$ it is symmetric, usc, continuous on
the complement of the diagonal, in $L_{loc}^{1}$ and the corresponding
integral operator yields a bijection, as in formula \ref{eq:bijection}
. To see this take any $\mu\in\mathcal{P}(S)$ and first observe that
\[
\int_{B}\psi_{\mu}dx=\int f\mu
\]
 for the continuous function $f:=\psi_{\nu}$ with $\nu:=1_{B}dx/\int_{B}dx$
(using the symmetry property \ref{eq:recipr}). Now decompose any
mean-normalized $\psi$ as
\[
\psi=\left(\psi+\int f\mu\right)-\int f\mu
\]
Since $\psi$ is normalized the bijection \ref{eq:bijection} shows
that the first term equals $\psi_{\mu},$ where $\mu:=\Delta\psi.$
Let now $\psi_{j}$ be a sequence in $\mathcal{L}_{S}(\R^{d})$ and
set $\mu_{j}:=\Delta\psi_{j}.$ Since $S$ is compact the space $\mathcal{P}(S)$
is also compact and hence there exists $\mu\in\mathcal{P}(S)$ such
that $\mu_{j}\rightarrow\mu$ in $\mathcal{P}(S),$ after perhaps
passing to a subsequence. All that remains is thus to verify that
$\psi_{\mu_{j}}\rightarrow\psi_{\mu}$ in $L_{loc}^{1}.$ To this
end first note that, since $-W_{\alpha}$ is usc, we have 
\[
\limsup_{j\rightarrow\infty}\psi_{\mu_{j}}\leq\psi
\]
 Moreover, by symmetry,
\[
\int_{B}\psi_{\mu_{j}}dx=\int f\mu_{j}\rightarrow\int f\mu:=\int_{B}\psi dx
\]
 since $f$ is continuous. But then it follows from general integration
theory that $\psi_{\mu_{j}}\rightarrow\psi$ in $L^{1}(B,dx).$ Finally,
the $L_{loc}^{1}$ convergence on the complement of $S$ follows directly
from the fact that $W_{\alpha}$ is continuous on the complement of
the diagonal. 

\emph{Step 2: the upper bound on $\sup_{K}\psi$}

By the previous step it will be enough to prove that the functional
$\psi\mapsto\sup_{K}\psi$ is usc on $\mathcal{L}_{S}(\R^{d}).$ To
this end first observe that, if $\mu_{j}\rightarrow\mu$ in $\mathcal{P}(S)$
and $x_{j}\rightarrow x$ then
\[
\limsup\psi_{\mu_{j}}(x_{j})\leq\psi_{\mu}(x),
\]
 using that the kernel $-W_{\alpha}$ is usc (see \cite[Thm 1.3]{la}).
As consequence, if $\psi_{j}\rightarrow\psi_{\infty}$ in $\mathcal{L}_{S}(\R^{d})$
then the previous inequality still holds if $\psi_{\mu_{j}}$ and
$\psi_{\mu}$ are replaced by $\psi_{j}$ and $\psi_{\infty},$ respectively
(using the decomposition argument in the previous step). Now, taking
$x_{j}$ so that $\psi_{j}(x_{j})=\sup_{K}\psi_{j}$ concludes the
proof of Step 2.
\end{proof}

\subsection{Energy functionals on $\mathcal{P}(S)$ and the equilibrium measure
of a weighted set $(S,\phi)$}

If $\mu$ has compact support and $\phi$ is a continuous function
on $\R^{d}$ we set

\begin{equation}
E_{\phi}(\mu):=-\frac{1}{2}\int\psi_{\mu}\mu+\int\phi\mu\in]-\infty,\infty]\label{eq:def of E phi text}
\end{equation}
The definition is made so that $E_{0}(\mu):=E(\mu)$ is the classical
energy of $\mu$ (formula \ref{eq:def of E in prel}).
\begin{prop}
\label{prop:conv of E}Let $S$ be a non-polar subset. Then the restriction
of $E$ to $\mathcal{P}(S)$ is lsc and strictly convex. Hence, so
is $E_{\phi}$ for any given continuous function $\phi.$
\end{prop}

\begin{proof}
This follows from (and is equivalent to) the positivity \ref{eq:pos def}.
\end{proof}
Thus, if $S$ is a non-polar compact subset, then $E_{\phi}$ admits
a unique minimizer $\mu_{(S,\phi)}$ on $\mathcal{P}(S),$ which is
called the \emph{equilibrium measure} of the \emph{weighted set} $(S,\phi).$ 

Given a non-polar compact weighted set $(S,\phi)$ it will also be
convenient to consider a normalized version of the functional $E_{\phi}$
on $\mathcal{P}(S)$ defined by 
\begin{equation}
E_{\omega_{\phi}}:=E_{\phi}-\inf_{\mathcal{P}(S)}E_{\phi},\label{eq:def of E omega phi}
\end{equation}
 which is invariant under $\phi\mapsto\phi+C$ for any constant $C.$ 
\begin{rem}
The notation is inspired by the corresponding complex-geometric setup,
using the notation in \cite{berm 1 komma 5,berm12}, applied to the
special case when $d=\alpha=2.$ In this case, when $\phi$ is expressed
as the restriction to $S$ of a function with logarithmic growth in
$\C,$ the symbol $\omega_{\phi}$ denotes the signed measure $\omega_{\phi}:=\Delta\phi$
(extended by zero from $\C$ to the Riemann sphere) and $E_{\omega_{\phi}}$
is the corresponding pluricomplex energy. Since the normalized energy,
defined by formula \ref{eq:def of E omega phi}, is invariant under
$\phi\mapsto\phi+C$ it depends, indeed, only on $\Delta\phi.$
\end{rem}

\begin{lem}
\label{lem:e-l eq for min}Given a weighted non-polar compact subset
$(S,\phi)$ the potential $\psi_{\mu}\in\mathcal{L}_{c}(\R^{d})$
of the equilibrium measure $\mu:=\mu_{(S,\phi)}$ has the following
property: there exists a constant $C$ such that 
\[
(i)\,\psi_{\mu}\leq\phi+C\,\,\,\,\text{q.e.\,on\,\ensuremath{S}}
\]
\[
(ii)\,\psi_{\mu}\geq\phi+C\,\,\,\,\text{on\,\ensuremath{S_{\mu},}}
\]
where $S_{\mu}$ denotes the support of $\mu.$
\end{lem}

\begin{proof}
This goes back to Frostman and is proved in \cite[Theorem 1.2]{c-g-z}
when $S=\R^{n}$ and in \cite{s-t} in the logarithmic case. The proof
in the general case is essentially the same and follows from rather
general variational considerations. Indeed, one first observes that
$u_{\mu}:=-(\psi_{\mu}-\phi)$ is a sub-gradient for the functional
$E_{\phi}(\mu).$ Hence, if $\mu$ minimizes $E_{\phi}$ on $\mathcal{P}(S)$,
then $\left\langle u_{\mu},\nu-\mu\right\rangle \geq0$ for any $\nu\in\mathcal{P}(S)$
of finite energy. Applying this inequality to $\nu=\mu(1+f)$ for
any $f\in L^{\infty}(\mu)$ such that $\int f\mu=0$ and $\left\Vert f\right\Vert _{L^{\infty}(\mu)}<1$
gives $\left\langle u_{\mu},f\mu\right\rangle \geq0.$ Replacing $f$
with $-f$ yields the reversed inequality, showing that $\left\langle u_{\mu},f\mu\right\rangle =0$
for any $f$ such that $\int f\mu=0.$ But this implies that $u_{\mu}=-C$
$\mu-$a.e. for some constant $C.$ Since, $u$ is lsc it follows
that $u_{\mu}\leq-C$ on $S_{\mu}.$ As a consequence, $\left\langle u_{\mu},\nu\right\rangle \geq-C$
for any $\nu\in\mathcal{P}(S)$ of finite energy, which implies $u_{\mu}\geq-C$
q.e. on $S$ (using that $u_{\mu}$ is lsc and that any subset of
positive capacity has a compact subset of positive capacity). 
\end{proof}
To a compact weighted set $(S,\phi)$ we now attach the following
function in $\mathcal{L}_{S}(\R^{d}):$

\begin{equation}
\psi_{(S,\phi)}:=\psi_{\mu_{(S,\phi)}}-C\label{eq:def of psi S phi as shifted eq}
\end{equation}
where $C$ is the constant appearing in the previous lemma.

\subsection{The projection operator $P_{S}$ }

In this section we assume that $\alpha\leq2$  (so that the domination
principle \ref{eq:dom princ} applies).

Now assume given a\emph{ weighted set} $(S,\phi),$ i.e. a subset
$S$ of $\R^{d}$ and a continuous function $\phi\in C(S).$ Consider
the function
\[
(\Pi_{S}\phi)(x):=\sup_{\mathcal{L}_{c}(\R^{d})}\{\psi(x):\,\,\,\psi\leq\phi\,\,\text{on \ensuremath{S}}\}
\]
Its upper semi-continuous regularization will be denoted by
\begin{equation}
P_{S}\phi:=(\Pi_{S}\phi)^{*}\label{eq:def of P S}
\end{equation}

\begin{prop}
\emph{\label{prop:P phi as equil potential and og relation}Let} $(S,\phi)$
be a weighted subset and assume that $S$ is compact and non-polar.
Then 
\begin{equation}
\psi_{(S,\phi)}=P_{S}\phi\label{eq:eq pot as proj}
\end{equation}
In particular, $\Delta(P_{S}\phi)$ is the unique minimizer of the
functional $E_{\phi}$ on $\mathcal{P}(S).$ Moreover, as a consequence
of \ref{eq:eq pot as proj},
\[
P_{S}\phi\in\mathcal{E}_{S}(\R^{d})
\]
 and the operator $P_{S}$ satisfies the following ``orthogonality
relation''
\begin{equation}
\int\left(\phi-P_{S}\phi\right)\Delta(P_{S}\phi)=0.\label{eq:og relation}
\end{equation}
More generally, if $S$ is bounded and equal to the union of increasing
compact subsets, then $P_{S}\phi\in\mathcal{E}_{\overline{S}}(\R^{d})$
 for any given continuous function $\phi$ on $\R^{d}.$ 
\end{prop}

\begin{proof}
Combining Lemma \ref{lem:e-l eq for min} with the domination principle
\ref{eq:dom princ} gives $\Pi_{S}\phi\leq\psi_{(S,\phi)}$ and hence
$P_{S}\phi\leq\psi_{(S,\phi)}.$ Indeed, let $\psi$ be a candidate
for the sup defining $\Pi_{S}\phi.$ By Lemma \ref{eq:def of Leg transform}
$\psi\leq\phi\leq\psi_{(S,\phi)}$ on the support of $\Delta(\psi_{(S,\phi)}).$
Hence, the domination principle implies that $\psi\leq\psi_{(S,\phi)}$
everywhere. Moreover, since, by Lemma \ref{lem:e-l eq for min}, $\psi_{(S,\phi)}\leq\phi$
on $S-N$ where $N$ is polar we also have $\psi_{(S,\phi)}\leq P_{S-N}\phi.$
The proof is thus concluded by invoking the fact that $P_{T\cup N}\phi=P_{T}\phi$
for any bounded Borel set $T$ and polar subset $N$ (applied to $T:=S-N).$
To see this first note that, trivially, $P_{T\cup N}\leq P_{T}.$
To prove the converse fix $\psi_{N}\in\mathcal{L}_{c}(\R^{d})$ such
that $\psi_{N}=-\infty$ on $N$ and $\psi_{N}\leq\phi$ on $T.$
The existence of $\psi_{N}$ follows form the fact that, by definition,
any polar subset $N$ is contained in the $-\infty-$locus of some
potential $\psi_{N}.$ Using the compactness of $S$ we can then arrange
that $\psi_{N}\in\mathcal{L}_{c}(\R^{d})$ and $\psi_{N}\leq\phi$
on $S$ and hence also $\psi_{N}\leq\phi$ on $T.$ Thus, for any
$\psi\in\mathcal{L}_{c}(\R^{d})$ such that $\psi\leq0$ on $T$ we
get $\psi_{\epsilon}:=(1-\epsilon)\psi+\epsilon\psi_{N}\leq\phi$
on $T\cup N.$ Hence, $\psi_{\epsilon}\leq\Pi_{T\cup N}\phi.$ Letting
$\epsilon\rightarrow0$ gives $\Pi_{T}\phi\leq\Pi_{T\cup N}\phi$
on the complement of the polar subset $\{\psi_{N}=-\infty\}$ and
hence $P_{T}\phi\leq P_{T\cup N}\phi$ everywhere, as desired (using
that if $\psi_{\mu}\leq\psi_{\nu}+C$ q.e. for a measure $\nu$ of
finite energy and a constant $C,$ then $\psi_{\mu}\leq\psi_{\nu}+C$
everywhere, as a special case of the domination principle, since $\nu$
does not charge polar subsets).

Finally, if $S$ it the union of increasing compact subsets $K_{i}$
then $P_{S}\phi\leq\psi_{i}:=P_{K_{i}}\phi$ for any $i.$ By the
previous step $\psi_{i}$ is a decreasing sequence in $\mathcal{E}_{\overline{S}}(\R^{d})$
and $\psi_{i}\leq\phi$ on $K_{i}-N_{i},$ where $N_{i}$ is polar.
Hence, by Prop \ref{prop:compact}, $\psi_{i}$ converges in $L_{loc}^{1}$
to $\psi_{\infty}\in\mathcal{E}_{\overline{S}}(\R^{d}),$ where $\psi_{\infty}\leq\phi$
on $S-N,$ where $N$ is the union of the polar sets $N_{i}$ and
hence polar (since the outer capacity is sub-additive). This means
that $P_{S}\phi\leq\psi_{\infty}$ and $\psi_{\infty}\leq P_{S-N}\phi.$
Since , as explained in the proof of the previous step, $P_{S-N}\phi=P_{S}\phi$
this concludes the proof. 
\end{proof}
\begin{lem}
\label{lem:The proj oper}Let $S$ be a compact subset. The operator
$P_{S}$ defines a (non-linear) increasing concave operator from $\mathcal{C}(S)$
onto $\mathcal{L}_{S}(\R^{d}).$ Moreover, for any $\psi\in\mathcal{L}_{c}(\R^{d})$
\[
P_{S}\psi\geq\psi
\]
with equality if $\psi\in\mathcal{L}_{S}(\R^{d}).$ Hence, $P_{S}$
is a projection operator, i.e. $P_{S}^{2}=P_{S}.$
\end{lem}

\begin{proof}
It follows directly from the definitions that $\phi_{0}\leq\phi_{1}$
implies that $\Pi_{S}\phi_{0}\leq\Pi_{S}\phi_{1}$ and hence also
$P_{S}\phi_{0}\leq P_{S}\phi_{1},$ i.e. $P_{S}$ is increasing. Concavity
of $\Pi_{S}$ follows directly from the definition as a sup of linear
functionals (defined by evaluation) and this implies the concavity
of $P_{S},$ as well. Next, if $\psi\in\mathcal{L}_{c}(\R^{d})$ then
$\Pi_{S}\psi\geq\psi$ (since $\psi$ is a candidate for the sup defining
$\Pi_{S}\psi).$ Since $\psi^{*}=\psi$ it follows that $P_{S}\psi\geq\psi.$
The fact that $P_{S}\psi\in\mathcal{L}_{S}(\R^{d})$ was proved in
the previous proposition. Conversely, if $\psi\in\mathcal{L}_{S}(\R^{d})$
then it follows, directly from the domination principle, that $\Pi_{S}\psi\leq\psi$
and hence also $P_{S}\psi\leq\psi\leq P_{S}\psi,$ which proves the
projection property in question. 
\end{proof}

\subsection{The primitive functional $\mathcal{E}$ on $\mathcal{L}_{c}(\R^{d})$
and its projection $\mathcal{F}$ to $C(S).$}

The operator $\Delta$ can be naturally identified with  a one-form
on the convex space $\mathcal{L}_{c}(\R^{d})\cap C(\R^{d}):$ 
\[
\left\langle \Delta\psi,v\right\rangle :=\int\Delta\psi v.
\]
According to the next proposition this one-form admits a primitive
that we shall denote by $\mathcal{E},$ i.e. a functional on $\mathcal{L}_{c}(\R^{d})\cap C(\R^{d})$
whose differential is the operator $\Delta:$
\[
(d\mathcal{E})(\psi)=\Delta\psi
\]
Since $\mathcal{L}_{c}(\R^{d})\cap C(\R^{d})$ is convex the primitive
$\mathcal{E}$ is uniquely determined up to an overall constant, which
may be fixed by imposing the normalization condition 
\[
\mathcal{E}(\psi_{0})=0
\]
 for a fixed reference element $\psi_{0}\in\mathcal{L}_{c}(\R^{d})\cap C(\R^{d}).$
We will sometimes use a subscript $\mathcal{E}_{\psi_{0}}$ to indicate
the dependence on the choice of $\psi_{0}.$ Integrating along an
affine line in $\mathcal{L}_{c}(\R^{d})\cap C(\R^{d})$ suggests the
following explicit formula, that we shall take as the definition of
$\mathcal{E}_{\psi_{0}}$ on the whole space $\mathcal{L}_{c}(\R^{d}):$ 

\[
\mathcal{E}_{\psi_{0}}(\psi):=\frac{1}{2}\int(\psi-\psi_{0})\left(\Delta\psi+\Delta\psi_{0}\right)\in[-\infty,\infty[
\]
Moreover, it will be convenient to allow the reference $\psi_{0}$
to be in $\mathcal{E}_{c}(\R^{d}).$
\begin{prop}
\label{prop:prop of beaut E}The functional $\mathcal{E}_{\psi_{0}}$
on $\mathcal{L}_{c}(\R^{d})$ has the following properties 
\begin{enumerate}
\item $\mathcal{E}_{\psi_{0}}(\psi)>-\infty$ iff $\psi\in\mathcal{E}_{c}(\R^{d}).$
Moreover, $\Delta\psi_{j}\rightarrow\Delta\psi$ in $\mathcal{P}(S)$
and $E(\Delta\psi_{j})\rightarrow E(\Delta\psi_{j})$ for $S$ compact
iff $\psi_{j}\rightarrow\psi$ in $L_{loc}^{1}$ and $\mathcal{E}_{\psi_{0}}(\psi_{j})\rightarrow\mathcal{E}_{\psi_{0}}(\psi)$
\item $\mathcal{E}_{\psi_{0}}$ is usc on $\mathcal{L}_{c}(\R^{d})$
\item Given $\psi_{1},\psi_{2}\in\mathcal{E}_{c}(\R^{d})$ we have
\begin{equation}
\frac{d\mathcal{E}_{\psi_{0}}(\psi_{1}+t(\psi_{2}-\psi_{1}))}{dt}_{|t=0}=\int\Delta\psi_{1}(\psi_{2}-\psi_{1})\label{eq:derivative of beaut E along affine}
\end{equation}
\item $\mathcal{E}_{\psi_{0}}$ is concave on $\mathcal{L}_{c}(\R^{d})$ 
\item $\mathcal{E}_{\psi_{0}}$ is strictly increasing on $\mathcal{E}_{c}(\R^{d}):$
$if$ $\psi\leq\Psi,$ then $\mathcal{E}_{\psi_{0}}(\psi)\leq\mathcal{E}_{\psi_{0}}(\Psi)$
with equality iff $\psi=\Psi.$ 
\item The following cocycle property holds: for any triple $\psi_{i}\in\mathcal{E}_{c}(\R^{d})$
the difference
\[
\mathcal{E}_{\psi_{0}}(\psi_{2})-\mathcal{E}_{\psi_{0}}(\psi_{1})
\]
 is independent of $\psi_{0}.$
\item For any $c\in\R$ we have $\mathcal{E}_{\psi_{0}}(\psi+c)=\mathcal{E}_{\psi_{0}}(\psi)+c$ 
\end{enumerate}
\end{prop}

\begin{proof}
To prove item 1 we may, by the cocycle property 6 proved below, assume
that $\Delta\psi_{0}=\rho dx$ for a continuous function $\rho$ with
compact support. Now decompose 
\[
\mathcal{E}_{\psi_{0}}(\psi)-\frac{1}{2}\int\psi\Delta\psi=-\frac{1}{2}\int\psi_{0}\Delta\psi+\frac{1}{2}\int\psi\Delta\psi_{0}+C_{0}
\]
Since $\psi_{0}$ is bounded on the support of $\Delta\psi$ the first
term in the rhs above is finite and so is the second one since $\psi\in L_{loc}^{1}$.
The same argument proves the convergence statement. Item 2 also follows
from the previous decomposition, using that $W$ is usc, just as in
the proof of Prop \ref{prop:compact}. As for the formula in item
3 it follows directly from the symmetry \ref{eq:recipr}. Similarly,
the concavity of $\mathcal{E}_{\psi_{0}}$ follows from the fact that
$\left\langle \Delta u,u\right\rangle \geq0$ if $u=\psi_{2}-\psi_{1}$
for $\psi_{i}\in\mathcal{E}_{c}(\R^{d})$ (by the positivity \ref{eq:pos def}).
That $\mathcal{E}_{\psi_{0}}$ is increasing follows directly from
item 3, since $\Delta\psi\geq0$ for any $\psi\in\mathcal{L}_{c}(\R^{d})$
and strictly increasing follows from the domination principle when
$\alpha\leq2.$ In the general case it follows from the strict concavity
of $\mathcal{E}_{\psi_{0}},$ which in turn follows from the strict
positivity in \ref{eq:pos def}). The cocycle property in item $6$
follows directly from expressing $\mathcal{E}_{\psi_{0}}(\psi_{2})-\mathcal{E}_{\psi_{0}}(\psi_{1})$
as an integral on $[0,1]$ of the derivative $d\mathcal{E}_{\psi_{0}}(\psi_{1}+t(\psi_{2}-\psi_{1}))/dt$
and noting that, by item $3,$ the derivative is independent of $\psi_{0}.$
Finally, item $7$ also follows directly from item 3.
\end{proof}
\begin{rem}
If $\psi_{0}$ is normalized so that $\psi_{0}=\psi_{\Delta\psi_{0}},$
then it follows directly from the symmetry \ref{eq:recipr} that $-\mathcal{E}_{\psi_{0}}(\psi_{\mu})=E(\mu)+C_{0}.$
However, it will be important to consider the functional $\mathcal{E}$
defined on all of $\mathcal{E}_{c}(\R^{d}).$  
\end{rem}

Next, assume that $\alpha\leq2$ (so that the domination principle
\ref{eq:dom princ} applies). Given a weighted compact and non-polar
set $(S,\phi),$ consider the following functional defined on $C(S):$

\[
-\mathcal{F}_{(S,\phi)}(u):=\mathcal{E}_{\psi_{0}}\circ P_{S}(\phi-u)-\mathcal{E}_{\psi_{0}}(P_{S}(\phi))
\]
Equivalently, this means, by the cocycle property in Prop \ref{prop:prop of beaut E},
that, if choose the particular \emph{canonical reference weight }
\begin{equation}
\psi_{0}:=P_{S}(\phi),\label{eq:reference psi zero adapted to weighted set}
\end{equation}
 (canonically attached to the weighted set $(S,\phi)),$ then 
\begin{equation}
\mathcal{F}_{(S,\phi)}(u)=-\mathcal{E}_{\psi_{0}}\circ P_{S}(\phi-u).\label{eq:def of beaut F}
\end{equation}
This choice of reference $\psi_{0}$ ensures the normalization $\mathcal{F}_{(S,\phi)}(0)=0.$
\begin{prop}
\label{prop:F beauti convex and diff}The functional $\mathcal{F}_{(S,\phi)}$
is convex and Gateaux differentiable on $\mathcal{C}(S)$ and its
differential at $u$ is represented by the measure $\Delta\left(P_{S}(\phi-u)\right),$
i.e. 

\[
\frac{d\mathcal{F}_{(S,\phi)}(u+tv)}{dt}_{|t=0}=\int v\Delta\left(P_{S}(\phi-u)\right)
\]
for any $v\in C(S).$ 
\end{prop}

\begin{proof}
This can be shown directly using the orthogonality relation \ref{eq:og relation}
(as in the complex geometric setting in \cite{b-b}, which covers
the logarithmic case $d=\alpha=2$). Alternatively, by Theorem \ref{thm:Legendre}
below $\mathcal{F}_{(S,\phi)}$ is the Legendre-Fenchel transform
of the strictly convex functional $E_{\omega_{\phi}}.$ Hence, by
basic convex duality theory $\mathcal{F}_{(S,\phi)}$ is Gateaux differentiable
and the differential at $u$ is the minimizer of $E_{\omega_{\phi-u}},$
i.e. $\mu_{\phi-u},$ which is equal to $\Delta P_{S}(\phi-u),$ by
Prop \ref{prop:P phi as equil potential and og relation}. 
\end{proof}

\subsection{Energy and Legendre transforms}

In this section we assume that $\alpha\leq2$ or (so that the domination
principle \ref{eq:dom princ} applies) and show that the functional
$\mathcal{F}_{(S,\phi)}$ can be viewed as a Legendre-Fenchel transform
of $E_{\omega_{\phi}}.$ 
\begin{lem}
Assume that, on $S,$ $\phi=\psi_{\nu}$, for a probability measure
$\nu$ on $S$ (which implies that the canonical reference weight
$\psi_{0}:=P_{S}\phi$ coincides with $\psi_{\nu}).$ Then, for any
$\mu\in\mathcal{P}_{S}(\R^{d})$ such that $E(\mu)<\infty,$ the corresponding
normalized energy functional is given by
\begin{equation}
E_{\omega_{\phi}}(\mu)=\mathcal{E}_{\psi_{0}}(\psi_{\mu})-\int(\psi_{\mu}-\psi_{0})\mu.\label{eq:E as realized leg tr in lemma}
\end{equation}
 
\end{lem}

\begin{proof}
First rewrite
\[
\mathcal{E}_{\psi_{0}}(\psi)=\frac{1}{2}\int(\psi-\psi_{0})\Delta(\psi-\psi_{0})+\int(\psi-\psi_{0})\Delta\psi_{0}
\]
Next, note that, in general, if $\mu$ and $\nu$ have finite energy,
then there exists a constant $C_{1},$ depending on $\nu,$ such that
\[
\frac{1}{2}\int(\psi_{\mu}-\psi_{\nu})\Delta(\psi_{\mu}-\psi_{\nu})=\frac{1}{2}\int\psi_{\mu}\Delta\psi_{\mu}-\int\psi_{\mu}\Delta\psi_{\nu}+C_{1}=:-E_{\phi}(\mu)+C_{1}
\]
 using the symmetry \ref{eq:recipr} in the first equality. This shows
that formula \ref{eq:E as realized leg tr in lemma} holds up to an
over all constant $C_{2}.$ But the rhs in the formula vanishes for
$\psi=\psi_{\nu}$ and so does $E_{\omega_{\phi}}(\nu),$ i.e. the
minimum of $E_{\omega_{\phi}}(\mu)$ is realized for $\mu=\nu,$ as
follows from ``completing the square''. Hence, $C_{2}=0,$ as desired.
\end{proof}
Before stating the next theorem we recall the general definition of
the Legendre-Fenchel transform. Let $f$ be a function on a topological
vector space $V.$ The \emph{Legendre-Fenchel transform} $\widehat{f}$
of $f$ is defined as following convex lower semi-continuous function
$\widehat{f}$ on the topological dual $V^{*}$ 
\begin{equation}
\widehat{f}(w):=\sup_{v\in V}\left\langle v,w\right\rangle -f(v)\label{eq:def of Leg transform}
\end{equation}
in terms of the canonical pairing between $V$ and $V^{*}.$ In the
present setting we will take $V=\mathcal{C}(S)$ and $V^{*}=\mathcal{M}(S),$
the space of all continuous functions and the space of all signed
Borel measures, respectively, on a compact topological space $S.$
Then the Legendre-Fenchel transform is involutive \cite{d-z}. Given
a compact subset $S$ of $\R^{d}$ we will denote by $\chi_{\mathcal{P}(S)}$
the lsc functional on the space of all signed measures $\mathcal{M}(S)$
on $S\Subset\R^{d}$ which is equal to $0$ on $\mathcal{P}(S)$ and
equal to $\infty$ on the complement of $\mathcal{P}(S)$ in $\mathcal{M}(S).$
The definition is made so that the functional $\chi_{\mathcal{P}(S)}+E_{(S,\omega_{\phi})}$
on $\mathcal{M}(S)$ is equal to $E_{(S,\omega_{\phi})}$ if $\mu\in\mathcal{P}(S)$
and otherwise equal to $\infty.$ 
\begin{thm}
\label{thm:Legendre}Let $S$ be a compact subset of $\R^{d}$ and
$\phi$ a continuous function on $S.$ Consider the functional $\chi_{\mathcal{P}(S)}+E_{\omega_{\phi}}$
on the space $\mathcal{M}(S)$ of all signed measures on $S.$ Its
Legendre-Fenchel transform is given by 
\begin{equation}
\widehat{\chi_{\mathcal{P}(S)}+E_{\omega_{\phi}}}=\mathcal{F}_{(S,\phi)},\label{eq:statement thm legendre first}
\end{equation}
 where $\mathcal{F}_{(S,\phi)}$ is the functional defined in formula
\ref{eq:def of beaut F}. Conversely, 
\begin{equation}
\chi_{\mathcal{P}(S)}+E_{\omega_{\phi}}=\widehat{\mathcal{F}_{(S,\phi)}}\label{eq:statement thm Legendre sec}
\end{equation}
Moreover, for any $\mu\in\mathcal{P}(S)$ such that $E(\mu)<\infty$
\begin{equation}
E_{\omega_{\phi}}(\mu)=\mathcal{E}_{P_{S}\phi}(\psi_{\mu})-\int(\psi_{\mu}-\phi)\mu\label{eq:general formula for E normalized in terms of potential}
\end{equation}
\end{thm}

\begin{proof}
In order to prove \ref{eq:statement thm legendre first} and \ref{eq:statement thm Legendre sec}
it is, since the Legendre-Fenchel transform is involutive on $\mathcal{M}(S),$
enough to prove \ref{eq:statement thm legendre first}, or equivalently
that, for any given $u\in\mathcal{C}(S),$
\begin{equation}
\inf_{\mathcal{P}(S)}\left(E_{\omega_{\phi}}(\mu)+\left\langle u,\mu\right\rangle \right)=\mathcal{E}_{P_{S}\phi}(P_{S}(\phi+u)).\label{eq:pf of Legen}
\end{equation}
Setting $\Phi=\phi+u$ and using that the inf above is attained at
$\mu=\Delta(P_{S}\Phi)$ (by Prop \ref{prop:P phi as equil potential and og relation})
 gives 
\[
\inf_{\mathcal{P}(S)}\left(E_{\phi}(\mu)+\left\langle u,\mu\right\rangle \right)=E_{\phi}(\Delta(P_{S}\Phi))+\left\langle \Phi-\phi,\Delta(P_{S}\Phi)\right\rangle .
\]
 Rewriting the first term in the rhs as 
\[
E_{\phi}(\Delta(P_{S}\Phi))=E_{P\phi}(\Delta(P_{S}\Phi))+\left\langle (\phi-P_{S}\phi),\Delta(P_{S}\Phi)\right\rangle ,
\]
 yields 
\[
\inf_{\mathcal{P}(S)}\left(E_{\omega_{\phi}}(\mu)+\left\langle u,\mu\right\rangle \right)=\left(E_{P_{S}\phi}(\Delta(P_{S}\Phi))-E_{\phi}((\Delta(P_{S}\phi))\right)+\left\langle \Phi-P_{S}\phi,\Delta(P_{S}\Phi)\right\rangle .
\]
Using the orthogonality relation in Prop \ref{prop:P phi as equil potential and og relation}
reveals that $E_{\phi}((\Delta(P_{S}\phi))=E_{P\phi}((\Delta(P_{S}\phi)).$
Hence, by the previous lemma the first term in the rhs of the previous
equation may be expressed as 
\[
E_{P_{S}\phi}(\Delta(P_{S}\Phi))=\mathcal{E}_{P_{S}\phi}(\Phi)-\left\langle (\Phi-P_{S}\phi),\Delta(P_{S}\Phi)\right\rangle 
\]
 Since the second term in the latter equation cancels the second term
in former equation we thus get
\[
\inf_{\mathcal{P}(S)}\left(E_{\omega_{\phi}}(\mu)+\left\langle u,\mu\right\rangle \right)=\mathcal{E}_{P_{S}\phi}(\Phi),
\]
 which coincides the right hand side in formula \ref{eq:pf of Legen}. 

To prove the final statement we proceed essentially as above. First
observe that, by definition,
\[
E_{\phi}(\mu)=E_{P_{S}\phi}(\mu)+\left\langle \phi-P_{S}\phi,\mu\right\rangle .
\]
Hence, applying the previous lemma to the weight $P_{S}\phi,$ shows
that there exists a constant $C$ (only depending on $\phi)$ such
that
\[
E_{\omega_{\phi}}(\mu)=\mathcal{E}_{P_{S}\phi}(\psi_{\mu})-\left\langle \psi_{\mu}-P_{S}\phi,\mu\right\rangle +\left\langle \phi-P_{S}\phi,\mu\right\rangle +C=
\]
\[
\mathcal{=E}_{P_{S}\phi}(\psi_{\mu})-\left\langle \psi_{\mu}-\phi,\mu\right\rangle +C.
\]
 Now, evaluating the previous equality for $\mu=\Delta(P_{S}\phi)$
and using that $\psi_{\mu}-P_{S}\phi$ is constant, by Prop \ref{prop:P phi as equil potential and og relation}
(so that we can replace $\psi_{\mu}$ with $P_{S}\phi$ in the rhs
of the previous equation, using item $7$ in Prop \ref{prop:prop of beaut E})
gives 
\[
0=0-\left\langle P_{S}\phi-\phi,\Delta(P_{S}\phi)\right\rangle +C.
\]
Finally, by the orthogonality relation in Prop \ref{prop:P phi as equil potential and og relation},
it follows that $C=0,$ which concludes the proof of the theorem.
\end{proof}

\subsection{\label{subsec:Regularity}Regularity }

In this section we assume that $\alpha\leq2.$ A weighted set $(S,\phi)$
will be said to be\emph{ regular} if $P_{S}\phi\leq\phi$ and a set
$S$ is said to be regular if $(S,0)$ is regular. A compact set $K$
is said to be \emph{locally regular} if it is regular at any point
$x\in K,$ i.e. if $(P_{K\cap U}0)(x)\leq0$ for any open ball $U$
centered at $x.$ In general, $P_{S}\phi\leq\phi$ always holds in
the interior of $S$ (see the beginning of the proof of Lemma \ref{lem:nonthin implies reg}).
\begin{lem}
\label{lem:reg is cont}A non-polar weighted compact set $(S,\phi)$
is regular iff $\sup_{S}(P_{S}\phi-\phi)=0$ iff $P_{S}\phi=\Pi_{S}\phi$
iff $P_{S}\phi$ is continuous.
\end{lem}

\begin{proof}
The first equivalence follows from the extremal definition of $P_{S}\phi,$
combined with Lemma \ref{lem:e-l eq for min} and Prop \ref{prop:P phi as equil potential and og relation}.
To prove the second equivalence we note that if $(S,\phi)$ is regular,
then $P_{S}\phi$ is a candidate for the sup defining $\Pi_{S}\phi$
and hence $P_{S}\phi\leq\Pi_{S}\phi.$ Since the reverse inequality
always holds we conclude that $P_{S}\phi=\Pi_{S}\phi.$ Finally, let
us show $(S,\phi)$ is regular iff $P_{S}\phi$ is continuous. First
assume that $(S,\phi)$ is regular. By the previous step $P_{S}\phi=\Pi_{S}\phi.$
Now, $P_{S}\phi$ is, by construction, usc. Hence, $\Pi_{S}\phi$
is continuous iff it is lsc. Accordingly, to prove that $\Pi_{S}\phi$
is continuous it is enough to show the following claim: the sup defining
$\Pi_{S}\phi$ can be taken over all \emph{continuous} $\psi\in\mathcal{L}_{c}(\R^{d})$
satisfying $\psi\leq\phi.$ To this end first note that there exists
a sequence $\psi_{j}\in\mathcal{L}_{c}(\R^{d})\cap C(\R^{d})$ such
that $\psi_{j}\rightarrow P_{S}\phi$ in $L_{loc}^{1}$ and such that
$\psi_{j}(x)\rightarrow P_{S}\phi(x)$ for any $x$ (as follows from
\cite[Thm 1.11 or Thm 3.7]{la}). Moreover, $\psi_{j}$ may be taken
to in $\mathcal{L}_{K}(\R^{d})$ for some compact set $K$ containing
$S.$ Now, since $(S,\phi)$ is regular Prop \ref{prop:det as BM}
gives that the functional $\psi\mapsto\sup_{S}(\psi-\phi)$ is continuous
on $\mathcal{L}_{K}(\R^{d}).$ Hence, replacing $\psi_{j}$ with $\tilde{\psi}_{j}:=\psi_{j}-\sup_{S}(\psi_{j}-\phi)$
and using that $\sup_{S}(P_{S}\phi-\phi)=0$ we may as well assume
that $\psi_{j}\leq\phi.$ But then the point-wise convergence of $\psi_{j}$
towards $P_{S}\phi$ proves the claim. Hence, $P_{S}\phi$ is continuous.
Conversely, if $P_{S}\phi$ is continuous, then $P_{S}\phi\leq\phi$
q.e on $S$ implies that $P_{S}\phi\leq\phi$ everywhere on $S,$
which means that $(S,\phi)$ is regular. 
\end{proof}
\begin{lem}
\label{lem:strong reg iff loc regular}Let $K$ be a non-polar compact
set $K.$ Then $(K,\phi)$ is regular for any $\phi\in C(K)$ iff
$K$ is locally regular.
\end{lem}

\begin{proof}
This is shown as in the complex setting \cite[Prop 6.1]{ng}. First
assume that $K$ is locally regular. Since $\phi$ is continuous we
have that $\phi\leq\phi(x)+\delta(\epsilon)$ on an open ball $B_{\epsilon}(x)$
of radius $\epsilon$ centered at $x,$ where $\delta(\epsilon)\rightarrow0$
as $\epsilon\rightarrow0.$ Hence, 
\[
P_{K}\phi\leq P_{K\cap B_{\epsilon}(x)}\phi\leq P_{K\cap B_{\epsilon}(x)}(\phi(x)+\delta(\epsilon))\leq P_{K\cap B_{\epsilon}(x)}0+\phi(x)+\delta(\epsilon).
\]
Letting $\epsilon\rightarrow0$ thus gives $(P_{K}\phi)(x)\leq\phi(x),$
showing that $(K,\phi)$ is regular. Conversely, assume that $(K,\phi)$
is regular for all $\phi\in\mathcal{C}(K).$ Take a point $x\in X$
and an open ball $B$ centered at $x.$ Define a function $\phi$
on $K$ by setting $\phi=0$ on $K\cap B$ and $\phi=P_{K\cap B}0$
on $K-B.$ The function $\phi$ is clearly usc and hence there exists
a sequence $\phi_{j}\in\mathcal{C}(K)$ decreasing to $\phi.$ Now,
if $\psi$ is candidate for the sup defining $\Pi_{K\cap B}0,$ then
$\psi\leq0$ on $K\cap B$ and hence $\psi\leq\Pi_{K\cap B}0$ everywhere.
As a consequence, $\psi\leq\phi$ on $K,$ which, in turn, implies
$\psi\leq P_{K}\phi_{j}\leq\phi_{j}$ for any $j,$ using in the last
equality that $(K,\phi_{j})$ is assumed regular. Hence, taking the
sup over all such $\psi$ and using that $\phi_{j}$ is continuous
on $K$ gives $P_{K\cap B}0\leq\phi_{j}.$ Finally, letting $j\rightarrow\infty$
we conclude that $P_{K\cap B}0\leq0$ on $K\cap B,$ as desired. 
\end{proof}

\subsubsection{Compact domains}

Now consider the case when $K$ is a compact domain, i.e. $K$ is
the closure of an open bounded set $\Omega.$ Following standard classical
terminology $\Omega$ is said to be\emph{ thin at $x_{0}\in\partial K:=K-\Omega$}
if there exists some potential $\psi_{\mu}$ such that $\limsup_{x\rightarrow x_{0}}\psi_{\mu}(x)<\psi_{\mu}(x_{0})$,
assuming $x\in\Omega$ (see the definition and discussion in \cite[page 307]{la}). 
\begin{lem}
\label{lem:nonthin implies reg}Let $K$ be a compact domain, i.e.
$K$ is the closure of an open bounded set $\Omega.$ If $\Omega$
is non-thin at all boundary points, then $(K,\phi)$ is regular for
any continuous $\phi.$ 
\end{lem}

\begin{proof}
Set $\psi:=P_{K}\phi.$ Then it follows from the continuity of $\phi$
that $\psi\leq\phi$ in $\Omega.$ Indeed, by definition $\psi$ is
the upper semi-continuous regularization $(\Pi_{K}\phi)^{*}$ of the
function $\Pi_{K}\phi,$ which satisfies $\Pi_{K}\phi\leq\phi$ on
$K.$ Thus, given $x\in\Omega$ and using that any ball $B_{\epsilon}(x)$
centered at $x$ of sufficiently small radius $\epsilon$ is contained
on $\Omega,$ we deduce that 
\[
\psi(x):=\lim_{\epsilon\rightarrow0}\sup_{B_{\epsilon}(x)}(\Pi_{K}\phi)\leq\lim_{\epsilon\rightarrow0}\sup_{B_{\epsilon}(x)}\phi=\phi(x),
\]
 since $\phi$ is continuous. Finally, given $x_{0}\in K-\Omega,$
the assumption of non-thinness implies that $\psi(x_{0})\leq\limsup_{x\rightarrow x_{0}}\psi(x)\leq\phi(x_{0})$
since $x\in\Omega$ and $\phi$ is continuous.
\end{proof}
The notion of thinness of a set $E$ at a point $x_{0}$ can equivalently
be formulated in terms of Wiener's capacity criterion (see \cite[Thm 5.2]{la}
and \cite[Thm 5.10]{la}), which, in turn, is equivalent to the following
capacity criterion (\cite[formula 5.1.7]{la}): for a given number
$q\in]0,1[,$
\begin{equation}
\sum_{m=1}^{\infty}\frac{\mathcal{C}_{\alpha}(E^{(m)})}{q^{m(d-\alpha)}}<\infty,\,\,\,E^{(m)}:=E\cap\left\{ x\in\R^{d}:|x-x_{0}|<q^{m}\right\} ,\label{eq:capacity criterion}
\end{equation}
 where $\mathcal{C}_{\alpha}$ denotes the capacity corresponding
to $\alpha$ (recalled in Section \ref{subsec:Capacities-and-determining}
in the appendix). Here we have assumed that we are not in the logarithmic
case $d=\alpha=2,$ where a similar capicity criterion applies (see
\cite[Thm 5.6]{la}). 
\begin{example}
In the three-dimensional Coulomb case, $d=3$ and $\alpha=2,$ algebraic
cusps are non-thin at the vertex, while Lebesgue cusps are thin at
the vertex \cite[page 287]{la}. These cusp as defined for a given
$m>0,$ by the surfaces of revolution in $\R^{3}$ where $\theta\leq r^{m}$
and $\theta\leq e^{-m/r}$ respectively (using planar polar coordinates).
See also \cite{is} for more general results in the Coulomb case. 
\end{example}

For general $\alpha\in]0,2]$ the capacity criterion above yields
the following
\begin{prop}
\label{prop:reg of smooth domains}Let $K$ be a compact domain in
$\R^{d}$ satisfying the \emph{interior cone condition}, i.e. for
any point $x_{0}\in\partial K$ there exists a cone contained in $K$
with a vertex at $x_{0}.$ Then $K$ is locally regular.
\end{prop}

\begin{proof}
By the previous lemma it is equivalent to show that $(P_{K}\phi)(x)\leq\phi(x)$
for any $\phi\in C(K)$ and $x\in K.$ Given $x_{0}\in\partial K$
denote by $C_{x_{0}}$ a cone in $K$ with vertex at $x_{0}.$ Note
that it is enough to verify that interior of a cone $C_{x_{0}}$ is
non-thin at the  vertex. Indeed, since $\phi$ is continuous $P_{K}\phi\leq0$
in the interior of $K$ and hence in the interior of $C_{x_{0}}.$
Taking a sequence of points $x_{i}$ in the interior of $C_{x_{0}}$
converging towards $x_{0}$ and exploiting that $C_{x_{0}}$ is non-thin
at $x_{0}$ it thus follows that $P_{S}\phi(x_{0})\leq\limsup_{x\rightarrow x_{0}}P_{S}\phi(x)\leq\phi(x_{0}),$
using in the last inequality that $\phi$ is continuous. Finally,
to verify that the interior of a compact cone $C_{x_{0}}$ is non-thin
at $x_{0}$ first observe that, since $W_{\alpha}$ is translationally
invariant, we may as well assume that $x_{0}$ is the origin. But
then $E:=C_{x_{0}}$ has the scaling property that $E^{(m)}=q^{m}\cdot E^{(1)}.$
It thus follows from the scaling properties of the kernel $W_{\alpha}$
that $\mathcal{C}_{\alpha}(E^{(m)})=q^{m(d-\alpha)}\mathcal{C}_{\alpha}(E^{(1)})$
and hence the capacity criterion \ref{eq:capacity criterion} is trivially
satisfied (for notational simplicity we have assued that we are not
in the logarithmic case $d=2=\alpha,$ where essentially the same
argument applies). 
\end{proof}

\subsection{Determining measures}

The definition of (strongly) determining measures was given in Section
\ref{subsec:Energy-approximation-and}. It may be equivalently formulated
as follows. A measure $\nu$ on $\R^{d}$ is said to be \emph{determining
for a weighted set $(S,\phi)$ }if for all $\psi\in\mathcal{L}_{c}(\R^{d})$
\[
\sup_{S}e^{\psi-\phi}=\left\Vert e^{\psi-\phi}\right\Vert _{L^{\infty}(S,\nu)}
\]
A measure $\nu$ is said to be\emph{ determining for $S$} if $\nu$
is determining for $(S,0)$ and\emph{ strongly determining} if $\nu$
is determining for $(S,\phi)$ for all $\phi\in C(S).$ Similarly
we will say that $\nu$ is (strongly) determining if it is (strongly
determining) for its support. 
\begin{prop}
\label{prop:determ implies reg}If $\mu_{0}$ does not charge polar
subsets, has compact support $S_{0}$ and is (strongly) determining,
then $S_{0}$ is (locally) regular. 
\end{prop}

\begin{proof}
Since $\psi_{\phi}:=P_{S_{0}}\phi\leq\phi$ q.e. and $\mu_{0}$ does
not charge polar subsets it follows that $\psi_{\phi}\leq\phi$ a.e.
wrt $\mu_{0}.$ By assumption this means that $\psi_{\phi}\leq\phi$
on $S_{0},$ i.e. $(S_{0},\phi)$ is regular, as desired. 
\end{proof}
Any compact weighted regular compact subset carries determining measures: 
\begin{prop}
\label{prop:eq meas is det}Let $(K,\phi)$ be a regular weighted
compact set. Then the corresponding equilibrium measure $\mu_{(K,\phi)}$
is determining for $(K,\phi).$ As a consequence, if $\mu_{0}$ has
the property that $\mu_{(K,\phi)}$ is absolutely continuous with
respect to $\mu_{0},$ then $\mu_{0}$ is also determining for $(K,\phi).$
\end{prop}

\begin{proof}
Assume that $\psi\leq\phi$ a.e. wrt $\mu_{(K,\phi)}.$ Recall that,
by Prop \ref{prop:P phi as equil potential and og relation}, $\mu_{(K,\phi)}=\Delta(P_{K}\phi)$
and $\phi=P_{K}\phi$ a.e. wrt to $\Delta(P_{K}\phi).$ Hence, by
the domination principle \ref{eq:dom princ}, $\psi\leq P_{K}\phi$
everywhere. Finally, since $(K,\phi)$ is assumed regular it thus
follows that $\psi\leq\phi$ everywhere, as desired.

This follows directly from the domination principle \ref{eq:dom princ}.
\end{proof}

\subsubsection{Compact domains}

We will next consider the special case when $S_{0}$ is a compact
domain (i.e. $S_{0}$ is the closure of an open bounded set), using
the following lemma:
\begin{lem}
Lebesgue measure $dx$ is strongly determining for any open subset
$U\subset\R^{d}.$ In other words, the measure $1_{U}dx$ is strongly
determining. 
\end{lem}

\begin{proof}
Fix a smooth compactly supported function $\rho$ such that $\rho dx\in\mathcal{P}(\R^{n})$
and set $\rho_{\delta}:=\delta^{n}\rho(x/\delta).$ Now, if $\psi_{\mu}\leq\phi$
a.e. on $U,$ then, for any given compact subset $K$ of $U,$ there
exists a sequence $\epsilon_{j}\rightarrow0$ such that $\psi_{j}:=\psi_{\mu}*\rho_{j^{-1}}\leq\phi+\epsilon_{j}$
on $K.$ But $\psi_{j}=\psi_{\mu*\rho_{j^{-1}}}$ and hence, by \ref{eq:limsup reg is psi},
\[
\psi(x)\leq\limsup_{j\rightarrow\infty}(\phi(x)+\epsilon_{j})=\phi(x)
\]
for any $x\in K$ and hence for any $x\in U.$ 
\end{proof}
Note that in the Coulomb case, $\alpha=2,$ the previous lemma follows
directly from the submean property of subharmonic functions. 
\begin{prop}
\label{prop:top domain}Let $S_{0}$ be a compact domain, i.e. $S_{0}=\bar{\Omega},$
where $\Omega$ is an open bounded set. If $\Omega$ is non-thin at
all boundary points, i.e. at all points in $X-\Omega,$ then $\mu_{0}:=1_{\Omega}dx$
is strongly determining for $S_{0}.$ In particular, if $S_{0}$ satisfies
the interior cone condition (appearing in Prop \ref{prop:reg of smooth domains}),
then $1_{\Omega}dx$ is strongly determining.
\end{prop}

\begin{proof}
Assume that $\psi_{\mu}\leq\phi$ a.e. wrt $1_{\Omega}dx.$ Then,
by the previous lemma, $\psi_{\mu}\leq\phi$ on $\Omega.$ Now take
$x_{0}\in S_{0}-\Omega.$ By the non-thinness assumption $\psi_{\mu}(x_{0})\leq\limsup_{x\rightarrow x_{0}}\psi_{\mu}(x)$
for any sequence of points $x\in\Omega$ converging towards $x_{0}.$
Since $\phi$ is continuous we deduce that $\psi_{\mu}(x_{0})\leq\phi(x_{0}),$
showing that $\mu_{0}:=1_{\Omega}dx$ is strongly determining. 
\end{proof}
In particular, $1_{\Omega}dx$ is strongly determining if $\Omega$
is a bounded Lipschitz domain. Moreover, as shown next, the Hausdorff
measure on the boundary $\partial\Omega$ of a bounded Lipschitz domain
is strongly determining for $\partial\Omega,$ in the Coulomb case.

\subsubsection{Lipschitz hypersurfaces}
\begin{thm}
\label{thm:Lip}Consider the Coulomb case $\alpha=2.$ The $(d-1)-$dimensional
Hausdorff measure $\mu_{0}$ on a Lipschitz hypersurface $K$ in $\R^{d}$
without boundary is strongly determining. 
\end{thm}

\begin{proof}
Denote by $T$ be the closure of a bounded tubular neighborhood of
$K$ and decompose it into two closed domains $T_{\pm}$ intersecting
along $K:$ 
\[
T=T_{-}\bigcup T_{+}.
\]
 By assumption, the domains $T_{\pm}$ may be taken to be compact
Lipschitz domains. Given a potential $\psi$ in $\R^{d}$ we fix a
constant $C$ such that $\psi\leq C$ on $T.$ Denote by $f_{\pm}$
the continuous function on $\partial T_{\pm}$ which is equal to a
given continuous function $\phi$ on $K$ and equal to $C$ on $\partial T.$
We denote by $h_{\pm}$ the harmonic extension of $f_{\pm}$ to $T_{\pm}.$
The function $h_{\pm}$ is in $\mathcal{C}(T_{\pm}),$ as follows
from the fact that $T_{\pm}$ satisfies the interior cone condition
(see \ref{prop:reg of smooth domains}). Now, by assumption, $\psi\leq f_{\pm}$
almost everywhere with respect to the Hausdorff measure $\sigma_{\pm}$
on $\partial T_{\pm}.$ But then it follows from \cite{da} that 
\begin{equation}
\psi\leq h_{\pm},\,\,\text{in the interior of\,\ensuremath{T_{\pm}}. }\label{eq:psi smaller than h plus minus}
\end{equation}
 Accepting, this for the moment and denoting by $h$ the continuous
function on $T$ which is equal to $h_{\pm}$ on $T_{\pm}$ we get
$\psi\leq h$ on $T.$ Hence, since $\psi$ is subharmonic, for any
$x\in K$ we have
\[
\psi(x)\leq\frac{1}{|B_{\delta}(x)|}\int_{B_{\delta}(x)}hdx
\]
Letting $\delta\rightarrow0$ and using that $h$ is continuous and
equal to $\phi(x)$ at $x$ we conclude that $\psi(x)\leq\phi(x),$
as desired. 

Finally, we note that the inequality \ref{eq:psi smaller than h plus minus}
is a standard consequence of the result in \cite{da}, saying, in
particular, that for a Lipschitz domain $D$ the harmonic measure
$\nu_{x}$ on $\partial D$ is absolutely continuous wrt the $(d-1)-$dimensional
Hausdorff measure $\sigma$ on $\partial D,$ for any $x\in\partial D.$
Indeed, by the standard maximum principle for subharmonic functions
\begin{equation}
\psi(x)\leq\int\nu_{x}\psi\label{eq:psi smaller than harmon ext}
\end{equation}
 if $x$ is in the interior of $D$ (this is immediate in the case
when $\psi$ is continuous in a neighborhood of $D$ and then general
case then follows writing $\psi$ as a decreasing limit of such functions).
Hence, applying \ref{eq:psi smaller than harmon ext} to $D=T_{\pm}$
and using that $\psi\leq f_{\pm}$ almost everywhere with respect
$\sigma_{\pm}$ gives
\[
\psi(x)\leq\int\nu_{x}f_{\pm}=h_{\pm},
\]
 proving \ref{eq:psi smaller than h plus minus}. 
\end{proof}
\begin{rem}
The method of proof can be adapted to many other situations. Indeed,
it only requires the existence of a neighborhood $T$ of $K$ such
that the harmonic measures on the corresponding boundaries of $T_{\pm}$
are absolutely continuous wrt the corresponding Hausdorff measures. 
\end{rem}

\section{\label{sec:Determining-measures-vs energ}Determining measures, Energy
approximation and Gamma-convergence }

Given a probability measure $\mu_{0}$ with compact support $S_{0}$
and a continuous function $\phi$ on $\R^{d}$ the corresponding \emph{free
energy functional $F_{\phi,\beta}$ at inverse temperature }$\beta\in]0,\infty]$
is defined by the following functional on $\mathcal{P}_{c}(\R^{d}):$

\begin{equation}
F_{\phi,\beta}(\mu)=E_{\phi}(\mu)+\frac{1}{\beta}D_{\mu_{0}}(\mu),\,\,\,E_{\phi}(\mu):=E(\mu)+\int\phi\mu\label{eq:def of free energy with phi text}
\end{equation}
 where $D_{\mu_{0}}$ denotes the \emph{entropy of $\mu$ relative
to $\mu_{0}$ }(also known as the\emph{ Kullback\textendash Leibler
Divergence}),\emph{ }i.e. 
\begin{equation}
D_{\mu_{0}}(\mu):=\int_{\R^{d}}\log\frac{\mu}{\mu_{0}}\mu,\label{eq:def of rel entropy}
\end{equation}
 when $\mu$ is absolutely continuous wrt to $\mu_{0}$ and otherwise
$D_{\mu_{0}}(\mu):=\infty.$ We define $F_{\phi,\infty}:=E_{\phi}.$
Note that 
\[
F_{\phi,\beta}\geq E_{\phi},
\]
 since $D_{\mu_{0}}\geq0.$ We also recall that the functional $D_{\mu_{0}}$
is lower semi-continuous (lsc) on $\mathcal{P}(K),$ for any given
compact subset $K$ \cite{d-z}. 

Similarly, when replacing the energy $E_{\phi}$ with its normalized
version $E_{\omega_{\phi}}$ \ref{eq:def of E omega phi} we will
write 
\[
F_{\omega_{\phi},\beta}:=E_{\omega_{\phi}}(\mu)+\frac{1}{\beta}D_{\mu_{0}}(\mu)=F_{\phi,\beta}-\inf_{\mathcal{P}(S_{0})}E_{\phi},
\]
 where $S_{0}$ denotes the support of $\mu_{0}.$ When $\phi=0$
we will simply use the notation $F_{\beta}:=F_{\phi,\beta}=F_{0,\beta}.$ 

\subsection{\label{subsec:The-Energy-Approximation}The Energy Approximation
property vs Gamma-convergence of free energies}

We recall the definition of \emph{Gamma-convergence,} introduced by
De Georgi (see the book \cite{bra} for background on Gamma-convergence):
\begin{defn}
\label{def:gamma}A family of functions $F_{\beta}$ on a topological
space \emph{$\mathcal{M}$ is said to Gamma-converge }to a function
$F$ on $\mathcal{M},$ as $\beta\rightarrow\infty,$ if 
\begin{equation}
\begin{array}{ccc}
\mu_{\beta}\rightarrow\mu\,\mbox{in\,}\mathcal{M} & \implies & \liminf_{\beta\rightarrow\infty}F_{\beta}(\mu_{\beta})\geq F(\mu)\\
\forall\mu & \exists\mu_{\beta}\rightarrow\mu\,\mbox{in\,}\mathcal{M}: & \lim_{\beta\rightarrow\infty}F_{\beta}(\mu_{\beta})=F(\mu)
\end{array}\label{eq:def of gamma conv}
\end{equation}
A sequence (family) $\mu_{\beta}$ as in the last point above is called
a\emph{ recovery sequence (family) }for $\mu.$ The limiting functional
$F_{\infty}$ is automatically lower semi-continuous on $\mathcal{M}.$ 
\end{defn}

We first make the following simple observation:
\begin{lem}
\label{lem:EA vs Gamma}A measure $\mu_{0}$ satisfies the Energy
Approximation Property (section \ref{subsec:Energy-approximation-and})
iff the free energy $F_{\beta}$ Gamma-converges towards the energy
$E$ on $\mathcal{P}(K).$
\end{lem}

\begin{proof}
First suppose that the Gamma-convergence holds. Given $\mu\in E(\mu)$
such that $E(\mu)<\infty$ we take a recovery family $\mu_{\beta},$
i.e. 
\[
E(\mu)\geq\limsup_{\beta\rightarrow\infty}F_{\beta}(\mu_{\beta}).
\]
 Since $F_{\beta}\geq E$ this directly implies the inequality \ref{eq:bound on limsup E}
and hence the Energy Approximation Property. To prove the converse
first observe that, since $E$ and $D_{\mu_{0}}$ are lsc on $\mathcal{P}(K)$
it is enough to show that for any $E(\mu)<\infty$ there exists a
recovery family, which, in turn, is equivalent to finding a family
$\mu_{\beta}$ such that $(i)\,\mu_{\beta}\rightarrow\mu$ in $\mathcal{P}(K),$
$(ii)\,E(\mu_{\beta})\rightarrow E(\mu),$ and $\beta^{-1}D_{\mu_{0}}(\mu_{\beta})\rightarrow0.$
Moreover, the latter condition may be replaced by the condition that
$(iii)\,D_{\mu_{0}}(\mu_{\beta})<\infty.$ Indeed, by relabeling the
family $\mu_{\beta}$ we can then arrange that $D_{\mu_{0}}(\mu_{\beta})\leq\beta^{1/2},$
say. Now, assuming that the energy approximation property holds, there
exists a family $\mu_{\beta}$ satisfying the conditions $(i)$ and
$(ii)$ and such that $\mu_{\beta}=\rho_{\beta}\mu_{0}$ for some
$\rho_{\beta}\in L^{1}(\mu_{0}).$ For any positive integer $j$ we
set 
\[
\mu_{\beta,j}:=\max(\rho_{\beta},j)\mu_{0}/\int\max(\rho_{\beta},j)\mu_{0}\in\mathcal{P}(K),
\]
 which satisfies $D_{\mu_{0}}(\mu_{\beta,j})<\infty.$ Moreover, by
the monotone convergence theorem, $E(\mu_{\beta,j})\rightarrow E(\mu_{\beta})$
and $\mu_{\beta,j}\rightarrow\mu_{\beta},$ as $j\rightarrow\infty.$
Hence, we can conclude using a standard diagonal argument. 
\end{proof}
Gamma-convergence is stable under addition by continuous functionals,
as follows directly from the definition. We will make use of the following
criterion for Gamma-convergence on $\mathcal{P}(K),$ formulated in
terms of the Legendre-Fenchel transform (Definition \ref{eq:def of Leg transform}):
\begin{prop}
\label{prop:crit for gamma conv}Let $F_{\beta}$ be a family of functions
on the space $\mathcal{P}(K)$ of all probability measures on a compact
space $K$ and extend $F_{\beta}$ by infinity to all of $\mathcal{M}(K).$
Assume that
\[
\lim_{\beta\rightarrow\infty}\widehat{F_{\beta}}(\phi)=f(\phi)
\]
for any $\phi\in C(X)$ and that $f$ defines a Gateaux differentiable
function on $\mathcal{C}(K).$ Then $F_{\beta}$ Gamma-converges to
$\widehat{f}$ on $\mathcal{M}(K)$ (the converse holds without any
differentiability assumption).
\end{prop}

See \cite{berm8} for the proof of the previous proposition. Unraveling
definitions reveals that, in the present setting, where $F_{\beta}$
is the free energy functional we have
\[
\lim_{\beta\rightarrow\infty}\widehat{F_{\beta}}(\phi)=\widehat{E}(\phi)
\]
 iff

\[
\lim_{\beta\rightarrow\infty}\inf_{\mathcal{P}(S_{0})}F_{\phi,\beta}=\inf_{\mathcal{P}(S_{0})}E_{\phi}
\]
Thus in order to establish the Energy Approximation property, or equivalently,
the Gamma-convergence of $F_{\beta}$ towards $E,$ it is equivalent
to establish the asymptotics above for the infima of $F_{\phi,\beta}.$
for all continuous weights $\phi.$ 
\begin{rem}
Lemma \ref{lem:EA vs Gamma} still holds if the entropy $D_{\mu_{0}}(\mu)$
is replaced by any lsc functional $\tilde{D}$ on $\mathcal{P}(K)$
with the property that $\tilde{D}(\mu)<\infty$ implies that $\mu$
is absolutely continuous with respect to $\mu_{0}$ and such that
$\tilde{D}$ is finite on $L^{\infty}(K)\mu_{0}$ (using the same
proof). But in the proof of Theorem \ref{thm:determ for phi iff inf converges}
we will (implicitly) exploit that the Legendre-Fenchel transform of
$\beta^{-1}D_{\mu_{0}}$ has good monotonicity and convergence properties
with respect to $\beta.$ Indeed, as is well-known, the Legendre-Fenchel
transform of $\beta^{-1}D_{\mu_{0}}$ (extended by $\infty$ to the
space $\mathcal{M}(S_{0})$ of all signed measures on $S_{0})$ is
the functional on $\mathcal{C}(S_{0})$ defined by $u\mapsto\beta^{-1}\log\int e^{\beta u}\mu_{0},$
which increases to $\sup_{X}u$ as $\beta\rightarrow\infty.$ But
the actual proof of Theorem \ref{thm:determ for phi iff inf converges}
does not explicitly invoke the Legendre-Fenchel transform, since we
will need to allow $u$ to be non-continuous, namely of the form $\psi-\phi,$
where $\psi$ is a potential.
\end{rem}

\subsection{\label{subsec:Determining-measures-vs}Determining measures vs Gamma-convergence
of the free energies }

In this section we will assume that $\alpha\leq2$ (so that the domination
principle \ref{eq:dom princ} applies). 
\begin{lem}
\label{lem:existence of psi beta}Assume given $\mu_{0}$ in $\mathcal{P}(\R^{d})$
not charging polar subsets and of compact support $S_{0}.$ Then,
for any $\phi\in C(\R^{d})$ and $\beta\in]0,\infty[,$ the corresponding
free energy functional $F_{\phi,\beta}$ on $\mathcal{P}(S_{0})$
admits a unique minimizer $\mu_{\phi,\beta}.$ Moreover, $\mu_{\phi,\beta}=\Delta\psi_{\phi,\beta},$
where $\psi_{\phi,\beta}$ is the unique solution in $\mathcal{E}_{S_{0}}(\R^{d})$
of the following equation: 
\begin{equation}
\Delta\psi=e^{\beta(\psi-\phi)}\mu_{0}\label{eq:eq for psi phi beta}
\end{equation}
 and we have 
\begin{equation}
\inf_{\mathcal{M}(S_{0})}F_{\omega_{\phi},\beta}=\sup_{\mathcal{E}_{c}(\R^{d})}\mathcal{G}_{\phi,\beta}=\sup_{\mathcal{E}_{S_{0}}(\R^{d})}\mathcal{G}_{\phi,\beta}=\mathcal{G}_{\phi,\beta}(\psi_{\phi,\beta})\label{eq:inf beta is sup G beta}
\end{equation}
 where $\mathcal{G}_{\phi,\beta}$ is the following functional on
$\mathcal{L}_{S}(\R^{d}),$ taking values in $[-\infty,\infty[:$
\begin{equation}
\mathcal{G}_{\phi,\beta}(\psi):=\mathcal{E}_{\psi_{0}}(\psi)-\mathcal{I}_{\beta}(\psi),\,\,\,\psi_{0}:=P_{S_{0}}(\phi)\label{eq:def of G beta}
\end{equation}
where 
\begin{equation}
\mathcal{I}_{\beta}(\psi):=\beta^{-1}\log\int e^{\beta(\psi-\phi)}\mu_{0}\label{eq:def of L beta}
\end{equation}
(where we have suppressed the dependence on $\phi$ in the notation
$\mathcal{I}_{\beta}).$
\end{lem}

\begin{proof}
\emph{Step 1: }$\sup_{\mathcal{E}_{c}(\R^{d})}\mathcal{G}_{\phi,\beta}=\sup_{\mathcal{E}_{S_{0}}(\R^{d})}\mathcal{G}_{\phi,\beta}=\mathcal{G}_{\phi,\beta}(\psi_{\phi,\beta})$

To simplify the notation we will write $\mathcal{E}_{\psi_{0}}=\mathcal{E}.$
First observe that $\mathcal{I}_{\beta}(\psi)>-\infty$ on $\mathcal{L}_{c}(\R^{d}).$
Indeed, if $\mathcal{I}_{\beta}(\psi)=-\infty$ then $\mu_{0}$ charges
the polar set $\{\psi=-\infty\},$ which contradicts the assumption
on $\mu_{0}.$ Now fix any compact set $S$ containing $S_{0}$ and
consider the functional $\mathcal{G}_{\phi,\beta}$ on $\mathcal{L}_{S}(\R^{d}).$
We note that $\mathcal{G}_{\beta}$ is usc. Indeed, by Prop \ref{prop:prop of beaut E}
$\mathcal{E}$ is usc and so is $-\mathcal{I}_{\beta},$ by Fatou's
lemma. Moreover, $\mathcal{G}_{\phi,\beta}(\psi+c)=\mathcal{G}_{\phi,\beta}(\psi)$
and hence it follows from the compactness in Prop \ref{prop:compact}
that $\mathcal{G}_{\beta}$ admits a maximizer $\psi_{\beta}.$ Since
$\mathcal{I}_{\beta}(\psi)>-\infty$ we have $\psi_{\beta}\in\mathcal{E}_{S}(\R^{d}).$
All that remains is to verify that $\psi_{\beta}$ satisfies the equation
\ref{eq:eq for psi phi beta} (after perhaps shifting $\psi_{\beta}$
by a constant). To this end fix a continuous bounded function $u,$
$u\in\mathcal{C}_{b}(\R^{d}),$ and set 
\[
g(t):=\mathcal{E}(P_{S}(\psi_{\beta}+tu))-\mathcal{I}_{\beta}(\psi_{\beta}+tu).
\]
 The maximum of the function $g$ is attained at $t=0.$ Indeed, for
any $\psi\in\mathcal{E}_{S}(\R^{d})+\mathcal{C}(S_{0})$ we have that
$\mathcal{I}_{\beta}(P_{S}\psi)\leq\mathcal{I}_{\beta}(\psi),$ using
that $P_{S}\psi\leq\psi$ q.e. on $S$ and hence a.e. with respect
to $\mu_{0}$ (since $\mu_{0}$ does not charge polar subsets). As
a consequence, 
\[
\mathcal{E}(P_{S}(\psi))-\mathcal{I}_{\beta}(\psi)\leq\mathcal{E}\left(P_{S}(\psi)\right)-\mathcal{I}_{\beta}\left(P_{S}(\psi)\right):=\mathcal{G}_{\beta}(P_{S}\psi)
\]
In particular, since $P_{S}\psi\in\mathcal{E}_{S}(\R^{d}),$
\[
\sup_{\psi\in\mathcal{E}_{S}(\R^{d})+C(S_{0})}\left(\mathcal{E}(P_{S}(\psi))-\mathcal{I}_{\beta}(\psi)\right)\leq\sup_{\psi\in\mathcal{E}_{S}(\R^{d})}\mathcal{G}_{\phi,\beta}=\mathcal{G}_{\phi,\beta}(\psi_{\phi,\beta}).
\]
 Thus, restricting $\psi$ in the lhs above to be in $\psi_{\beta}+\R u$
shows that the maximum of the function $g$ is, indeed, attained at
$t=0.$ Moreover, by Prop \ref{prop:F beauti convex and diff} $g(t)$
is differentiable and hence $g'(0)=0$ shows, using that $P_{S}\psi=\psi,$
that the equation \ref{eq:eq for psi phi beta}  holds when integrated
against any $u\in\mathcal{C}_{b}(\R^{d}).$ Since a probability measure
is uniquely determined by its action on $C_{b}(\R^{d})$ this conclude
the proof of Step $1.$

\emph{Step 2:} $\inf_{\mathcal{M}(S_{0})}F_{\omega_{\phi},\beta}=\sup_{\mathcal{E}_{S_{0}}(\R^{d})}\mathcal{G}_{\phi,\beta},$
where the infimum is realized precisely at $\Delta\psi_{\phi,\beta}$

Since $E_{\omega_{\phi}}$ is convex and $D_{\mu_{0}}$ is strictly
convex on the subset $\{D_{\mu_{0}}<\infty\}\subset\mathcal{P}(S_{0})$
(by Jensen's inequality) the functional $F_{\omega_{\phi},\beta}$
has at most one minimizer. By the previous step it will thus be enough
to show that $\mu_{\phi,\beta}:=\Delta\psi_{\phi,\beta}$ minimizes
$F_{\omega_{\phi},\beta}.$ But this follows directly from the fact
that $-(\psi_{\mu}-\phi)$ is a subgradient for $E_{\phi}$ and $\log(\mu/\mu_{0})$
is a subgradient for $D_{\mu_{0}}$ (by convexity). Finally, evaluating
$F_{\omega_{\phi},\beta}$ at $\mu_{\phi,\beta}$ and using formula
\ref{eq:general formula for E normalized in terms of potential} gives
\[
F_{\omega_{\phi},\beta}(\mu_{\phi,\beta})=\mathcal{E}_{P_{S}\phi}(\psi_{\phi,\beta})-\int(\psi_{\phi,\beta}-\phi)\mu+\frac{1}{\beta}\int\log e^{\beta(\psi_{\phi,\beta}-\phi)}=\mathcal{E}_{P_{S}\phi}(\psi_{\phi,\beta}).
\]
 Finally, since $e^{\beta(\psi_{\phi,\beta}-\phi)}(=\Delta\psi_{\phi,\beta})$
is a probability measure we have that $\mathcal{I}_{\beta}(\psi_{\beta,\phi})=0$
and hence $\mathcal{E}_{P_{S}\phi}(\psi_{\phi,\beta})=\mathcal{G}_{\phi,\beta}(\psi_{\phi,\beta}),$
which concludes the proof of Step $2.$ 

\emph{Step 3:} The solution $\psi_{\phi,\beta}$ of equation \ref{eq:eq for psi phi beta}
is uniquely determined.

By the previous step $\Delta(\psi_{\phi,\beta})$ is the unique minimizer
of $F_{\omega_{\phi},\beta}$ on $\mathcal{P}(S_{0}).$ Hence, $\psi_{\phi,\beta}$
is uniquely determined up to an additive constant. But, since, as
explained in the previous step, $\mathcal{I}_{\beta}(\psi_{\beta,\phi})=0$
the constant in question vanishes.
\end{proof}
We next establish an approximate reversed Hölder type inequality for
measures $\mu_{0}$ not charging polar subsets. The result mimics
the logarithmic case, which is covered by the complex-geometric setting
in \cite[Thm 1.14]{b-b-w} and shows that $\mu_{0}$ is determining
iff $\mu_{0}$ satisfies a potential-theoretic analog of the Bernstein-Markov
inequality for polynomials:
\begin{prop}
\label{prop:det as BM}Assume that $\mu_{0}$ has compact support
$S_{0}$ and does not charge polar subsets. Then the following is
equivalent for a given continuous function $\phi:$
\begin{itemize}
\item $\mu_{0}$ is determining for $(S_{0},\phi)$
\item For all $\epsilon>0$ there exist a constant $C$ such that 
\begin{equation}
\sup_{S_{0}}e^{\psi-\phi}\leq C^{1/p}e^{\epsilon}\left\Vert e^{\psi-\phi}\right\Vert _{L^{p}(S_{0},\mu_{0})}\label{eq:transc BM}
\end{equation}
for any $\mathcal{\psi\in L}_{K}(\R^{d})$ and $p>0.$ 
\end{itemize}
As a consequence, if $K$ is compact and $(K,\phi)$ is regular then
the functional 
\[
\mathcal{L}_{K}(\psi):=\sup_{K}(\psi-\phi)
\]
 is continuous on $\mathcal{L}_{S}(\R^{d})$ for any given compact
set $S.$
\end{prop}

\begin{proof}
Given the general properties recalled in Section \ref{subsec:Potential-theoretic-preliminarie}
and the compactness result in Prop \ref{prop:compact} the proof follows,
more or less verbatim, from the proof of the corresponding result
in \cite[Thm 1.14]{b-b-w}. For completeness we provide the argument
here. 

\emph{Step 1: The functional $\mathcal{L}_{K}$ is usc on $\mathcal{L}_{K}(\R^{n})$
for any compact set $K.$}

This is shown exactly as in the case $\phi=0$ appearing in the proof
of Step 2 in Prop \ref{prop:compact}. 

\emph{Step 2: If $\mu_{0}$ does not charge polar sets, then the functional
$\mathcal{I}_{p}$ (formula \ref{eq:def of L beta}) is continuous
on $\mathcal{L}_{K}(\R^{n})$ for any $p>0.$}

If $\psi_{j}$ is a sequence of functions in $\mathcal{L}_{K}(\R^{n})$
converging in $L_{loc}^{1}$ towards $\psi,$ then, by \ref{eq:bound on sup psi in Prop comp}
and \ref{eq:limsup is psi qe}
\[
(i)\,\sup_{K}\psi_{j}\leq C,\,\,\,\,\limsup\psi_{j}=\psi\,\,\,\mu_{0}-\text{a.e}
\]
since $\mu_{0}$ does not charge polar sets. The continuity of the
functional $\mathcal{I}_{p}$ now follows from a Hilbert space argument
using convex combinations of $f_{j}:=e^{\psi_{j}-\phi},$ by repeating
the argument in the proof of \cite[Thm 1.14]{b-b-w} word by word. 

\emph{Step 3:In general, $\mathcal{I}_{p}$ is increasing in $p$
and }
\[
\lim_{p\rightarrow\infty}\mathcal{I}_{p}(\psi)=\mathcal{L}_{\infty}(\psi):=\log\left\Vert e^{\psi-\phi}\right\Vert _{L^{\infty}(S,\mu_{0})}
\]

Indeed, this follows from Hölder's inequality and standard integration
theory. 

Now,\emph{ if }$\mu_{0}$ does not charge polar sets, then, combining
Step 2 and Step 3, reveals that the functional $\mathcal{L}_{\infty}$
is lsc. If $\mu_{0}$ is moreover determining then $\mathcal{L}_{\infty}=\mathcal{L}_{S_{0}}$
and hence $\mathcal{L}_{\infty}$ is also usc continuous by Step 1
and hence continuous. To conclude the proof of the inequality \ref{eq:transc BM}
it will be enough to show that $f_{p}:=\mathcal{I}_{p}-\mathcal{L}_{\infty}$
converges uniformly to $0$ on $\mathcal{L}_{S_{0}}(\R^{d}).$ Since
$f_{p}(\psi+c)=f_{p}(\psi)$ it is enough to prove this on the subspace
of all mean-normalized $\psi.$ But since the latter space if compact
(Prop \ref{prop:compact}) the uniform convergence in question follows
from Step 3, using Dini's lemma. Conversely, if the inequality \ref{eq:transc BM}
holds, then letting $p\rightarrow\infty$ gives $\mathcal{L}_{S_{0}}\leq\mathcal{L}_{\infty}$
on $\mathcal{L}_{S_{0}}(\R^{d}),$ i.e. $\mu_{0}$ is determining
(since trivially $\mathcal{L}_{\infty}\leq\mathcal{L}_{S_{0}}).$ 

Finally, the last statement in the proposition is obtained by taking
$\mu_{0}$ to be the equilibrium measure of $(K,\phi)$ and using
Prop \ref{prop:eq meas is det}. 
\end{proof}
We note that for any measurable function $u$ on a measure space $(S,\mu_{0})$
\[
\sup_{\mu_{0}}u:=\log\left\Vert e^{u}\right\Vert _{L^{\infty}(S,\mu_{0})},
\]
 is called the \emph{essential sup} of $u$ on $(S,\mu_{0}).$ Given
a measure $\mu_{0}$ we now define the following function on $\R^{d},$
taking values in $]0,\infty]:$
\begin{equation}
(\Pi_{\mu_{0}}\phi)(x):=\sup_{\mathcal{L}_{S_{0}}(\R^{d})}\{\psi(x):\,\,\,\sup_{\mu_{0}}(\psi-\phi)\leq0\},\label{eq:def of Pi mu}
\end{equation}
 where $S_{0}$ denotes the support of $\mu_{0}.$ Its upper semi-continuous
regularization is denoted by 
\[
P_{\mu_{0}}\phi:=(\Pi_{\mu_{0}}\phi)^{*}
\]
This definition should be compared with definition of $P_{S}\phi,$
in formula \ref{eq:def of P S}. However, in general, $P_{\mu_{0}}\phi$
can be different from $P_{S_{0}}\phi$, for the support $S_{0}$ of
$\mu_{0}$ (unless $\mu_{0}$ is determining).
\begin{rem}
In the logarithmic case $d=\alpha=2$ the function $P_{\mu_{0}}\phi$
coincides with the minimal carrier Green function \cite{st-t} when
$\phi=0.$ For a general $\phi$ it coincides with the quasi-plurisubharmonic
envelope on Kähler manifolds $X$ introduced in \cite{glz} (specialized
to the case when $X$ is the Riemann sphere).
\end{rem}

\begin{lem}
\label{lem:prop of P mu}Let $\mu_{0}$ be a measure on $\R^{d}$
which does not charge polar subsets and with compact support $S_{0}$
and $\phi$ a continuous function on $\R^{d}.$ Then 
\[
P_{\mu_{0}}\phi\in\mathcal{E}_{S_{0}}(\R^{d})
\]
and 
\[
\sup_{\mu_{0}}(P_{\mu_{0}}\phi-\phi)=0
\]
\end{lem}

\begin{proof}
\emph{Step1: $\Pi_{\mu}\psi$ is locally bounded from above}

Given a large ball $B$ it is enough to show the existence of a constant
$C$ such that
\[
\delta(\psi):=\sup_{B}(\psi-\phi)-\sup_{\mu}(\psi-\phi)\leq C.
\]
 By Step 1 in the proof of Prop \ref{prop:det as BM} the first functional
in the lhs above is usc on $\mathcal{L}_{S_{0}}(\R^{d})$ for any
compact set $B.$ Moreover, as explained in the proof of Prop \ref{prop:det as BM}
the second functional is lsc for any measure $\mu_{0}$ not charging
polar subsets. This means that the functional $\delta(\psi)$ is usc
on $\mathcal{L}_{S_{0}}(\R^{d})$ and satisfies $\delta(\psi+c)=\delta(\psi)$
for any $c\in\R.$ By the compactness of the subspace of $\mathcal{L}_{S_{0}}(\R^{d})$
consisting of mean-normalized functions this yields the existence
of a constant $C$ as above.

\emph{Step 2: $P_{\mu_{0}}\phi\in\mathcal{E}_{S_{0}}(\R^{d})$ and
$\sup_{\mu_{0}}(P_{\mu_{0}}\phi-\phi)=0$}

First we recall ``Choquet's lemma'': let $\{u_{\alpha}\}_{\alpha\in A}$
be a family of real valued functions on a metric separable space $X$
(that we shall take to be $\R^{d}).$ Suppose furthermore that this
family is locally bounded from above. Then there exists a countable
subset $B$ of $A$ such that 
\[
(\sup\{u_{\beta}\}_{\beta\in B})^{*}=(\sup\{u_{\alpha}\}_{\alpha\in A})^{*},
\]
 where $\sup\{u_{\beta}\}$ denotes the function on $\R^{d}$ defined
as the point-wise sup. Thus, by Choquet's lemma, there exists a countable
family of functions $\psi_{i},$ which are candidates for the sup
defining $P_{\mu}\phi,$ satisfying 
\[
(\sup\{\psi_{i}\})^{*}=(P_{\mu}\phi).
\]
Recall that, in general, a Borel subset $S\subset\R^{d}$ is called
a\emph{ $\mu_{0}-$carrier} of a measure $\mu_{0}$ on $\R^{d}$ if
$\mu(S)=\mu(\R^{d}),$ i.e. if $\mu_{0}(\R^{d}-S)=0.$ Since $\sup_{\mu_{0}}(P_{\mu_{0}}\psi_{i}-\psi_{i})=0$
we have that $\psi_{i}\leq\phi$ on a $\mu_{0}-$carrier $S_{i}.$
Denote by $S$ the intersection of all $S_{i}.$ Then $S$ is also
a $\mu_{0}-$carrier. Take a subset $K_{\sigma}\Subset S$ which is
a union of increasing compact subsets of $S$ such that $\mu_{0}(K_{\sigma})=\mu_{0}(S)$
(the existence of $K_{\sigma}$ follows from the fact that a Borel
measure $\mu_{0}$ is, in particular, interior regular). Since $\psi_{i}\leq\phi$
on $K_{\sigma}$ we have $\psi_{i}\leq P_{K_{\sigma}}\phi.$ Moreover,
by Prop \ref{prop:P phi as equil potential and og relation}, $P_{K_{\sigma}}\phi\in\mathcal{L}_{S_{0}}(\R^{d})$
and $P_{K_{\sigma}}\phi\leq\phi$ q.e. on the $\mu_{0}-$carrier $K_{\sigma}.$
Hence, $P_{\mu}\phi\leq P_{K_{\sigma}}\phi\leq P_{\mu}\phi,$ using
in the last inequality that $P_{K_{\sigma}}\phi\leq\phi$ $\mu_{0}-$almost
everywhere, since $\mu_{0}$ does not charge polar sets. This shows
that $P_{\mu}\phi\in\mathcal{E}_{S_{0}}(\R^{d})$ and \emph{$\sup_{\mu_{0}}(P_{\mu_{0}}\phi-\phi)\leq0.$
}But then the extremal definition of $\Pi_{\mu_{0}}$ forces $\sup_{\mu_{0}}(P_{\mu_{0}}\phi-\phi)=0.$ 
\end{proof}
We shall also need the following 
\begin{lem}
Suppose that $K$ is compact and $(K,\phi)$ is regular. Then 
\begin{equation}
\inf_{\mathcal{P}(K)}E_{\omega_{\phi}}=\sup_{\psi\in\mathcal{L}_{c}(\R^{d})}\mathcal{G}_{K}=\sup_{\psi\in\mathcal{L}_{K}(\R^{d})}\mathcal{G}_{K},\label{eq:inf E as sup G in lemma}
\end{equation}
 where 
\begin{equation}
\mathcal{G}_{K}(\psi):=\mathcal{E}(\psi)-\sup_{K}(\psi-\phi).\label{eq:def of G K in lemma}
\end{equation}
Moreover, $P_{K}\phi$ is the unique maximizer of the functional $\mathcal{G}_{K}$
subject to the normalization $\sup_{K}(\psi-\phi)=0.$ Similarly,
if $\mu_{0}$ has compact support and does not charge polar subsets,
then $P_{\mu_{0}}\phi$ is the unique maximizer of the functional
\begin{equation}
\mathcal{G}_{\infty}(\psi):=\mathcal{E}(\psi)-\sup_{\mu_{0}}(\psi-\phi),\label{eq:def of G inft in lemma}
\end{equation}
 subject to the normalization $\sup_{\mu_{0}}(\psi-\phi)=0.$ 
\end{lem}

\begin{proof}
By Theorem \ref{thm:Legendre} the lhs in formula \ref{eq:inf E as sup G in lemma}
is given by $\mathcal{E}(P_{K}\phi)$ which in turn is given by $\mathcal{G}_{K}(P_{K}\phi),$
by the regularity assumption. Moreover, if $\psi\in\mathcal{L}_{c}(\R^{d})$
and $\sup_{K}(\psi-\phi)=0,$ then $\psi\leq P_{K}\phi$ (by the very
definition of $P_{K}\phi)$ and hence $\mathcal{G}_{K}(\psi)\leq\mathcal{G}_{K}(P_{K}\phi),$
since $\mathcal{E}$ is increasing (Prop\ref{prop:prop of beaut E}).
Moreover, the uniqueness in question follows from the fact that $\mathcal{E}$
is strictly increasing$.$ The corresponding results for $\mathcal{G}_{\infty}(\psi)$
are shown in a similar way, now using that $\sup_{\mu_{0}}(P_{\mu_{0}}\phi-\phi)=0,$
by the previous lemma.
\end{proof}
We are now ready for the proof of the core analytic result of the
present paper:
\begin{thm}
\label{thm:determ for phi iff inf converges}Let $\mu_{0}$ be a measure
on $\R^{d}$ which does not charge polar subsets and assume that $\mu_{0}$
has compact support $S_{0}.$ Given a continuous function $\phi$
on $\R^{d}$ the following is equivalent: 
\begin{enumerate}
\item $\mu_{0}$ is determining for $(S_{0},\phi)$ 
\item $(S_{0},\phi)$ is regular and 
\begin{equation}
\lim_{\beta\rightarrow\infty}\inf_{\mathcal{P}(S_{0})}F_{\phi,\beta}=\inf_{\mathcal{P}(S_{0})}E_{\phi}\label{eq:lim inf F =0000E4r inf E in theorem text}
\end{equation}
\item $(S_{0},\phi)$ is regular and the minimizers $\mu_{\phi,\beta}$
of $F_{\phi,\beta}$ converge weakly towards the minimizer $\mu_{(S_{0},\phi)}$
of $E_{\phi}$ as $\beta\rightarrow\infty$ (and then convergence
in energy automatically holds)
\item $(S_{0},\phi)$ is regular and the solution $\psi_{\beta,\phi}$ of
the equation \ref{eq:eq for psi phi beta} converges towards $P_{S_{0}}\phi$
in energy, as $\beta\rightarrow\infty.$ 
\end{enumerate}
\end{thm}

\begin{proof}
First note that since $F_{\phi,\beta}$ coincides with $F_{\omega_{\phi},\beta}$
up to an additive constant, which is independent of $\beta,$ it is
equivalent to prove the theorem with $F_{\phi,\beta}$ and $E_{\phi}$
replaced by $F_{\omega_{\phi},\beta}$ and $E_{\omega_{\phi}},$ respectively. 

\emph{Step 1: $1$ implies $2$ and $3$ and $4.$}

By Lemma \ref{lem:existence of psi beta}
\[
\inf_{\mathcal{M}(S_{0})}F_{\omega_{\phi},\beta}=\sup_{\mathcal{E}_{S_{0}}(\R^{d})}\mathcal{G}_{\phi,\beta}
\]
and hence, by the previous lemma, the convergence \ref{eq:lim inf F =0000E4r inf E in theorem text}
may be reformulated as
\begin{equation}
\lim_{\beta\rightarrow\infty}\sup_{\mathcal{L}_{S_{0}}(\R^{d})}\mathcal{G}_{\beta}=\sup_{\psi\in\mathcal{L}_{S_{0}}(\R^{d})}\mathcal{G}_{S_{0}}\label{eq:conv of sup G beta in pf}
\end{equation}
Now, if $\mu_{0}$ is determining for $(S,\phi_{0}),$ then, by Prop
\ref{prop:det as BM}, for any $\epsilon>0$ there exists a constant
$C$ such that
\begin{equation}
\sup_{S_{0}}(\psi-\phi)-C/\beta-\epsilon\leq\mathcal{I}_{\beta}(\psi)\leq\sup_{S_{0}}(\psi-\phi)+C/\beta\label{eq:L beta in terms of sup}
\end{equation}
Since the functional $\mathcal{E}$ is usc this immediately implies
the convergence in item $2.$ Moreover, by compactness (Prop \ref{prop:compact})
we may, after perhaps passing to a subsequence, assume that the maximizer
$\psi_{\beta,\phi}$ of $\mathcal{G}_{\phi,\beta}$ converges towards
a maximizer of $\mathcal{G}_{S_{0}}.$ Hence, by the previous lemma,
$\psi_{\beta,\phi}$ converges towards $P_{S_{0}}\phi,$ which combined
with \ref{eq:L beta in terms of sup} gives $\mathcal{E}(\psi_{\beta,\phi})\rightarrow\mathcal{E}(P_{S_{0}}\phi).$
This implies (Prop \ref{prop:prop of beaut E}) that $E_{\omega_{\phi}}(\mu_{\beta,\phi})\rightarrow E_{\omega_{\phi}}(\Delta P_{S_{0}}\phi)=E_{\omega}(\mu_{(S_{0},\phi})$
and hence $2,3$ and $4$ follow.

\emph{Step 2: $2$ implies $1$ }

First note that, since trivially, $\mathcal{I}_{\beta}\leq\mathcal{L}_{\infty}\leq\mathcal{L}_{S_{0}}$
we have that
\begin{equation}
\liminf_{\beta\rightarrow\infty}\sup_{\mathcal{L}_{S_{0}}(\R^{d})}\mathcal{G}_{\beta}\geq\sup_{\psi\in\mathcal{L}_{S_{0}}(\R^{d})}\mathcal{G}_{\infty}\geq\sup_{\psi\in\mathcal{L}_{S_{0}}(\R^{d})}\mathcal{G}_{S_{0}}\label{eq:lower bd on G beta in pf}
\end{equation}
Combined with \ref{eq:conv of sup G beta in pf} this means that if
item $2$ holds, then the inequalities above must be equalities and
hence
\begin{equation}
\sup_{\mathcal{L}_{S_{0}}(\R^{d})}\mathcal{G}_{\infty}=\sup_{\psi\in\mathcal{L}_{S_{0}}(\R^{d})}\mathcal{G}_{S_{0}}\label{eq:sup G infy is sup G S in pf}
\end{equation}
But this implies that $P_{\mu_{0}}\phi=P_{S_{0}}\phi.$ Indeed, by
definition, we have $P_{\mu_{0}}\phi\geq P_{S_{0}}\phi$ and since
$\mathcal{L}_{S_{0}}(P_{S_{0}}\phi)=0=\mathcal{L}_{\infty}(P_{\mu_{0}}\phi)$
the equality \ref{eq:sup G infy is sup G S in pf} forces $\mathcal{E}(P_{\mu_{0}}\phi)\geq\mathcal{E}(P_{S_{0}}\phi).$
By the strict monotonicity of $\mathcal{E}$ this means that $P_{\mu_{0}}\phi=P_{S_{0}}\phi.$
Since, by definition, $P_{\mu_{0}}\phi$ and $P_{S_{0}}\phi$ are
defined as the upper semi-continuous regularizations of $\Pi_{\mu_{0}}\phi$
and $\Pi_{S_{0}}\phi$, respectively, it thus follows that 
\[
\Pi_{\mu_{0}}\phi\leq(\Pi_{\mu_{0}}\phi){}^{*}=(\Pi_{S_{0}}\phi)^{*}\leq\phi
\]
 using that $(S_{0},\phi)$ is assumed regular in the last equality.
Hence, $\mu_{0}$ is determining for $(S_{0},\phi).$

\emph{Step 3: The weak convergence in item $3$ implies convergence
in energy and items $4$ and $2$}

Assume that $\mu_{\phi,\beta}$ converges towards $\mu_{(S_{0},\phi)}.$
By compactness (Prop \ref{prop:compact}) this means that there exist
constants $C_{\beta}$ such that 
\[
\psi_{\beta,\phi}+C_{\beta}\rightarrow P_{S_{0}}\phi
\]
 in $L_{loc}^{1}.$ Since $\mathcal{L}_{\beta,\phi}(\psi_{\beta,\phi})=0$
it follows that 
\[
\lim_{\beta\rightarrow\infty}C_{\beta}=\lim_{\beta\rightarrow\infty}\mathcal{L}_{\beta,\phi}(P_{S_{0}}\phi)=\mathcal{L}_{\infty,\phi}(P_{S_{0}}\phi)=\mathcal{L}_{S_{0},\phi}(P_{S_{0}}\phi)=0
\]
 using in the next to last equality  that $P_{S_{0}}\phi$ is continuous
(by Lemma \ref{lem:reg is cont}), since $(S_{0},\phi)$ is assumed
regular. Hence, $\psi_{\beta,\phi}$ converges towards $P_{S_{0}}\phi$
in $L_{loc}^{1}$ and the lower bound \ref{eq:lower bd on G beta in pf}
gives 
\[
\liminf_{\beta\rightarrow\infty}\mathcal{E}(\psi_{\beta,\phi})\geq\mathcal{E}(P_{S_{0}}\phi).
\]
 Since $\mathcal{E}$ is usc this shows that, in fact, 
\begin{equation}
\mathcal{E}(\psi_{\beta,\phi})\rightarrow\mathcal{E}(P_{S_{0}}\phi).\label{eq:conv of beaut E in pf}
\end{equation}
Hence, item $4$ holds. Now, by Lemma \ref{lem:existence of psi beta},
\[
\beta^{-1}D_{\mu_{0}}(\mu_{\beta,\phi})=\int(\psi_{\beta,\phi}-\phi)\mu_{\beta,\phi}\rightarrow\int(P_{S_{0}}\phi-\phi)\Delta(P_{S_{0}}\phi)=0
\]
 using \ref{eq:conv of beaut E in pf} in the convergence step and
the orthogonality relation \ref{eq:og relation} in the last equality.
All in all this means the weak convergence in $3$ implies the convergence
in energy of $\mu_{\beta,\phi},$ as well as the convergence of free
energies in item $2.$
\end{proof}
Finally, combining the previous theorem with Prop \ref{prop:crit for gamma conv}
(and the subsequent discussion) and Lemma \ref{lem:EA vs Gamma} we
arrive at the following result, which contains, in particular, Theorem
\ref{thm:EA vs det intr} and Theorem \ref{thm:det vs gamma intro}
stated in the introduction, except the LDP statement proved in Section
\ref{subsec:The-case-of the Ries}.
\begin{thm}
\label{thm:det vs gamma vs ea}Let $\mu_{0}$ be a measure on $\R^{d}$
which does not charge polar subsets and assume that the support $S_{0}$
of $\mu_{0}$ is compact. Then the following is equivalent: 
\begin{enumerate}
\item The measure $\mu_{0}$ is strongly determining 
\item $S_{0}$ is locally regular and $\inf_{\mathcal{P}(S_{0})}F_{\phi,\beta}\rightarrow\inf_{\mathcal{P}(S_{0})}E_{\phi},$
as $\beta\rightarrow\infty,$ for any given $\phi\in C(S_{0}).$
\item $S_{0}$ is locally regular and the functional $F_{\beta}$ converges
towards $E,$ as $\beta\rightarrow\infty,$ in the sense of Gamma-convergence.
\item $S_{0}$ is locally regular and the measure $\mu_{0}$ has the Energy
Approximation Property.
\item $S_{0}$ is locally regular and for any given $\phi\in\mathcal{C}(S_{0})$
the measures $\mu_{\phi,\beta}$ converge weakly towards $\mu_{(S_{0},\phi)},$
as $\beta\rightarrow\infty$ (and then convergence in energy automatically
holds)
\end{enumerate}
\end{thm}

\subsection{\label{subsec:Constructive-approximations-usin}Quasi-explicit approximations
using finite energy weights $\phi$}

In this section we provide a constructive procedure for obtaining
the approximation in Theorem \ref{thm:EA vs det intr}. To this end
we assume, as before, that $\alpha\leq2$ and consider\emph{ generalized
weights}
\[
\phi\in\mathcal{C}(S)+\mathcal{E}_{S}(\R^{d})-\mathcal{E}_{S}(\R^{d}).
\]
(for the construction in question it is enough to work with $\mathcal{C}(S)+\mathcal{E}_{S}(\R^{d}),$
but since it requires not extra effort we will consider the more general
setting). First recall that by basic Hilbert space duality theory
we have the following
\begin{lem}
\label{lem:duality}Let $S$ be a compact subspace of $\R^{d}.$ Then
a function $\phi$ is in $\mathcal{E}_{S}(\R^{d})$ iff $|\left\langle \phi,\mu\right\rangle |<\infty$
for all measures $\mu\in\mathcal{P}(S)$ satisfying $E(\mu)<\infty.$
Moreover, if $\phi\in\mathcal{E}_{S}(\R^{d}),$ then the functional
$\left\langle \phi,\cdot\right\rangle $ is continuous wrt the weak
topology on any sublevel set $\{E\leq C\}$ in $\mathcal{P}(S).$ 
\end{lem}

Now, given a generalized weight $\phi\in\mathcal{C}(S)+\mathcal{E}_{S}(\R^{d})-\mathcal{E}_{S}(\R^{d})$
we define the corresponding weighted energy by
\[
E_{\phi}(\mu):=E(\mu)+\left\langle \phi,\mu\right\rangle 
\]
 if $E(\mu)<\infty$ and otherwise $E_{\phi}(\mu):=\infty.$ 
\begin{lem}
Let $S$ be a non-polar subset. Then the restriction of $E_{\phi}$
to $\mathcal{P}(S)$ is lsc and strictly convex. 
\end{lem}

\begin{proof}
This is shown using Hilbert space theory exactly as in the proof of
\cite[Thm 3.21]{berm13} (which concerns the case when $d=\alpha=2).$ 
\end{proof}
In particular, if $S$ is a non-polar compact set, then $E_{\phi}$
admits a unique minimizer on $\mathcal{P}(X)$ that we shall call
denote, as before, by $\mu_{(S,\phi)}.$ Combining the previous two
lemmas yields the following generalization of the convergence in item
5 of Theorem \ref{thm:det vs gamma vs ea} to generalized weights
$\phi:$
\begin{prop}
Let $\mu_{0}$ be a measure in $\R^{d}$ with compact support $S_{0}$
such that $\mu_{0}$ does not charge polar sets and $\mu_{0}$ is
determining and assume that 
\[
\phi\in\mathcal{\mathcal{C}}(S_{0})+\mathcal{E}_{S_{0}}(\R^{d})-\mathcal{E}_{S_{0}}(\R^{d}).
\]
Then the corresponding free energies $F_{\beta,\phi}$ converge, as
$\beta\rightarrow\infty,$ to $E_{\phi}$ in the sense of Gamma-convergence.
As a consequence, the minimizer $\mu_{\beta,\phi}$ of $F_{\beta,\phi}$
converges in energy towards the weighted equilibrium measure $\mu_{(S_{0},\phi)}$
of $S_{0}.$
\end{prop}

\begin{proof}
Since $D_{\mu_{0}}\geq0$ the lower bound in the Gamma-convergence
follows directly from the lower semi-continuity of $E_{\phi}$ in
the previous lemma. The reconstruction property then follows from
the reconstruction property in the case when $\phi=0$ using the continuity
statement in Lemma\ref{lem:duality}. That is to say that any reconstruction
sequence for $E(\mu)$ is also a reconstruction sequence for $E_{\phi}(\mu).$

We thus arrive at the following constructive version of the approximation
in Theorem \ref{thm:EA vs det intr}:
\end{proof}
\begin{thm}
Let $\mu_{0}$ be a measure in $\R^{d}$ with compact support $S_{0}$
such that $\mu_{0}$ does not charge polar sets and $\mu_{0}$ is
determining. Given $\mu\in\mathcal{P}(S_{0})$ such that $E(\mu)<\infty,$
let $\mu_{\beta}$ be the minimizer of the free energy functional
$F_{\beta,\psi_{\mu}}$ on $\mathcal{P}(S_{0}),$ i.e. $F_{\beta,\psi_{\mu}}:=E_{\psi}+\beta^{-1}D_{\mu_{0}}.$
Then $\mu_{\beta}$ converges in energy towards $\mu,$ as $\beta\rightarrow\infty.$
\end{thm}

\begin{proof}
By definition $\mu=\Delta\psi_{\mu}$ and $\mu$ minimizes $E_{\psi_{\mu}}$
on $\mathcal{P}(S_{0}).$ Hence, the convergence follows directly
from the convergence in the previous proposition.
\end{proof}
This is a constructive approximation in the sense that there are quasi-explicit
ways of approximating the minimizer $\mu_{\beta,\psi}$ of $F_{\beta,\psi},$
for a given $\psi\in\mathcal{E}_{S_{0}}(\R^{d}).$ For example, when
$\psi\in\mathcal{E}_{S_{0}}(\R^{d})$ is assumed bounded on $S_{0}$
it follows from Cor \ref{cor:conv in energy of exp} below that the
measure $\mu_{N,\beta}$ defined as the expectations $\E(\delta_{N})$
of the empirical measure of the Riesz gas associated to the finite
measure $e^{-\beta\psi}\mu_{0},$ converges in energy towards $\mu_{\beta}:$
\[
\mu_{N,\beta}:=\E(\delta_{N})=\frac{\int_{(\R^{d})^{N-1}}e^{-\beta H(x,x_{2},...x_{N})}\left(e^{-\beta\psi}\mu_{0}\right)^{\otimes N-1}}{\int_{(\R^{d})^{N}}e^{-\beta H(x_{1},...x_{N})}\left(e^{-\beta\psi}\mu_{0}\right)^{\otimes N}}e^{-\beta\psi(x)}\mu_{0},
\]
 where $H$ denotes the Hamiltonian of the Riesz gas. In the general
case we can simply replace $\psi$ with $\max\{\psi,-R\}$ for a given
parameter $R>0$ and obtain $E(\mu_{\beta})$ from the double limit
where first $N\rightarrow\infty$ and then $R\rightarrow\infty.$
By a diagonal argument this yields a sequence of measures $\mu_{N,R_{N}},$
absolutely continuous with respect to $\mu_{0}$ and converging in
energy towards a given measure $\mu\in\mathcal{P}(S_{0})$ of finite
energy. This is a quasi-explicit approximation in the sense that $\mu_{N,\beta}$
is given by a quotient of two integrals, whose integrands are explicitly
given.

\subsection{\label{subsec:The-limit beta zero}The limit $\beta\rightarrow0$}

Before turning to large deviations we note that  in the opposite (infinite
temperature) limit $\beta\rightarrow0$ the minimizers $\mu_{\beta}$
always converge towards the reference measure $\mu_{0}.$
\begin{prop}
Let $\mu_{0}$ be a measure on $\R^{d}$ which does not charge polar
subsets and of compact support $S_{0}$ and assume that $\phi\in\mathcal{C}(S_{0}).$
Then $\beta F_{\beta,\phi}\rightarrow D_{\mu_{0}}$ as $\beta\rightarrow0$
in the sense of Gamma-convergence on $\mathcal{P}(S_{0}).$ In particular,
the minimizer $\mu_{\beta}$ of $F_{\beta}$ converges weakly towards
$\mu_{0}.$ 
\end{prop}

\begin{proof}
First observe that, since $D_{\mu_{0}}\leq\beta F_{\beta}:=D_{\mu_{0}}+\beta E_{\phi}$
and $D_{\mu_{0}}$ is lsc, the lower bound in the definition of Gamma-convergence
is satisfied. All that remains is thus to show that for any $\mu\in\mathcal{P}(S_{0})$
such that $D_{\mu_{0}}(\mu)<\infty$ there exists a recovery family
$\mu_{\beta}.$ In the case when $E(\mu)<\infty$ we can trivially
take $\mu_{\beta}=\mu.$ Next, note that there exists a sequence $\nu_{j}\in\mathcal{P}(S_{0})$
such that $E(\nu_{j})<\infty$ and $D_{\mu_{0}}(\nu_{j})\rightarrow D_{\mu_{0}}(\mu_{0})$
in $\mathcal{P}(S_{0}),$ as $j\rightarrow\infty.$ Indeed applying
Lemma \ref{lem:existence of psi beta} to $\phi=0$ and a fixed $\beta,$
say $\beta=1,$ shows that there exists a measure $\nu$ of finite
energy which is absolutely continuous wrt $\mu_{0}:$
\[
\nu=\rho\mu_{0},\,\,\,\rho\in L^{1}(\mu_{0})
\]
Hence, the truncated sequence $\nu_{j}:=\max(\rho,j)/\int\nu_{j}$
has the required properties. Using a truncation argument again and
the monotone convergence theorem then shows that any $\mu$ such that
$D_{\mu_{0}}(\mu)<\infty$ has the property that there exists a sequence
$\mu_{j}\in\mathcal{P}(S_{0})$ such that $E(\mu_{j})<\infty$ and
$D_{\mu_{0}}(\mu_{j})\rightarrow D_{\mu_{0}}(\mu).$ But then the
recovery property for any $\mu$ follows by a simple diagonal argument. 
\end{proof}

\section{\label{sec:Large-deviations}Large deviations}

We start with the following general setup. Let $X$ be a compact topological
space and $W$ a symmetric proper lsc function on $X\times X$ called
the\emph{ pair interaction potential.} Given  a probability measure
$\mu_{0}$ with support $X$ the corresponding \emph{Gibbs measures}
at inverse temperature $\beta_{N}\in]0,\infty[$ are defined as the
following sequence of symmetric probability measures on $X^{N}:$
\[
\mu_{\beta_{N}}^{(N)}:=\frac{1}{Z_{N,\beta_{N}}}e^{-\beta_{N}H^{(N)}}\mu_{0}^{\otimes N},
\]
 where 

\begin{equation}
H^{(N)}(x_{1},...x_{N}):=\frac{1}{(N-1)}\frac{1}{2}\sum_{i\neq j}W(x_{i},x_{j})\label{eq:def of Hamilt with W genera}
\end{equation}
and the normalization constant $Z_{N,\beta_{N}}$ is assumed to be
non-zero (it is automatically finite, since $W$ is lsc and $X$ is
compact). We also assume that the following limit exists: 
\[
\beta:=\lim_{N\rightarrow\infty}\beta_{N}\in]0,\infty]
\]
Setting
\[
E(\mu):=\frac{1}{2}\int_{X^{2}}W\mu^{\otimes2},
\]
the corresponding free energy functional $F_{\beta}$ on $\mathcal{P}(X)$
is defined as in formula \ref{eq:def of free energy with phi text}
(with $\phi=0).$ The empirical measure $\delta_{N}$ (formula \ref{eq:def of empir measur})
defines a $\mathcal{P}(X)-$valued random variable on $(X^{N},\mu_{\beta_{N}}^{(N)}).$
By definition, its law is the probability measure 
\begin{equation}
\Gamma_{N}:=(\delta_{N})_{*}\mu_{\beta_{N}}^{(N)}\label{eq:law intro}
\end{equation}
on $\mathcal{P}(X).$

We recall the general definition of a \emph{Large Deviation Principle
(LDP)} for a sequence of measures \cite{d-z}, which is modeled on
the classical Laplace steepest descent principle for integrals:
\begin{defn}
\label{def:large dev}Let $Y$ be a compact Polish space, i.e. a compact
complete separable metric space.

$(i)$ A function $I:\,Y\rightarrow]-\infty,\infty]$ is a \emph{rate
function} if it is lower semi-continuous and $\inf_{Y}I=0$

$(ii)$ A sequence $\Gamma_{N}$ of measures on $Y$ satisfies a \emph{large
deviation principle} with \emph{speed} $r_{N}$ and \emph{rate function}
$I$ if

\[
\limsup_{N\rightarrow\infty}\frac{1}{r_{N}}\log\Gamma_{N}(\mathcal{F})\leq-\inf_{\mu\in\mathcal{F}}I(\mu)
\]
 for any closed subset $\mathcal{F}$ of $Y$ and 
\[
\liminf_{N\rightarrow\infty}\frac{1}{r_{N}}\log\Gamma_{N}(\mathcal{G})\geq-\inf_{\mu\in\mathcal{G}}I(\mu)
\]
 for any open subset $\mathcal{G}$ of $Y.$ 
\end{defn}

Fixing a metric on $Y$ The LDP may also be equivalently expressed
in terms of $\Gamma_{N}(B_{\epsilon}(\mu)),$ where $B_{\epsilon}(\mu)$
denotes the closed ball of radius $\epsilon$ centered at $\mu\in Y.$
For example, if $\Gamma_{N}$ is the law of the empirical measure
$\delta_{N}$ of a random point process, then the LDP is equivalent
\cite[Theorems 4.1.11 , 4.1.18 ]{d-z} to
\[
\lim_{\epsilon\rightarrow0}\liminf_{N\rightarrow\infty}\frac{1}{r_{N}}\log\P\left(\frac{1}{N}\sum_{i=1}^{N}\delta_{x_{i}}\in B_{\epsilon}(\mu)\right)=\lim_{\epsilon\rightarrow0}\limsup_{N\rightarrow\infty}\frac{1}{r_{N}}\log\P\left(\frac{1}{N}\sum_{i=1}^{N}\delta_{x_{i}}\in B_{\epsilon}(\mu)\right)=-I(\mu)
\]
for some functional $I(\mu)$ (which, as a consequence, thus has to
be lower semi-continuous).

Given $W$ and $\mu_{0}$ and a sequence $\beta_{N}$ as above we
will say that \emph{the corresponding LDP holds at inverse temperature
$\beta$ }if the Gibbs measures $\mu_{\beta_{N}}^{(N)}$ are well-defined
and the laws $\Gamma_{N}$ of the corresponding empirical measures
on $X^{N}$ satisfy a LDP. 
\begin{thm}
\label{thm:LDP general}Assume given a proper lsc pair interaction
potential $W$ and a measure $\mu_{0}$ with compact support $X.$ 
\begin{itemize}
\item When $\beta\in]0,\infty[$ the corresponding LDP holds with speed
$\beta N$ iff the functional $F_{\beta}$ is proper lsc on $\mathcal{P}(S_{0})$
iff there exists a measure of finite energy, which is absolutely continuous
wrt $\mu_{0}.$ Then the rate functional is given by $F_{\beta}-\inf_{\mathcal{P}(X)}F_{\beta}.$
\item When $\beta=\infty$ the corresponding LDP holds with speed $N\beta_{N}$
if $F_{\beta}$ is proper lsc for all $\beta\in]0,\infty]$ and Gamma-continuous
as $\beta\rightarrow\infty.$ The rate functional is then given by
$E-\inf_{\mathcal{P}(X)}E.$
\end{itemize}
\end{thm}

\begin{proof}
This result is essentially contained in \cite{d-l-r,berm10,gz}. But
for completeness we provide some details. First observe that since
$X$ is compact $F_{\beta}$ is proper lsc iff $\inf_{\mathcal{P}(S_{0})}F_{\beta}<\infty.$
Since $E$ is bounded from below on $\mathcal{P}(S_{0}),$ the latter
condition immediately implies the existence of a measure $\mu$ of
finite energy and which is absolutely continuous wrt $\mu_{0}.$ Conversely,
if such a measure $\mu$ exists then writing $\mu=\rho\mu_{0}$ and
setting $\nu:=\max\{1,\rho\}\mu_{0}/C,$ where $C$ ensures that $\nu\in\mathcal{P}(K)$
gives $F_{\beta}(\nu)<\infty.$ Indeed, $D_{\mu_{0}}(\nu)<\infty$
and $E(\nu)<\infty,$ using that $W$ is bounded from below on $S_{0}\times S_{0}.$ 

Next, if $\inf_{\mathcal{P}(S_{0})}F_{\beta}<\infty$ then the LDP
for $\beta<\infty$ essentially follows from the results in\cite{d-l-r,berm10,gz}
(the converse is trivial since the rate functional of an LDP is proper
lsc). For completeness let us recall the argument given in \cite{berm10},
which builds on the variational approach introduced in \cite{m-s}
(see also \cite{k2,clmp} for similar results). Fix a continuous functional
$\Phi$ on $\mathcal{P}(S_{0})$ and set $H_{\Phi}^{(N)}:=H^{(N)}+N\delta_{N}^{*}\Phi,$
$F_{\beta,\Phi}:=F_{\beta}+\Phi$ and 
\[
\mathcal{F}_{\beta_{N}}^{(N)}[\Phi]:=-\frac{1}{N\beta_{N}}\log\int e^{-\beta_{N}H_{\Phi}^{(N)}}\mu_{0}^{\otimes N}
\]
Using Bryc's criterion for a LDP it is, as explained in \cite{berm10},
enough to prove that 

\begin{equation}
\lim_{N\rightarrow\infty}\mathcal{F}_{\beta_{N}}^{(N)}[\Phi]=\inf_{\mathcal{P}(S_{0})}F_{\beta,\Phi}\label{eq:lim F Phi is inf in pf LDP}
\end{equation}
Note if $F_{\beta}$ is proper lsc, then so is $F_{\beta,\Phi}.$
The starting point of the proof of the asymptotics \ref{eq:lim F Phi is inf in pf LDP}
is \emph{Gibbs variational principle }(which follows from Jensen's
inequality): 
\begin{equation}
\mathcal{F}_{\beta_{N}}^{(N)}[\Phi]=N^{-1}\inf_{\mu_{N}}\left(\int_{X^{N}}H_{\Phi}^{(N)}\mu_{N}+D_{\mu_{0}^{\otimes N}}(\mu_{N})\right)\label{eq:Gibbs var pr}
\end{equation}
It implies, using that $W$ is lsc (to handle the energy term) and
the sub-additivity of the entropy (see \cite{berm10}), the lower
bound 
\begin{equation}
\inf_{\mathcal{P}(S_{0})}F_{\beta,\Phi}\leq\liminf_{N\rightarrow\infty}\mathcal{F}_{\beta_{N}}^{(N)}[\Phi]\label{eq:lower bound on free energy Phi in pf}
\end{equation}
As for the corresponding upper bound 
\begin{equation}
\limsup_{N\rightarrow\infty}\mathcal{F}_{\beta_{N}}^{(N)}[\Phi]\leq\inf_{\mathcal{P}(S_{0})}F_{\beta,\Phi}\label{eq:upper bound of F Phi in pf LDP}
\end{equation}
 it is shown by taking $\mu_{N}=\mu^{\otimes N}$ in the rhs of formula
\ref{eq:Gibbs var pr}, where $\mu$ realizes the infimum of the proper
lsc functional $F_{\beta,\Phi}$ using that 
\[
N^{-1}\int_{X^{N}}H^{(N)}\mu^{\otimes N}+\beta^{-1}N^{-1}D_{\mu_{0}^{\otimes N}}(\mu^{\otimes N})=E(\mu)+\beta^{-1}D_{\mu}(\mu),
\]
 if $E(\mu)<\infty$ (by the Fubini-Tonelli theorem) together with
the basic fact $(\delta_{N})(\mu^{\otimes N})\rightarrow\delta_{\mu}$
weakly on $\mathcal{P}(X)$ to handle the term depending on $\Phi.$ 

Next consider the case when $\beta=\infty.$ As pointed out above,
in order to establish the LDP in question, it is enough to show that
the limit \ref{eq:lim F Phi is inf in pf LDP} also holds for $\beta=\infty.$
To this end first observe that the corresponding lower bound is easier
since the entropy term can be dropped. Moreover, to prove the corresponding
upper bound fix $\beta>0$ and note that, by Hölder's inequality,
\[
\mathcal{F}_{\beta_{N}}^{(N)}[\Phi]\leq\mathcal{F}_{\beta}^{(N)}[\Phi]
\]
 for $N$ sufficiently large. Hence, the upper bound \ref{eq:upper bound of F Phi in pf LDP}
for $\beta=\infty$ is obtained by first letting $N\rightarrow\infty,$
then using the corresponding upper bound for $\beta<\infty$ and finally
letting $\beta\rightarrow\infty$ and using the assumed Gamma-convergence
of $F_{\beta}$ towards $F_{\infty}.$ 
\end{proof}
\begin{example}
\label{exa:counter with V}Even if $W$ is assumed bounded, the LDP
may hold at $\beta=\infty$ with a rate functional which is different
than $E-\inf E.$ A simple such example is obtained by taking $X=[0,1],$
$\mu_{0}=dx$ and $W(x,y):=V(x)+V(y)$ where $V$ is the proper lsc
function defined by $V(x)=0$ for $x\neq0$ and $V(0)=-1,$ say. Since
$V=0$ a.e. wrt $dx$ we have that $\mu_{\beta}^{(N)}=dx^{\otimes N}.$
But if the LDP would hold with a rate functional $E-\inf E,$ then
\[
\lim_{N\rightarrow\infty}-\frac{1}{N\beta_{N}}\log Z_{N}=\inf_{\mathcal{P}([0,1])}E=\inf_{\mathcal{P}([0,1])}\int_{[0,1]}V\mu=\inf_{[0,1]}V=-1,
\]
 which contradicts $Z_{N}=\int_{[0,1]}dx=1.$ This example also illustrates
that the expectations of the empirical measure $\delta_{N}$ (which
here equals $dx)$ may, in general, not converge to a minimizer of
$E$ (which here equals $\delta_{0}).$ Also note that in this example,
the measure $\delta_{0}$ does not have the Energy Approximation property
(since $E(\delta_{0})=-1,$ while $E(\mu)=0$ if $\mu=\rho dx).$
Similarly, $F_{\beta}$ Gamma-converges to the constant functional
$0.$ Moreover, in this setting the measure $dx$ is not determining,
since setting $\psi:=\psi_{\delta_{0}}-1$ gives $\psi=-V$ and hence
$\psi=0$ a.e. $dx,$ while $\psi(0)>0.$ Moreover, the Bernstein-Markov
inequality fails in this example (with $\phi=0),$ as is seen by taking
$\mu=\delta_{0}.$
\end{example}

Modifying the previous example we also have the following example
involving the Coulomb gas in the plane $\C$ subject to an exterior
potential $\phi,$ showing that the corresponding LDP does not always
hold at $\beta=\infty$ if $\phi$ is lsc, but not continuous.
\begin{example}
\label{exa:counter 2D coul with weight}Set $W(z,w):=-\log|z-w|^{2}+\phi(z)+\phi(w)$
for a given function $\phi$ in $\C$ and denote by $H_{\phi}^{(N)}$
and $E_{\phi}$ the corresponding $N-$particle Hamiltonian and energy
functional, respectively. Consider the corresponding Gibbs measure
at $\beta_{N}=N,$ say, with $\mu_{0}$ given by Lebesgue measure
on a given compact subset $X$ in $\C.$ Fix a continuous function
$\phi_{0}$ on $X$ and $t\geq0$ and set $\phi:=\phi_{0}(z)-t\chi_{S^{1}}(z)$
in $\C,$ where $\chi=1$ on the unit-circle $S^{1}$ and $\chi=0$
on the complement of $S^{1}.$ Assume to get a contradiction that
\begin{equation}
\liminf_{N\rightarrow\infty}N^{-2}\log Z_{N}[\phi]\geq-\inf_{\mathcal{P}(X)}E_{\phi},\,\,\,Z_{N}[\phi]:=\frac{1}{Z_{N,\beta_{N}}}\int e^{-\beta_{N}H^{(N)}}\mu_{0}^{\otimes N}\label{eq:lower bound Z contr}
\end{equation}
 Now, since the measure $\nu$ defined by the uniform measure on $S^{1}$
is a candidate for the inf of $E_{\phi},$ the rhs in formula \ref{eq:lower bound Z contr}
is bounded from below by $-E_{0}(\nu)-\phi_{0}(r)+t$ and $E_{0}(\nu)$
is finite. But $\phi_{t}=\phi_{0}$ a.e. wrt $\mu_{0}$ and hence
$Z_{N}[\phi]=Z_{N}[\phi_{0}],$ which implies that the lhs in formula
\ref{eq:lower bound Z contr} is uniformly bounded from above by a
finite constant $C_{0}.$ Taking $t$ sufficiently large thus gives
the desired contradiction. Finally, note that the same example applies
in the non-compact case where $X=\C$ if $\phi_{0}(z)\geq(1+\epsilon)\log(|1+|z|^{2})-C$
for some $\epsilon>0$ and $C>0.$ 
\end{example}

As we will show in Section \ref{subsec:The-2d-Coulomb} the Gamma-continuity
assumption on $F_{\beta}$ is not necessary for the existence of a
LDP at $\beta=\infty$ with rate functional $F_{\beta}-\inf F_{\beta}.$
On the other hand, by the previous example it it is not enough to
assume that $F_{\infty}$ is proper lsc. In the case of the 2d Coulomb
gas, this will be illustrated using the well-known notion of Bernstein-Markov
inequalities. This notion can be extended to a general pair interaction
potential $W$ as follows (the case of the Riesz gas was introduced
in \cite{b-l-w}).
\begin{defn}
\label{def:bm general}Given $\phi\in\mathcal{C}(X)$ we will say
that a measure $\mu_{0}$ satisfies the \emph{weighted Bernstein-Markov
inequality (wrt the pair interaction $W$) if for any $\epsilon>0$
there exists a constant $C>0$ such that for any $p>0$}
\begin{equation}
\sup_{X}e^{\psi_{\mu}-\phi}\leq C^{1/p}e^{\epsilon}\left\Vert e^{\psi_{\mu}-\phi}\right\Vert _{L^{p}(S_{0},\mu_{0})}\label{eq:BM ineq in def}
\end{equation}
for all discrete measures $\mu$ of the form $\mu=N^{-1}\sum_{i=1}^{N}\delta_{x_{i}},$
for some $x_{i}\in X.$ We say that $\mu_{0}$ satisfies the \emph{strong
Bernstein-Markov property} if it satisfies the weighted Bernstein-Markov
inequality for all $\phi\in\mathcal{C}(X).$

The following result shows, in particular, that the Bernstein-Markov
inequality is a sufficient condition for the LDP to hold at zero-temperature
if $E$ is proper lsc and strictly convex (see \cite{berm 1 komma 5,b-l}
for the logarithmic case and complex case and \cite{b-l-w} for the
case of the Riesz gas).
\end{defn}

\begin{thm}
\label{thm:LDP for BM}Assume that $E$ is proper lsc on $\mathcal{P}(X).$ 
\begin{itemize}
\item If $\mu_{0}$ has support $X$ and satisfies the Bernstein-Markov
inequality, then 
\begin{equation}
-\lim_{N\rightarrow\infty}\frac{1}{N\beta_{N}}\log Z_{N}=\inf_{\mathcal{P}(X)}E\label{eq:conv of log ZN in Thm LDP for BM}
\end{equation}
and the following concentration property holds: any limit point $\Gamma$
in $\mathcal{P}(\mathcal{P}(X))$ of the law $\Gamma_{N}$ of the
empirical measure $\delta_{N}$ is supported in $\text{arg}\inf_{\mathcal{P}(X)}E.$
In particular, if $E$ admits a unique minimizer $\mu,$ then $\delta_{N}$
converges in law towards $\mu.$ 
\item If $\mu_{0}$ has the strong Bernstein-Markov property and $E$ is
strictly convex on $\mathcal{P}(X),$ then the LDP holds at a speed
$\beta_{N}N$ and with rate functional $E-\inf_{\mathcal{P}(X)}E.$
\end{itemize}
\end{thm}

\begin{proof}
Set $\Phi(\mu)=\left\langle \mu,\phi\right\rangle $ for a given $\phi\in C(X)$
and assume that $\mu_{0}$ satisfies the weighted Bernstein-Markov-inequality
for the weight $\phi.$ Then, 

\begin{equation}
\limsup_{N\rightarrow\infty}\mathcal{F}_{\beta_{N}}^{(N)}[\Phi]\leq\inf_{\mathcal{P}(X)}(E+\Phi)\label{eq:limsup of F Phi using BM}
\end{equation}

To see this, first observe that the function $\psi$ on $X$ obtained
by freezing all but one of arguments in $H^{(N)}(x_{1},x_{2},...,x_{N})$
is of the form $\psi_{\mu}$ for $\mu$ a discrete measure of the
form appearing in the definition of the Bernstein-Markov-inequality.
Hence, using the weighted Bernstein-Markov-inequality $N$ times gives
\[
\limsup_{N\rightarrow\infty}\mathcal{F}_{\beta_{N}}^{(N)}[\Phi]\leq\limsup_{N\rightarrow\infty}N^{-1}\inf_{X^{N}}H_{\Phi}^{(N)}
\]
The bound \ref{eq:limsup of F Phi using BM} now follows from the
following fact, which holds for any $\Phi\in C(\mathcal{P}(X)):$
\begin{equation}
\lim_{N\rightarrow\infty}N^{-1}\inf_{X^{N}}H_{\Phi}^{(N)}=\inf_{\mathcal{P}(X)}(E+\Phi)\label{eq:classical}
\end{equation}
This is essentially well-known and classical (a proof is provided
below). Now, combining the upper bound \ref{eq:limsup of F Phi using BM}
with  the corresponding lower bound \ref{eq:lower bound on free energy Phi in pf}
(which always holds) gives 
\begin{equation}
\lim_{N\rightarrow\infty}\mathcal{F}_{\beta_{N}}^{(N)}[\Phi]=\inf_{\mathcal{P}(X)}(E+\Phi)\label{eq:conv of F Phi in pf LDP BM}
\end{equation}
 for all linear and continuous $\Phi.$ In particular, specializing
to $\Phi=0$ proves \ref{eq:conv of log ZN in Thm LDP for BM}. To
prove the concentration property in the first point we note that the
lower bound \ref{eq:lower bound on free energy Phi in pf} can be
refined to give 
\[
\int_{\mathcal{P}(X)}E\Gamma\leq\liminf_{N\rightarrow\infty}\mathcal{F}_{\beta_{N}}^{(N)}[0]
\]
 Combining this inequality with \ref{eq:conv of F Phi in pf LDP BM}
(for $\Phi=0)$ and using that $E$ is lsc gives the concentration
property in question.

Finally, if the Bernstein-Markov-property holds for all $\phi,$ then
the asymptotics \ref{eq:conv of F Phi in pf LDP BM} hold for all
linear bounded functionals $\Phi.$ Hence, if $E$ is strictly convex
the LDP in question follows from the Gärtner-Ellis theorem (see \cite[Lemma 4.7]{berm8}
for a convenient reformulation of the Gärtner-Ellis theorem).

\emph{Proof of the asymptotics \ref{eq:classical}:}

We follow the argument in the proof of Theorem \ref{thm:LDP general}.
By \ref{eq:lower bound on free energy Phi in pf} it is enough to
prove the corresponding upper bound. To this end fix $\beta>0$ and
note that, since, $\inf_{X^{N}}H_{\Phi}^{(N)}$ is trivially bounded
from above by $\mathcal{F}_{\beta}^{(N)}[\Phi]$ the upper bound \ref{eq:upper bound of F Phi in pf LDP}
gives, for a fixed $\mu_{0}$ on $X,$
\[
\limsup_{N\rightarrow\infty}N^{-1}\inf_{X^{N}}H_{\Phi}^{(N)}\leq\inf_{\mathcal{P}(X)}\left(E_{\Phi}+\beta^{-1}D_{\mu_{0}}\right)
\]
Thus, the upper bound in question is obtained by taking $\mu_{0}$
as the minimizer of $E_{\Phi}$ (using that $D_{\mu_{0}}(\mu_{0})=0).$
We note that this proof of\emph{ \ref{eq:classical}} is closely related
to the proof of the result in \cite[Cor 1.6]{berm10}, saying that
$N^{-1}H^{(N)},$ identified with a functional on $\mathcal{P}(X),$
Gamma-converges towards $E$ (which implies\emph{ \ref{eq:classical}}
and is, in fact, equivalent to\emph{ \ref{eq:classical}} for all
$\Phi$).
\end{proof}
\begin{rem}
The proof of the first point is similar to the proof of the corresponding
result in \cite{ki-s}, which is claimed without any assumptions on
$\mu_{0}$ (see the discussion in Section \ref{subsec:General-pair-interactions}).
The main difference is that the Bernstein-Markov-property of $\mu_{0}$
is used here to justify the upper bound in \cite[Lemma 4]{ki-s},
which does not hold for a general $\mu_{0}$ (by Example \ref{exa:counter with V})
and which corresponds to\ref{eq:limsup of F Phi using BM} here. See
also \cite[Section 4]{b-l-w} for another approach based on the Bernstein-Markov-property. 
\end{rem}

\subsection{\label{subsec:The-case-of the Ries}The case of the Riesz gas }

Let us now specialize to the case of the Riesz gas, i.e. the case
when the pair interaction potential $W(x,y)$ is taken as the Riesz
kernel $W_{\alpha}$ (section \ref{sec:Weighted-potential-theory}).
The following result contains, in particular, the LDP for the Coulomb
gas $(\alpha=2)$ appearing in Theorem \ref{thm:det vs gamma intro}
in the introduction.
\begin{thm}
\label{thm:Riesz gas}Assume that $\alpha\in]0,d[.$ Given a measure
$\mu_{0}$ with compact support $S_{0},$ not charging polar subsets,
the following holds for the corresponding Riesz gas:
\begin{itemize}
\item For any $\beta\in]0,\infty[,$ the LDP holds with speed $\beta N$
and rate functional $F_{\beta}-\inf_{\mathcal{P}(X)}F_{\beta}.$ 
\item If $\alpha\leq2,$ then the LDP holds for $\beta\in]0,\infty]$ at
a speed $\beta_{N}N$ with a rate functional which is continuous wrt
Gamma-convergence iff $\mu_{0}$ is strongly determining. 
\end{itemize}
\end{thm}

\begin{proof}
Combining Theorems \ref{thm:det vs gamma vs ea}, \ref{thm:LDP general}
we just have to verify that if $\mu_{0}$ does not charge polar subsets,
then the assumption in the first point of Theorem \ref{thm:LDP general}
is satisfied. But this follows from Lemma \ref{lem:existence of psi beta},
by taking $\mu=\mu_{\beta}.$ 
\end{proof}
\begin{cor}
\label{cor:conv in energy of exp}Assume that $\alpha\in]0,d[$ and
let $\mu_{0}$ be as in the previous theorem and denote by $\mu_{N,\beta}$
the expectation of the empirical measure of the corresponding Riesz
gas. In other words, $\mu_{N,\beta}$ is the push-forward to $\R^{d}$
of the Gibbs measure defining the Riesz gas. Then $\mu_{N,\beta}$
converges in energy towards the minimizer $\mu_{\beta}$ of the free
energy functional $F_{\beta}.$
\end{cor}

\begin{proof}
The weak convergence of $\mu_{N,\beta}$ towards $\mu_{\beta}$ follows
directly form the LDP in the previous theorem. To prove that $E(\mu_{N,\beta})\rightarrow E(\mu_{\beta})$
it is, by basic integration theory, enough to show that there exists
a constant $C_{\beta}$ such that
\begin{equation}
\mu_{N,\beta}\leq C_{\beta}\mu_{0}\label{eq:ineq in applic of BM}
\end{equation}
 To prove this inequality first observe that
\[
\frac{\mu_{N,\beta}}{\mu_{0}}\leq\left\{ \sup_{S_{0}}e^{\beta\psi}:\,\,\psi\in\mathcal{L}_{S_{0}}(\R):\,\,\int e^{\beta\psi}\mu_{0}=1\right\} 
\]
using that $H(x,x_{2},...x_{N})$ is in $\mathcal{L}_{S_{0}}(\R)$
for any fixed $(x_{2},...,x_{N})$ and integrating over $S_{0}^{N-1}.$
By Prop \ref{prop:det as BM} thus shows that, for any given $\epsilon>0$
there exists a constant $C_{\epsilon}$ such that $\mu_{N,\beta}\leq C_{\epsilon}e^{\epsilon\beta}\mu_{0}$
which, in particular, implies the inequality \ref{eq:ineq in applic of BM}. 
\end{proof}
\begin{rem}
It follows from the LDP above, when $\alpha\leq2$ (and its proof),
that the functional $\mathcal{F}_{(S,\phi)}$ defined by formula \ref{eq:def of beaut F}
can be expressed in terms of the moment generating function of the
corresponding empirical measure:
\begin{equation}
\mathcal{F}_{(S_{0},\phi)}(u)=\lim_{N\rightarrow\infty}\frac{1}{\beta_{N}N}\log\E\left(e^{N\beta_{N}\sum_{i=1}^{N}u(x_{i})}\right)\label{eq:beaut F as momen generating}
\end{equation}
 In the complex-geometric setting in \cite{berm 1 komma 5} (which
covers in particular the case when $d=\alpha=2)$ the proof of the
corresponding LDP goes the other way around:\emph{ }first the analog
of \ref{eq:beaut F as momen generating} is established and then the
LDP is deduced from the Gärtner-Ellis theorem.
\end{rem}

According to Theorem \ref{thm:LDP for BM} the Bernstein-Markov-property
of a measure $\mu_{0}$ is a sufficient criterion for the LDP to hold
at $\beta=\infty.$ However, in general, the corresponding rate functional
is not Gamma-continuous up to $\beta=\infty,$ even if $\mu_{0}$
is assumed to be absolutely continuous wrt $dx.$ This will be exemplified
in the following section. 

\subsection{\label{subsec:The-2d-Coulomb}The 2d Coulomb gas and orthogonal polynomials
on the real line}

Now consider the ``logarithmic case'' $\alpha=2=d,$ i.e. the Coulomb
gas on a measure $\mu_{0}$ in $\R^{2}$ that we shall identify with
$\C.$ In this section we assume that the support $S_{0}$ of $\mu_{0}$
is contained in $\R\subset\C.$ 
\begin{lem}
Assume that $\mu_{0}$ has compact support $S_{0}$ contained in $\R.$
Then it has the Bernstein-Markov inequality iff it satisfies the weighted
Bernstein-Markov inequality for all weights $\phi$ (i.e. iff it has
the strong Bernstein-Markov property). Similarly, $\mu_{0}$ is determining
iff it is strongly determining.
\end{lem}

\begin{proof}
This is well-known, but for completeness we recall the argument. First
assume that the Bernstein-Markov inequality holds in the non-weighted
case, $\phi=0.$ Now take a general continuous function $\phi$ on
$\R.$ In a neighborhood of $S_{0}$ we can express $-\phi$ as the
uniform limit of $\log(|q_{k}|^{2})$ for some polynomials $q_{k}$
on $\C$ of degree $k$ (using the Stone-Weierstrass theorem). The
Bernstein-Markov inequality wrt $\phi$ then follows from the non-weighted
one by replacing $p_{k}$ in formula \ref{eq:BM prop intro} with
$p_{k}q_{k}.$ Similarly, if $\mu_{0}$ is determining for $(S_{0},0),$
then it is also determining for $(S_{0},\phi),$ as shown by replacing
$\psi$ in formula \ref{eq:det intro} with $\psi+k^{-1}\log(|q_{k}|^{2}).$ 
\end{proof}
Combining Theorem \ref{thm:LDP for BM} with Proposition \ref{prop:totik}
below now gives the following characterization of measures $\mu_{0}$
on $\R$ such that the corresponding LDP holds at $T=0:$ 
\begin{thm}
\label{thm:BM iff LDP on R}Let $\mu_{0}$ be a measure whose support
is a compact regular subset $S_{0}$ of $\R$ and such that $\mu_{0}$
does not charge polar subsets. Then the LDP for the corresponding
Coulomb gas at $T=0$ holds with rate functional $E-\inf E$ iff $\mu_{0}$
satisfies the Bernstein-Markov-inequality. As a consequence, it is
not enough to assume that $E$ is proper lsc (i.e. that $S_{0}$ is
non-polar) for the LDP to hold at $T=0.$ More precisely, there exists
a measure $\mu_{0}$ with support $[-1,1],$ which is absolutely continuous
wrt $dx$ and such that the corresponding expectations $\E(\delta_{N_{k}})$
do not converge towards the equilibrium measure of $[-1,1]$ when
$\beta_{N}=N-1$ and $N\rightarrow\infty.$ 
\end{thm}

\begin{proof}
To prove the ``only if'' direction we set $\beta_{N}=N-1$ (and
hence $\beta=\infty)$ and note that $p_{N}=2$ in formula \ref{eq:Gibbs as Vanderm intro}.
This means that the corresponding Coulomb gas in $\C$ defines a determinantal
point process with correlation kernel $K_{k}(z,w),$ where $K_{k}$
is the integral kernel of the orthogonal projection from $L^{2}(\C,\mu_{0})$
onto the space $\mathcal{P}_{k}(\C)$ of all polynomials $p_{k}(z)$
on $\C$ of degree at most $k:=N-1:$ 
\begin{equation}
K_{k}(z,w)=\sum_{j=0}^{k}p_{j}(z)\overline{p_{j}(w),}\label{eq:Bergman kernel in pf}
\end{equation}
for an orthonormal base $p_{j}$ in $\mathcal{P}_{k}(\C)$ (known
as the Christoffel-Darboux kernel in the literature on orthogonal
polynomials and the Bergman kernel in the complex analysis literature).
In fact, this is the case for any measure $\mu_{0}$ on $\C$ not
charging polar subsets (see, for example, \cite{berm12}). Accordingly,
it follows from general properties of determinantal point processes
that 
\begin{equation}
\E(\delta_{N_{k}})=\frac{1}{k+1}K_{k}(x,x)\mu_{0}\label{eq:Exp of delta is K}
\end{equation}
Now, if the LDP holds at $T=0$ with rate functional $E,$ then it
follows, in particular, that $\E(\delta_{N_{k}})$ converges towards
the equilibrium measure $\mu_{S_{0}}.$ But combining formula \ref{eq:Exp of delta is K}
with Prop \ref{prop:totik} below then implies that $\mu_{0}$ satisfies
the Bernstein-Markov-inequality. For the last statement it is enough
to construct a measure $\mu_{0}$ on $\R$ not charging polar subsets
and not satisfying the BM-inequality. The existence of such a measure
is without doubt well-known to experts, but for completeness a concrete
such measure is constructed in the appendix.
\end{proof}
The following proposition, used in the proof of the previous theorem,
is an unpublished result of Totik (thanks to Norman Levenberg for
pointing this out).
\begin{prop}
\label{prop:totik}Let $\mu_{0}$ be a measure whose support is a
compact regular subset $S_{0}$ of $\R$ and such that $\mu_{0}$
does not charge polar subsets. Denote by $K_{k}$ the corresponding
kernel defined by formula \ref{eq:Bergman kernel in pf}. If $\frac{1}{k+1}K_{k}(x,x)\mu_{0}$
converges weakly towards the equilibrium measure $\mu_{S_{0}},$ then
$\mu_{0}$ satisfies the Bernstein-Markov inequality. 
\end{prop}

\begin{proof}
Let us explain how to deduce this from the results in \cite{b-l-p-w}
concerning measures $\mu_{0}$ with compact support $S_{0}\subset\R.$
We denote by $p_{k}$ the sequence of orthonormal polynomials in $L^{2}(\mu_{0})$
associated to $\mu_{0}$ of degree $k,$ by $\gamma_{k}$ the positive
non-vanishing leading coefficient of $p_{k},$ i.e. $p_{k}=\gamma_{k}x^{k}+O(x^{k-1})$
and by $\nu_{k}$ the empirical measure on the zeroes of $p_{k}.$
The proposition then follows directly from combining the following
three results proved in \cite[Thm 3.2.3]{st-t} ,\cite[Thm 13.1]{si}
and \cite[Cor 2.2.3]{st-t}, respectively:
\begin{enumerate}
\item If $S_{0}$ is regular, then $\mu_{0}$ satisfies the BM-inequality
iff $\mu_{0}$ is regular in the sense of Saff-Totik i.e. 
\[
\lim_{m\rightarrow\infty}m^{-1}\log\gamma_{m}=\inf_{S_{0}}E
\]
 (the lower bound holds for any $\mu_{0})$
\item $\frac{1}{k+1}K_{k}(x,x)\mu_{0}$ converges weakly towards $\mu\in\mathcal{P}(S)$
iff $\nu_{k}$ converges weakly towards $\mu\in\mathcal{P}(S)$ 
\item If $S_{0}$ has non-zero capacity (i.e $\inf_{S_{0}}E$ is finite)
and $\nu_{k}$ converges weakly towards the equilibrium measure $\mu_{S_{0}},$
then either $\mu_{0}$ is regular or there exists a polar Borel subset
$C\subset S_{0}$ such that $\mu(C)=\mu_{0}(S_{0}).$ 
\end{enumerate}
\end{proof}
The proof of the previous proposition relies on special properties
of orthonormal polynomials on subsets of real line, not shared by
general orthonormal polynomials on subsets of $\C.$ Accordingly,
the equivalence in Theorem \ref{thm:BM iff LDP on R} is widely open
in the general logarithmic setting in $\C$ (as well as in higher
dimensions). This being said, Theorem \ref{thm:det vs gamma intro}
can be viewed as a general variant of Theorem \ref{thm:BM iff LDP on R}
where the property of being Bernstein-Markov property is replaced
by the stronger property of being determining (and then the conclusion
is also stronger). By Prop \ref{prop:det as BM} this amounts to demanding
that the Bernstein-Markov inequality \ref{eq:BM ineq in def} holds
for \emph{all} potentials $\psi_{\mu}.$ 

\subsection{Proof of Cor \ref{cor:phase intro} }

By Theorem \ref{thm:det vs gamma intro} we just have to provide a
measure $\mu_{0}$ with support $K\subset\C,$ which is absolutely
continuous wrt Lebesgue measure (and, hence does not charge polar
subsets) with the BM-property, but which is not determining. When
$K=[-1,1]$ such an example has been constructed by Totik (reported
in \cite{b-l-p-w}) and as indicated in \cite{b-l-p-w}, the general
case is similar (for completeness a proof is provided in the appendix).

\section{\label{sec:Relations-to-phase}Relations to the Ehrenfest classification
of phase transitions }

\subsection{The general setting}

Let us start by recalling the classical Ehrenfest classification of
phase transitions in a general statistical mechanical setting, where
the Hamiltonian $H^{(N)}$ in formula \ref{eq:def of Hamilt with W genera}
is replaced by a general measurable (not necessarily symmetric) function
on $(X^{N},\mu_{0}^{\otimes N})$. The corresponding \emph{free energy
at temperature $T_{N}$ }is defined by 
\[
F_{N,T_{N}}=-\frac{T_{N}}{N}\log\int_{X_{N}}e^{-\frac{1}{T_{N}}H^{(N)}}\mu_{0}^{\otimes N},
\]
assuming that it is finite. By definition, there is a \emph{phase
transition of order $m$ }at temperature $T\in]0,\infty]$ if, for
any sequence $T_{N}\rightarrow T$ the limit 
\begin{equation}
f(T):=\lim_{N\rightarrow\infty}F_{N,T_{N}}\label{eq:f of T as limit}
\end{equation}
 exists and the derivatives of order $j=1,...,m$ exist at $T,$ but
not the derivative of order $m+1.$ It should be stressed that the
notion of a phase transitions is often used in a broader sense (as
discussed in the case of the Coulomb case in \cite{st}), but here
we shall be concerned only with the Ehrenfest classification. 

We recall that phase transitions have been studied extensively in
the setting of spin models, such as the\emph{ Ising} and\emph{ Potts}
models on graphs, where the space $X$ is finite. For example, on
the complete graph with $N$ nodes the (ferromagnetic) Potts model
is defined by the Hamiltonian $H^{(N)}$ of the form \ref{eq:def of Hamilt with W genera}
with pair-interaction $W(x,y)=-x\cdot y$ and space $X=\{1,2,..q\}$
for a given integer $q\geq2,$ endowed with the counting measure $\mu_{0}.$
The case $q=2$ is the Ising model on the complete graph (known as
the \emph{Curie-Weiss model} for magnetism). As is well-known, there
is a critical critical temperature $T_{c}\in]0,\infty[$ such that
$f(T)$ is smooth for $T>T_{c}$ and a phase transition occurs at
$T=T_{c},$ which is of order two when $q=2$ and order one when $q\geq3$
\cite{w}. Moreover, according to the ``mean-field philosophy''
this implies phase transitions for the Ising and Potts model on $\Z^{d},$
when $d$ is sufficiently large \cite{b-c}.

However, by the following basic lemma, there are no zeroth-order phase
transitions when $T>0,$ i.e. no points where $f$ is discontinuous:
\begin{lem}
If the limit \ref{eq:f of T as limit} exists for any $T\in]0,\infty[,$
then $f$ is concave and increasing on $]0,\infty[$ and, in particular,
continuous. 
\end{lem}

\begin{proof}
If the limits exists then we can take $T_{N}=T$ for all $T$ and
observe that $T\mapsto F_{N,T}$ is concave and increasing (as follows,
for example, from Gibbs variational principle \ref{eq:Gibbs var pr}).
Since these properties are preserved by point-wise convergence the
lemma follows.
\end{proof}
Moreover, the following lemma explains why zeroth-order phase transitions
do not appear, even at $T=0,$ in the spin models discussed above.
\begin{lem}
For a Hamiltonian of the form \ref{eq:def of Hamilt with W genera},
with lower semi-continuous pair interaction potential $W,$ there
is no zeroth-order phase transition under the following condition:
\[
\lim_{T\rightarrow0}\inf_{\mathcal{P}(X)}\left(E+TD_{\mu_{0}}\right)=\inf_{\mathcal{P}(X)}E
\]
In particular, this is the case if the pair interaction potential
$W$ is continuous.
\end{lem}

\begin{proof}
By \ref{eq:lower bound on free energy Phi in pf} (for $\Phi=0)$
\[
\inf_{\mathcal{P}(X)}E\leq f(0).
\]
 Since $f(0)\leq f(T)$ letting $T\rightarrow0$ it follows from the
assumption that $f$ is continuous at $T=0,$ as desired. To prove
the last statement note that, since $D$ is lsc we have, in general,
that 
\[
\lim_{T\rightarrow0}\inf_{\mathcal{P}(X)}\left(E+TD_{\mu_{0}}\right)=\lim_{T\rightarrow0}\left(E(\mu_{T})+TD_{\mu_{0}}(\mu_{T})\right)\leq\liminf_{T\rightarrow0}E(\mu_{T})
\]
But if $E$ is continuous, then it follows from the compactness of
$\mathcal{P}(X)$ that the rhs above is equal to the infimum of $E.$ 
\end{proof}
Finally, we make the following observation (which applies in particular
to Riesz interactions when $\mu_{0}$ satisfies a Bernstein-Markov
inequality):
\begin{lem}
\label{lem:commute}Assume that $e^{-H^{(N)}}$is continuous on $X^{N}$
and that there exists a sequence $\epsilon_{N}$ in $\R,$ tending
to zero, such that
\[
F_{N,T_{N}}\leq\inf_{X^{N}}\frac{1}{N}H^{(N)}+\epsilon_{N}
\]
Then there is a zeroth-order phase transition at $T=0$ iff 
\begin{equation}
\lim_{T\rightarrow0}\lim_{N\rightarrow\infty}F_{N,T}\neq\lim_{N\rightarrow\infty}\lim_{T\rightarrow0}F_{N,T}\label{eq:neq}
\end{equation}
\end{lem}

\begin{proof}
By the continuity assumption $\lim_{T\rightarrow0}F_{N,T}=\inf_{X^{N}}\frac{1}{N}H^{(N)}.$
Indeed, in general, the $L^{p}(\mu_{0})-$norms of a bounded function
$f$ on a compact set $K$ converge, as $p\rightarrow\infty,$ to
the essential sup $\left\Vert f\right\Vert _{L^{\infty}(K,\mu_{0})}$
(compare Step 3 in the proof of Prop \ref{prop:det as BM}). In the
present case $f=e^{-H^{(N)}}$ is continuous and hence the essential
sup coincides with the ordinary sup. Moreover, if the inequality in
the lemma holds then necessarily $f(0)=\lim_{N\rightarrow\infty}\inf_{X^{N}}\frac{1}{N}H^{(N)}.$
Hence, the rhs in \ref{eq:neq} is equal to $f(0),$ while the lhs
is equal to $\lim_{T\rightarrow0}f(T).$ 
\end{proof}
While there is an abundance of first and second order phase transitions
in the physics and mathematics literature, zeroth-order phase transition
appear to be of a rather pathological nature. Still, there has been
some speculations on zeroth-order phase transitions in the physics
literature in the context of superfluidity (see \cite{ma}) and black
holes \cite{g-k-m}. To the best of the authors knowledge there are,
however, no previous examples of zeroth-order phase transitions in
the rigorous sense described above.

\subsection{Phase transitions for the 2d Coulomb gas}

Now consider the setting of the Coulomb gas in $\R^{2}$ with a given
exterior continuous potential $\phi$ and fix a measure $\mu_{0}$
on $\R^{2}$ which has the Bernstein-Markov property. Then the corresponding
free energy $f_{\phi}(T)$ exists for all $T\in[0,\infty[$ (by Theorem
\ref{thm:LDP for BM}). Any measure $\mu_{0}$ as in Corollary \ref{cor:phase intro}
provides an example of such a measure, for which the corresponding
Coulomb gas has a zeroth-order phase transition. 

We recall that phase transitions are also frequently studied as the
strength of $\phi$ is varied (in the standard case of spin systems
$\phi(x):=-x)$. This means that $\phi$ is replaced by 
\[
\phi_{h}:=\phi_{0}+h\phi
\]
 for a given parameter $h\in\R$ and continuous functions $\phi_{0}$
and $\phi.$ We then set
\[
f(T,h):=f_{\phi_{h}}(T)
\]
 for $(T,h)\in[0,\infty[\times\R.$ Set $T=0$ and consider the function
$h\mapsto f(T,h).$ By Prop \ref{prop:F beauti convex and diff} there
is no zeroth or first order phase transitions. A third order phase
transition was discovered by Gross-Witten in the context of lattice
gauge theories and unitary random matrices \cite{g-w} (and used in
\cite{jo-2} to study the expected length of the longest increasing
subsequence in a random permutation). This phase transition concerns
the case when $\mu_{0}$ is the invariant measure on the unit-circle
$S^{1}$ in $\C,$ $\phi_{0}=0$ and $\phi(z)$ is half the real part
of $z,$ i.e. equal to $\text{\ensuremath{\cos\theta}}$ on $S^{1}$
(the phase transition appears at $h=2).$ See \cite{ma-s} or a general
discussion about third-order phase transitions for 2d Coulomb gases.
Here we give simple examples of \emph{second order }phase transitions
on the unit-disc.
\begin{prop}
Consider the Coulomb gas in $\R^{2}$ and let $\mu_{0}$ be normalized
Lebesgue measure on the closed unit-disc $K$ and $\phi$ a non-constant
radial subharmonic function $\phi$ on a neighborhood of $K.$ Set
$\phi_{h}=h\phi.$ Then the corresponding function $h\mapsto f(0,h)$
is differentiable at $h=0,$ but not two times differentiable.
\end{prop}

\begin{proof}
In order to use standard complex analytic normalizations it will be
convenient to use a normalization where $\Delta:=\frac{1}{4\pi}(\partial_{x}^{2}+\partial_{y}^{2}).$
These normalizations ensure that $\Delta\log|z|^{2}$ is the uniform
probability measure on $S^{1}.$ First observe that without loss of
generality we may, by replacing $\phi$ by $A\phi+B$ assume that
$\int\Delta\phi\leq1$ on $K$ and $\phi=0$ on $\partial K.$ By
the maximum principle it then follows that $\phi\leq0$ in $K.$ Set
$\psi_{h}:=\log|z|^{2}$ when $|z|\geq1.$ For $|z|\leq1$ we set
$\psi_{h}=h\phi$ when $h\geq0$ and $\psi_{h}=0$ if $h<0$ and make
the following 
\[
\text{Claim:\,}\psi_{h}=P_{K}(h\phi).
\]
First observe that $\psi_{h}(z)$ is subharmonic. Indeed, writing
$\psi_{h}(z)=\Phi(x)$ for $x:=\log|z|^{2}$ the subharmonicity of
$\psi_{h}$ is equivalent to the convexity of $\Phi,$ which in turn
follows from noting that $\Phi(x)=x$ when $x\geq0$ and when $x\leq0$
we have $\partial_{x}^{2}\Phi\geq0$ and 
\[
\partial_{x}\Phi(0)=\int_{-\infty}^{0}\partial_{x}^{2}\Phi=\int_{K}\Delta\phi\leq1
\]
Moreover, this implies that, when $h\geq0,$ 
\[
\Delta\psi_{h}=(1-ch)\delta_{\partial K}+h\Delta\phi,\,\,\,c=\int_{K}\Delta\phi(=1-\partial_{x}\Phi(0))
\]
where $\delta_{\partial K}$ denotes the uniform measure on the unit-circle
$\partial K.$ Moreover, when $h<0,$
\[
\Delta\psi_{h}=\delta_{\partial K}
\]
Hence, $\psi_{h}\leq h\phi$ on $K$ and $\psi_{h}=h\phi$ almost
everywhere with respect to $\Delta\psi_{h}.$ The claim above thus
follows from the domination principle. Now, by Prop \ref{prop:F beauti convex and diff}
we have that 
\[
\frac{df(0,h)}{dh}=\int_{K}\phi\Delta\psi_{h}=\int_{|z|<1}\phi\Delta\psi_{h},
\]
 using that $\phi=0$ on the boundary of the unit-disc $K.$ By the
previous discussion this means that $\frac{df(0,h)}{dh}$ vanishes
identically when $h<0$ and is equal to $h$ times the non-zero number
$\int_{|z|<1}\phi\Delta\phi$ when $h\geq0.$ Hence, $\frac{df(0,h)}{dh}$
is not differentiable at $h=0,$ as desired. 
\end{proof}
\begin{rem}
The second order phase transition above can be contributed to the
fact that the support of the weighted equilibrium measure $\mu_{h}$
changes drastically at $h=0:$ for $h>0$ it contains a disc inside
$K,$ which disappears when $h\leq0.$
\end{rem}

In particular, if $\phi=|z|^{2},$ say, and if $\mu_{0}$ is taken
as the measure whose support is the unit-disc, provided by Corollary
\ref{cor:phase intro}, then the corresponding Coulomb gas exhibits
a rather peculiar phase diagram in the $(T,h)-$plane. Indeed, for
any fixed $h$ there is a zeroth-order phase transition as $T\rightarrow0^{+}$
and moreover, when $T=0$ there is a second order phase transition
as $h\rightarrow0.$ Let us also remark that, comparing with standard
physics terminology, the measure 
\[
\mu_{h,T}:=\lim_{N\rightarrow\infty}\E(\delta_{N})
\]
(which minimizes the corresponding free energy functional) plays the
role of an \emph{order parameter }for the phase transitions (which
usually appears as a physical observable). By Theorem \ref{thm:determ for phi iff inf converges},
a zeroth-order phase transition at $T=0$ is equivalent to the discontinuity
of $T\mapsto\mu_{T,0},$ viewed as a curve in $\mathcal{P}(S_{0}).$
Equivalently, this means that there exists some (smooth) exterior
potential $\phi$ such that the corresponding free energy $f(T,h)$
satisfies 
\[
\lim_{T\rightarrow0^{+}}\frac{\partial f(T,0)}{\partial h}\neq\frac{\partial f(0,0)}{\partial h}
\]

\section{Appendix}

\subsection{\label{subsec:Capacities-and-determining}Capacities and determining
measures}

We start by recalling the notion of (non-weighted) capacity, mainly
following \cite{la}. Given a parameter $\alpha\in]0,d[$ the corresponding\emph{
capacity }of a compact set $K\subset\R^{d}$ is defined by 

\[
\mathcal{C}_{\alpha}(K):=1/\inf_{\mu\in\mathcal{P}(K)}E(\mu),
\]
 where $E$ is the energy of $\mu.$ The\emph{ inner capacity} of
a general bounded set $S\subset\R^{d}$ is defined by 
\[
\mathcal{\mathcal{C}_{\alpha}}(S)_{*}=\sup_{K\subset S}\mathcal{\mathcal{C}_{\alpha}}(K)
\]
where the sup ranges over all compact subsets $K$ of $S.$ Similarly,
the \emph{outer capacity} is defined by 
\[
\mathcal{\mathcal{C}_{\alpha}}(S)^{*}:=\inf_{S\subset U}\mathcal{\mathcal{C}_{\alpha}}(U),
\]
 where the sup ranges over all bounded open sets $U$ containing $S.$
A bounded subset $S$ is said to be\emph{ polar} if $\mathcal{C}_{\alpha}(S)^{*}=0.$
This equivalently means that there exists a potential $\psi$ such
that $S\Subset\{\psi=-\infty\}.$ A subset $S$ is said to be \emph{capacitable
}if $\mathcal{\mathcal{C}_{\alpha}}(S)_{*}=\mathcal{C}_{\alpha}(S)^{*}.$
Any bounded Borel set $S$ is capacitable. The set functional $\mathcal{\mathcal{C}_{\alpha}}$
is \emph{invariant under translations }and satisfies Choquet's capacity
axioms on Borel sets:
\begin{itemize}
\item (monotonicity) If\emph{ $E\subset F$ then $\mathcal{C}_{\alpha}(E)\leq\mathcal{C}_{\alpha}(F).$}
\item (inner continuity) If $S_{i}$ is a sequence of sets increasing to
$S$ and $S=\bigcup_{i}S_{i},$ then $\mathcal{C}_{\alpha}(S_{i})\rightarrow\mathcal{C}_{\alpha}(S)$
\item (outer continuity). If $K_{i}$ is a sequence of compact sets decreasing
to the compact set $K,$ then $\mathcal{C}_{\alpha}(K_{i})\rightarrow\mathcal{C}_{\alpha}(K)$
\end{itemize}
Moreover, $\mathcal{C}_{\alpha}$ is\emph{ sub-additive:} given a
sequence of compact subset $K_{j}$ 
\begin{equation}
\mathcal{C}_{\alpha}(\bigcup_{j}K_{j})\leq\sum_{j}\mathcal{C}_{\alpha}(K_{j}),\label{eq:sub-ad}
\end{equation}
 assuming in the case $d=\alpha=2$ that the diameter of $K$ is at
most one \cite[Thm 5.1.4 a]{ra} (then $\mathcal{C}_{\alpha}$ is
usually called the Wiener capacity).
\begin{example}
\label{exa:capacity-of ball}The capacity of a ball $B_{r}$ of radius
$r$ centered at $x\in\R^{d}$ is given by $\mathcal{C}_{\alpha}(B_{r}(x))=A(d,\alpha)r^{d-\alpha}$
for an explicit constant $A(d,\alpha)$ \cite{la}, unless $\alpha=d=2,$
in which case $\mathcal{C}_{\alpha}(B_{r}(x))=-1/\log r.$ In particular,
$\mathcal{C}_{\alpha}(B_{r}(x))$ decreases to $0(=\mathcal{C}_{\alpha}(\{x\})$
as $r\rightarrow0$ (which is consistent, as it must with outer continuity).
\end{example}

Similarly, we define the \emph{weighted capacity} $\mathcal{C}_{\alpha}(K,\phi)$
of a compact weighted set $(K,\phi)$ by replacing the energy $E(\mu)$
with its weighted analog $E_{\phi}(\mu).$ Inner and outer weighted
capacities are then defined just as before. It follows from the previous
case $\phi=0$ that $S\mapsto\mathcal{C}_{\alpha}(S,\phi)$ satisfies
Choquet's axioms on bounded Borel sets, for any given continuous function
$\phi$ on $\R^{d}.$ 

Next recall that if $\mu$ is supported on a compact set $K,$ then
a Borel subset $C$ of $K$ is said to be a\emph{ $\mu-$carrier}
if $\mu(C)=\mu(K).$ 
\begin{prop}
\label{prop:capac crit for det}Assume that $\alpha\leq2.$ Then $\mu$
is determining for a regular compact weighted set $(K,\phi)$ iff
for any $\mu-$carrier $C$ of $K$ which is the union of increasing
compact subsets of $K,$ 
\begin{equation}
\mathcal{C}_{\alpha}(C,\phi)=\mathcal{C}_{\alpha}(K,\phi)\label{eq:cap of C is Cap of K in prop}
\end{equation}
\end{prop}

\begin{proof}
Assume first that $\mu$ is determining and let $C$ be a $\mu-$carrier
$C,$ which is the union of increasing compact subsets of $K_{i}$
of $K.$ Since $\mu$ is assumed determining we have $P_{C}\phi=P_{K}\phi$
and by the outer continuity of $\mathcal{C}$ we have $\mathcal{\mathcal{C}_{\alpha}}(K_{j},\phi)\rightarrow\mathcal{C}_{\alpha}(C,\phi).$
But, by Prop\ref{prop:P phi as equil potential and og relation} and
its proof) $P_{K_{i}}\phi$ decreases to $P_{C}\phi.$ As a consequence,
$\mathcal{C}_{\alpha}(K_{j},\phi)^{-1}:=E(\Delta(P_{K_{i}}\phi))\rightarrow E(\Delta(P_{C}\phi))=E(\Delta(P_{K}\phi))=:\mathcal{C}_{\alpha}(K,\phi)^{-1}.$
Hence, \ref{eq:cap of C is Cap of K in prop} holds. Conversely, assume
that \ref{eq:cap of C is Cap of K in prop} holds for any carrier
$C$ as above. By Lemma \ref{lem:prop of P mu} (and its proof) there
exists such a carrier $C$ with the property that $P_{C}\phi=P_{\mu}\phi.$
Hence, by the previous argument $E(\Delta(P_{\mu}\phi))$ is equal
to the infimum of $E_{\phi}$ on $\mathcal{P}(K).$ This means, by
uniqueness of minimizers, that $\Delta(P_{\mu}\phi)=\Delta(P_{K}\phi)$
and hence there exists a constant $c$ such that $P_{C}\phi+c=P_{K}\phi.$
But, then it follows from Lemma\ref{lem:prop of P mu} that $c=\sup_{\mu}(P_{K}\phi-\phi)$
and since $P_{K}\phi$ is continuous (by Lemma \ref{lem:reg is cont})
this means that $c=\sup_{K}(P_{K}\phi-\phi)=0.$ Hence, $P_{\mu}\phi=P_{K}\phi,$
which implies that $\mu$ is determining for $(K,\phi)$ (just as
in the proof of Theorem \ref{thm:determ for phi iff inf converges}). 
\end{proof}
\begin{rem}
The capacity criterion above goes back to Ullman in the case when
$d=\alpha=2$ and $\phi=0$ (see \cite[Thm 2]{uI}) and is usually
called\emph{ Ullman's criterion} in the theory of orthogonal polynomials
on the real line \cite{st-t}.
\end{rem}

\subsection{Explicit construction of the measures in Corollary \ref{cor:phase intro}
and Theorem \ref{thm:BM iff LDP on R}}

Following Totik's example for $K=[-1,1]$ (reported in \cite{b-l-p-w})
and the general discussion in \cite{b-l-p-w}, the idea of the construction
is to start with a sufficiently dense set of points on $K$ and then
replacing them by balls of sufficiently small radius, ensuring that
the corresponding measure $\mu_{0}$ is carried by a measure which
has small capacity. Since it requires no more effort we will consider
the general setting in $\R^{d}$ and the Riesz gas with $\alpha\leq2,$
using the general notion of Bernstein-Markov inequalities (definition
\ref{def:bm general}).
\begin{lem}
Let $K$ be a compact domain in $\R^{d}$ and fix $\alpha\leq2.$
For any $\phi\in\mathcal{C}(K),$ there exists a measure $\mu_{0}$
with support $K$ such that $\mu_{0}$ is absolutely continuous wrt
$dx$ and satisfies the strong Bernstein-Markov property, but $\mu_{0}$
is not determining for $(K,\phi).$
\end{lem}

\begin{proof}
We will use the following sufficient criterion for a measure $\mu$
whose support $K$ is assumed locally regular to have the strong Bernstein-Markov-property:
there exists $r_{0},a,C>0$ such that for any $z\in K$ and $r\in[0,r_{0}]$
\begin{equation}
\mu(B_{R}(z))\geq Cr^{a}\label{eq:mass crit}
\end{equation}
(see \cite{b-l-p-w} for the case $d=\alpha=2$ and \cite{b-l-w}
for the case of a general $\alpha).$ This will be contrasted with
the capacity criterion in Prop \ref{prop:capac crit for det}. Fix
a positive integer $k$ and consider the ``grid'' $K\cap(\Z k^{-1})^{d}.$
We let $\Lambda_{k}$ be the finite set contained in the interior
of $K$ obtained by removing from $K\cap(\Z k^{-1})^{d}$ all points
with distance less than $k^{-1}$ to $\partial K$ and denote by $\nu_{k}$
the empirical measure on $\Lambda_{k}.$ Next take a sequence $\lambda_{k}$
with polynomial decay such that $\sum_{k=1}^{\infty}\lambda_{k}<\infty,$
say $\lambda_{k}=k^{-2},$ and set 
\begin{equation}
\nu=\sum_{k=1}^{\infty}\lambda_{k}\nu_{k}\label{eq:decomp of nu}
\end{equation}
Then, for $k$ sufficiently large the mass criterion \ref{eq:mass crit}
is satisfied. Indeed, if $k^{-1}\leq10r$ say, then \ref{eq:mass crit}
 holds for $\mu_{k}$ with with $a=d$ and a constant $C$ independent
of $k.$ Hence, 
\begin{equation}
\nu(B_{R}(z))\geq\sum_{k^{-1}\leq10r}\lambda_{k}\nu_{k}\geq Cr^{d}\sum_{k^{-1}\leq10r}\lambda_{k}\geq C'r^{d+1},\label{eq:mass of nu}
\end{equation}
 showing that $\nu$ satisfies the mass criterion \ref{eq:mass crit}.
Next, we will modify the construction to get a measure $\mu$ not
charging polar subsets. To this end fix a sequence $\epsilon_{k}$
of positive numbers such that $\epsilon_{k}<k^{-1}$ and define $\mu_{k}$
as the measure obtained by replacing each Dirac mass at a point $x$
in the definition of $\nu_{k}$ by the normalized Lebesgue measure
on a ball or radius $\epsilon_{k},$ centered at $x.$ Equivalently,
this means that 
\[
\mu_{k}:=\int_{|s|\leq\epsilon_{k}}(T_{s})_{*}\nu_{k}ds,
\]
 where $T_{s}$ denote the translation map $x\mapsto x+s$ for a given
$s\in\R^{d}.$ We then define $\mu$ as in the decomposition in \ref{eq:decomp of nu}.
By translation invariance the same estimate \ref{eq:mass of nu} holds
for $\mu$ and hence $\mu$ satisfies the strong Bernstein-Markov-inequality,
according to the  mass criterion \ref{eq:mass crit}. Moreover, since
$\mu_{k}$ is absolutely continuous wrt Lebesgue measure so is $\mu.$
Hence, $\mu$ does not charge polar subsets and clearly has support
$K.$ Finally, we note that if $\epsilon_{k}$ is sufficiently small,
then the capacity criterion \ref{eq:cap of C is Cap of K in prop}
is not satisfied and hence $\mu$ is not determining for $(K,\phi)$.
Indeed, by construction, the set 
\[
C:=\bigcup_{k}B_{\epsilon_{k}}(\Lambda_{k})
\]
 is a carrier for $\mu,$ where $B_{\epsilon_{k}}(\Lambda_{k})$ denotes
an $\epsilon_{k}-$neighborhood of $\Lambda_{k},$ i.e a disjoint
union of $M_{k}$ balls in $K$ of radius $\epsilon_{k},$ where $M_{k}\sim k^{d}.$
After a harmless scaling we may as well assume that the diameter of
$K$ is equal to one. Then, as recalled in the previous section, $\mathcal{C}_{\alpha}$
is sub-additive  and invariant under translations. Hence, 
\[
\mathcal{C}_{\alpha}(C)\leq\sum_{k}M_{k}\mathcal{C}_{\alpha}(B_{\epsilon_{k}})\leq\sum_{k}Ck^{d}\mathcal{C}_{\alpha}(B_{\epsilon_{k}})
\]
 where $B_{\epsilon_{k}}$ the closed ball of radius $\epsilon_{k}$
centered at $0.$ Since $\epsilon\mapsto\mathcal{C}_{\alpha}(B_{\epsilon})$
strictly decreases to $0$ as $\epsilon\rightarrow0$ (see Example
\ref{exa:capacity-of ball}) this means that, given a continuous function
$\phi$ on $\R^{d}$ and a positive number $\delta>0$ we can take
$\epsilon_{k}$ sufficiently small to ensure that $\mathcal{C}_{\alpha}(C,\phi)<\delta.$
In particular, taking $\delta=\mathcal{C}_{\alpha}(K,\phi)$ the capacity
criterion \ref{eq:cap of C is Cap of K in prop} is violated and hence
$\nu$ is not determining for $(K,\phi).$
\end{proof}
A similar construction yields the following
\begin{lem}
There exists a measure $\mu_{0}$ with support $[0,1]$ such that
$\mu_{0}$ is absolutely continuous wrt $dx,$ but which does not
satisfy the Bernstein-Markov-inequality.
\end{lem}

\begin{proof}
We recall the following necessary condition for a measure $\mu_{0}$
on $[0,1]$ to satisfy the Bernstein-Markov-inequality \cite[Thm 4.2.8]{st-t}:
for any $\eta>0$
\begin{equation}
\lim_{r\rightarrow\infty}\mathcal{C}_{\alpha}\left(\left\{ x\in[0,1]:\,\mu(B_{r}(x))\geq e^{-\eta r^{-1}}\right\} \right)=\mathcal{C}_{\alpha}([0,1])\label{eq:nec cond for bm}
\end{equation}
Denote by $k$ an integer of the form $k=2^{m}$ for some positive
integer $m.$ We will show that the necessary condition above is not
satisfied for a measure of the form
\[
\mu=\sum_{k}\lambda_{k}\mu_{k}
\]
 as defined in the previous construction, if $\epsilon_{k}$ and $\lambda_{k}$
are both taken sufficiently small. More precisely, we will show that
this happens if $\lambda_{k}=e^{-\epsilon_{k}^{-2}}$ and $\epsilon_{k}$
satisfies
\[
k\mathcal{C}_{\alpha}(B_{2\epsilon_{k}})<\mathcal{C}_{\alpha}([0,1])/2,
\]
 say. To see this fix $r>0$ and first note that for any $x\in[0,1]$
\begin{equation}
\left(\sum_{k:\,\epsilon_{k}\leq r}\lambda_{k}\mu_{k}\right)(B_{r}(x))\leq Ce^{-r^{-2}/2}\label{eq:pf of constr not bm}
\end{equation}
Next, consider the set $A_{k}$ defined as an $\epsilon_{k}-$neighborhood
of the support $B_{\epsilon_{k}}(\Lambda_{k})$ of $\mu_{k}$ (which
contains the support of $\mu_{j}$ for $j\leq k),$ i.e. $A_{k}=B_{2\epsilon_{k}}(\Lambda_{k}).$
The definition is made so that, if $r<\epsilon_{k}$ then the $r-$neighborhood
of $([0,1]-A_{k})$ does not intersect $B_{\epsilon_{k}}(\Lambda_{k}).$
Hence, 
\[
x\in([0,1]-A_{k})\implies\left(\sum_{k:\,\epsilon_{k}>r}\lambda_{k}\mu_{k}\right)(B_{r}(x))=0,
\]
which, combined with \ref{eq:pf of constr not bm}, means that the
inequality in condition \ref{eq:nec cond for bm} fails when $x\in[0,1]-A_{k}.$
But if $\epsilon_{k}$ is sufficiently small, then we get, by the
sub-additive of the capacity (just as in the previous construction)
that 
\[
\mathcal{C}_{\alpha}\left(A_{k}\right)<\mathcal{C}_{\alpha}([0,1])/2
\]
say, for all $k.$ Hence the capacity condition \ref{eq:cap of C is Cap of K in prop}
is violated, showing that $\mu$ does not satisfy the Bernstein-Markov-inequality. 
\end{proof}

\end{document}